\documentclass[11pt]{article}
\usepackage[utf8]{inputenc}
\usepackage[T1]{fontenc}
\usepackage{lmodern}
\usepackage[letterpaper,margin=1in]{geometry}
\usepackage{amsmath,amssymb}
\usepackage{graphicx,booktabs,array,longtable,float,needspace}
\usepackage[font=small,labelfont=bf]{caption}
\usepackage[numbers,sort&compress]{natbib}
\usepackage{microtype}
\usepackage{placeins}
\usepackage{xurl}
\usepackage[hidelinks]{hyperref}
\usepackage{fancyhdr}
\title{\LARGE\bfseries Predicting Cognitive Tasks from Single-Channel EEG through Task-Semantic Alignment}
\author{Ryota Kanai\\[0.3em]Araya, Inc.}

\date{10 October 2026}
\hypersetup{pdfauthor={Ryota Kanai},pdftitle={Predicting Cognitive Tasks from Single-Channel EEG through Task-Semantic Alignment},pdfsubject={Completed-study preprint archived version; no new external-model or controlled-data experiments},pdfkeywords={EEG,task-semantic alignment,zero-shot,few-shot,transfer,data integration}}
\begin{document}
\maketitle
\vspace{-1.2em}
\begin{abstract}
Studies of data scaling in electroencephalography (EEG) and related noninvasive recordings suggest that simple wearable devices could eventually reveal more about cognitive states. However, publicly available EEG data are difficult to combine. Different laboratories record different tasks, participants, and devices. Task-agnostic self-supervised learning can exploit such data, but representations learned only with EEG signals have not consistently produced feature representations that are directly effective for decoding the cognitive state. We therefore aimed to incorporate task information by embedding experimental instructions, stimuli and response demands in a common semantic space, and trained an EEG encoder to align signals from a single anterior--posterior differential channel with those task vectors. This framework supports zero-shot prediction without target-condition calibration and few-shot prediction using labeled samples. On a reserved ten-class evaluation, the highest observed zero-shot balanced accuracy was 30.13\% (chance 10\%). In a separate developmental few-shot experiment, balanced accuracy reached 34.47\% using 16 labeled EEG examples per class from another participant, and 74.99\% using examples from the test participant's own task recording. Our experiments suggest that training models on datasets with multiple tasks per participant appeared useful, whereas adding datasets containing only one or a few tasks sometimes reduced established performance despite broader task coverage. These exploratory results suggest that EEG datasets that repeat varied cognitive tasks and everyday activities within the same individuals across days, devices and independent studies would be particularly useful for training an EEG foundation model for predicting cognitive states.
\end{abstract}\noindent{\small\textit{Keywords: electroencephalography; task-semantic alignment; zero-shot decoding; few-shot calibration; data integration; cognitive tasks.}}\par\medskip

\section*{1. Introduction}\addcontentsline{toc}{section}{1. Introduction}

The amount of neural data available for machine learning is increasing, and recent decoding studies in electroencephalography (EEG) show measurable benefits from larger training sets. Sato and colleagues recorded 175 hours of speech-production EEG from one person: in their zero-shot speech-segment retrieval task, Top-1 accuracy increased from approximately 2.5\% with about ten training hours to 48\% with the larger dataset \citep{sato2024speech}. A subsequent report extended the recording and calibration analysis \citep{sato2026speech}. Inoue and colleagues showed that heterogeneous electrode configurations can contribute to speech decoding \citep{inoue2025heterogeneous}, then examined training-data expansion for a wearable, around-ear EEG system \citep{inoue2026aroundear}. Meta's research has similarly combined public electroencephalography and magnetoencephalography recordings to decode perceived speech \citep{defossez2023speech}, and studied typed-sentence decoding using EEG and magnetoencephalography \citep{levy2026typed}. A recent magnetoencephalography-only extension reports further gains with recording volume \citep{zhang2026megscaling}. Related scaling trends have been documented for decoding visual stimuli from brain activity \citep{banville2025scaling}. These findings motivate the possibility that once powerful foundation models are established for EEG, simpler, consumer-oriented devices may also become more useful for decoding the cognitive states from EEG. 

The principal obstacle to increasing task-labeled EEG data is heterogeneity. Publicly available EEG datasets were acquired for different scientific questions, with different participants, electrodes, reference choices and environments. Their labels also describe different things: a stimulus type, a response rule, a memory operation or a particular period of a task. Consequently, pooling waveforms and treating all study-specific labels as a single categorical target does not produce a coherent learning problem. The Cognitive Atlas distinguishes experimental tasks from the psychological constructs they are intended to engage; text--neuroimaging resources describe related cross-study associations \citep{poldrack2011atlas,yarkoni2011,dockes2020neuroquery}.

In the present study, we use \emph{task semantics} to make these differences explicit. Each source task is described using its instructions, sensory material, required response and evaluated phase. A sentence encoder was used for mapping the detailed and structured descriptions to a shared task vector. An EEG encoder is then trained to align the corresponding signal segments with these vectors. Given a bank of candidate descriptions, the same encoder can rank tasks without fitting a new classifier for every source. We call this \emph{zero-shot prediction} when no labeled EEG from the target evaluation condition is used to calibrate the readout. This does not necessarily imply that the participant, study and cognitive operation are all unfamiliar. We examine those forms of novelty separately. \emph{Few-shot prediction} uses a small set of labeled EEG examples to adapt the task readout. Our few-shot experiments test utilized data from other participants or from the test participant's own recording to adapt the task readout for specific test datasets or participants.

Self-supervised learning (SSL) offers a complementary route to integrating EEG without relying on task annotations. BENDR, LaBraM, CBraMod and REVE demonstrate the value of pretraining under different downstream protocols \citep{kostas2021,jiang2024labram,wang2025cbramod,elouahidi2025reve}. SingLEM encodes each channel separately and combines the resulting features in downstream classification \citep{sukhbaatar2026singlem}. However, the advantage of frozen representations under simple linear probing varies across datasets and adaptation methods \citep{yang2026worth,liu2026compass}. Work on speech decoding from magnetoencephalography demonstrates that self-supervised representations can also transfer across people and recording contexts \citep{jayalath2025speechssl}. Information retained by a signal-level objective need not be the information required to separate cognitive tasks; recent work also documents spectral and participant-related biases in some reconstruction-based models \citep{kommineni2026bias}. This is a motivation to test explicit task supervision, rather than evidence that self-supervised approaches are ineffective.

Our initial masked-latent SSL pilots used approximately 585 hours of unique EEG recordings. A supervised ten-task evaluation changed from 34.41\% to 35.41\% balanced accuracy after pretraining, whereas a separate six-task evaluation changed from 20.86\% to 20.09\%. These pilot results are summarized in Supplement S0. Since we did not obtain clear benefits from pretraining with SSL, we trained the task-semantic models from fresh initialization, while recognizing that signal-derived and task-derived objectives could also be combined \citep{an2026stem}.

Language provides a practical interface for this supervision. Contrastive Language--Image Pre-training (CLIP) learned image representations from natural-language supervision \citep{radford2021clip}. EEG-CLIP uses clinical descriptions; NeuroLM, LEAF and EEG-PRIME use language alignment or instruction to support different EEG decoding settings \citep{ndir2025eegclip,jiang2025neurolm,jiang2026leaf,zhang2026prime}. The EEG Foundation Challenge explicitly tests transfer across tasks and people \citep{aristimunha2025challenge}. Our focus is the integration of diverse cognitive-task protocols under a physical one-channel constraint. Table 1 distinguishes this problem from these related approaches.

For this research, the key question is what kind of additional training data improves transfer. More EEG hours, more people, new task types and repetitions of the same task in different laboratories are distinct resources. A multi-task experiment can provide several task contrasts from the same participants under approximately shared acquisition conditions. By contrast, a study devoted to a single task may expand the vocabulary of operations represented during training, but task identity can also become inseparable from that study's participants and recording setup. We therefore tested both an expanded task vocabulary and mixtures anchored in multi-task datasets. This approach asks whether cross-study task-semantic learning can improve zero-shot performance, and whether the representation is useful when a few labeled target-context examples are available. The experiments investigate these hypotheses without fitting a universal EEG scaling exponent.

A common output space also creates an interpretive problem. If each study contributes a unique label, source recognition may be sufficient for correct classification. Within a study, task blocks can differ in stimuli, responses and recording time \citep{li2021perils}. Decoding a label consequently need not identify a selective cognitive process \citep{poldrack2006,geirhos2020}. Candidate-set composition, model selection and repeated inspection of evaluation data further affect the conclusion \citep{xian2019,gulrajani2021,koh2021wilds,cawley2010,damour2022,dwork2015}. We examine both the gains obtained by pooling tasks and the distinctions those gains support. We report developmental scaling and calibration results, a model panel fixed before scoring, and subsequent diagnostic analyses. 

\Needspace{105pt}
\par\medskip\noindent{\small\textbf{Table 1. Prior work and the transfer question actually evaluated.}}
{\fontsize{8.8}{11.2}\selectfont\setlength{\tabcolsep}{3pt}\renewcommand{\arraystretch}{1.18}
\begin{longtable}{>{\raggedright\arraybackslash}p{94.25pt}>{\raggedright\arraybackslash}p{133.25pt}>{\raggedright\arraybackslash}p{224.25pt}}
\toprule
\textbf{Work} & \textbf{Relevant contribution} & \textbf{Relation to novelty axes / present scope} \\
\midrule\endfirsthead
\multicolumn{3}{l}{\textit{Table 1 (continued)}}\\
\toprule
\textbf{Work} & \textbf{Relevant contribution} & \textbf{Relation to novelty axes / present scope} \\
\midrule\endhead
\midrule\multicolumn{3}{r}{\textit{Continued on next page}}\\\endfoot
\bottomrule\endlastfoot
BENDR; LaBraM; CBraMod; REVE \citep{kostas2021,jiang2024labram,wang2025cbramod,elouahidi2025reve} & Reusable EEG representations; diverse adaptation protocols. & Task-specific probing or adaptation is different from a fixed text-candidate decoder. No head-to-head comparison is claimed here. \\
SingLEM \citep{sukhbaatar2026singlem} & Each EEG channel is encoded independently; downstream classifiers may concatenate channels. & Leakage-controlled pretrained representation evaluated with supervised leave-one-subject-out classification. Single-channel input is not unique to this study. The evaluation encoder was pretrained on 68 datasets, excluding the three downstream source datasets; a separate 71-dataset encoder is offered for general reuse. \\
EEG-CLIP; NeuroLM \citep{ndir2025eegclip,jiang2025neurolm} & EEG--text alignment and language-based multitask interfaces. & Clinical-description or instruction-transfer questions; not automatically equivalent to an unobserved cognitive-operation test. \\
LEAF \citep{jiang2026leaf} & Instruction-conditioned EEG features aligned with label-text representations. & Instruction effects on representation geometry motivate text-side diagnostics; reported downstream protocols must be retained. \\
EEG-PRIME \citep{zhang2026prime} & Masked pretraining followed by prototype-aligned instruction tuning; held-out motor-imagery evaluation. & Closest to unseen-study transfer at the broad task-family level for the held-out motor-imagery experiments. Exact definitions and input conditions differ. \\
EEG Foundation Challenge \citep{aristimunha2025challenge} & Cross-task and cross-subject evaluation using Healthy Brain Network EEG. & Prior formulation of these transfer axes; behavioral targets and task banks differ from the present mixed operational labels. \\
STEM \citep{an2026stem} & Subject/task supervision integrated with contrastive self-supervision. & Supports using available task information; does not test this paper's physical input or task-description bank. \\
EEG-FM-Compass; Yang et al. \citep{yang2026worth,liu2026compass} & Systematic evaluation of pretrained EEG models and conventional decoders. & Motivates protocol-matched controls. The current nine-condition panel is not a comparison of these public models. \\
Araya speech EEG/EMG \citep{sato2024speech,sato2026speech,inoue2025heterogeneous,inoue2026aroundear} & Larger within-person recordings, cross-device speech training, and around-ear speech decoding. & Strong motivation for data scaling and wearability; speech segments and word vocabularies are different targets from cross-study cognitive-task identification. \\
Meta language decoding \citep{defossez2023speech,levy2026typed,zhang2026megscaling} & Speech perception across public datasets and typed-sentence decoding from EEG/MEG; later MEG scaling. & Demonstrates the value of alignment and larger datasets, but does not establish this study's one-channel or unseen-task results. \\
Brain-image scaling \citep{banville2025scaling} & Empirical scaling in stimulus-image decoding across EEG, magnetoencephalography and functional magnetic resonance imaging. & Different target from pooling instructed cognitive tasks; scaling in EEG is not claimed as a first. \\
\end{longtable}}
\vspace{-0.6em}\noindent{\footnotesize Comparisons concern evaluation protocols, not numerical performance across different datasets. No public-model benchmark is reported here.}\par\medskip

\section*{2. Study scope and evaluation structure}\addcontentsline{toc}{section}{2. Study scope and evaluation structure}

\subsection*{2.1 The task space is the common interface}

Each source task has a free-form \emph{rich description} and a field-structured \emph{canonical description}. The all-MiniLM-L6-v2 sentence encoder maps each to 384 coordinates \citep{reimers2019,wang2020minilm,allminilmcard}. The primary \emph{centroid readout} unit-normalizes the two vectors, sums them and normalizes the result. A learned projection maps the EEG representation into the same space; prediction selects the candidate with greatest cosine similarity. Methods 5.2--5.5 give the transformations and objectives.

The task register links source-native labels through \emph{verified functional task} (VFT) identifiers. These are protocol-based equivalence assignments, not validated latent cognitive factors. They determine which EEG examples share a target during training. The continuous task vectors additionally define relations among targets and permit a new candidate bank for inference. Dataset identifiers serve as provenance fields rather than intended input to the text encoder.

\begin{figure}[!htbp]
\centering
\includegraphics[width=\linewidth,height=0.62\textheight,keepaspectratio]{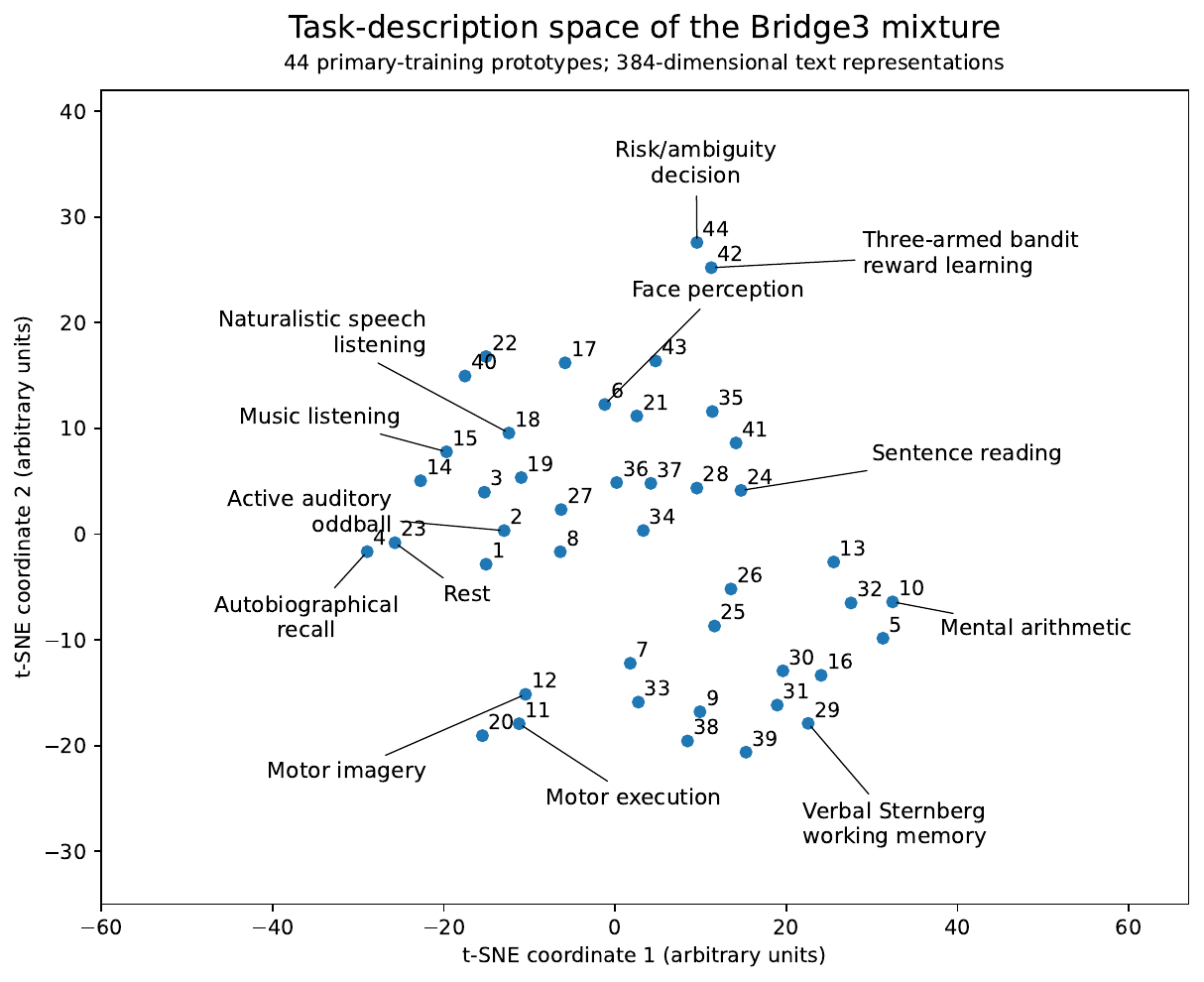}
\caption{\textbf{Task descriptions define the common output space.} The 44 represented primary-training functional-task prototypes of the Bridge3 mixture, projected from 384 dimensions using t-SNE (cosine distance, perplexity 10, random initialization, seed 42). Each point is the normalized rich/canonical centroid for one task. Numbers identify the task names in Table S13. The original-space similarity matrix is shown in Figure S8.}
\label{fig:taskspace}
\end{figure}

Figure~\ref{fig:taskspace} displays the 44 primary-training task prototypes represented in the Bridge3 mixture, defined in Section 2.2, from its frozen bank using t-distributed stochastic neighbor embedding (t-SNE). The original-space cosine similarities include 0.918 for motor execution and motor imagery, 0.960 for verbal memory manipulation and retention, and 0.884 for active and passive auditory oddball tasks. These values describe text representations: proximity does not establish neural similarity or successful EEG discrimination. The full cosine matrix, nearest-neighbor table and projection sensitivity analysis are in Supplement S8. The coordinates are used only for display.

\subsection*{2.2 Model names and input conventions}

The input is an \emph{anterior--posterior two-electrode difference} (AP2E): one signal from two actually recorded scalp electrodes. Primary inputs last eight seconds. \emph{RAW} denotes the waveform encoder; \emph{LATE} denotes late fusion of waveform and multi-resolution spectral features before projection into the task space. Both use the same one-channel input.

The experiments below used several training configurations, each designed to address a different question about how to build a generalizable EEG model. They differ either in the composition of the training data, the amount of training, or the method used to align EEG representations with
semantic descriptions of cognitive tasks.

We first investigated whether adding datasets covering a wider range of cognitive tasks would improve generalization to unseen studies and tasks. The \emph{ST19-MATCHED} configuration (abbreviated as ST19 in figures) combined an existing core of heterogeneous EEG recordings, including studies in which the same participants performed multiple tasks, with 19 additional datasets, many containing only one or a few tasks. This configuration comprised 38 dataset identifiers in total. 

We then extended this approach to \emph{ST40}, which included 40 additional datasets relative to the original core, giving 59 dataset identifiers. The motivation was to increase the diversity of cognitive tasks represented during training. However, adding datasets in which each study contributes only a small number of tasks can also strengthen the association between task identity and study-specific recording characteristics. The labels ST19 and ST40 refer to the number of added
datasets.

A separate \emph{Full-Mixture} (FM) experiment addressed a different question of whether performance would continue to improve with more optimization when the available training data remained unchanged. Instead of progressively adding datasets, this experiment used a fixed mixture of tasks and studies and extended the training duration. This allowed us to examine the effect of additional optimization steps without simultaneously changing the composition of the training data.

The \emph{Bridge3} configuration was designed to explore a more selective alternative to broad dataset expansion. Starting from the ST19-MATCHED mixture, which had shown promising performance during development, we added three datasets chosen to extend cognitive-task coverage while limiting disruption to the original training distribution. During training, approximately 95\% of primary EEG
samples were drawn from the ST19-MATCHED mixture and 5\% from the three additional datasets. Bridge3 therefore tested whether a small, targeted expansion could retain the benefits of the original multi-task-rich mixture while introducing additional task types.

Finally, we compared two approaches to learning EEG representations using the Bridge3 data. In the
\emph{fixed-text} approach, semantic task vectors derived from task descriptions remained fixed, and the EEG encoder was trained to map EEG segments into this predefined task space. In the \emph{CLIP-X} approach, the same Bridge3 primary-data mixture and native four-second auxiliary data were used, but the semantic representations were also adapted during training. Specifically, CLIP-X introduced a residual text adapter, a learned similarity scale, and regularization terms encouraging preservation of task-space geometry and consistency across studies. Both approaches used bidirectional, multi-positive contrastive alignment between EEG and task representations;
CLIP-X was therefore a broader modification of the learning procedure, rather than simply the introduction of a contrastive objective. The suffix X denotes the additional cross-study consistency term in the original implementation.

Throughout the paper, checkpoint labels 16k, 64k, and 256k indicate 16,000, 64,000, and 256,000
optimizer updates, respectively.

\subsection*{2.3 Development, reserved evaluation and source structure}

We began by developing and comparing models using two publicly available EEG datasets that contain several experimental task conditions. The first, EEG-Asymmetries (ds007172) \citep{data_ds007172,reinke2025asymmetries}, includes different perceptual and cognitive tasks. The second, ERP CORE \citep{kappenman2021}, contains recordings from several established experimental paradigms. Together, these datasets formed our initial development benchmark, which we call \emph{D1}. Using datasets with multiple task conditions was particularly useful because it allowed us to compare EEG recorded during different tasks within the same data source, rather than relying entirely on differences between studies. Early evaluations distinguished five conditions from EEG-Asymmetries and six from ERP CORE. In later experiments, we combined two closely related face-target conditions into one category, giving ten task classes in total: four from EEG-Asymmetries and six from ERP CORE.

Although these two datasets provided a useful starting point, they covered only a limited range of cognitive tasks. To test whether the models could generalize to a broader variety of activities, we assembled a second development benchmark from six additional studies. Each study contributed one evaluated task, including activities not represented in the original benchmark. We call this extension \emph{D1-ST}, where ST refers to the single-task sources used in this evaluation. Including these studies allowed us to test more diverse cognitive tasks, but also introduced an important difficulty: because each task came from a different source, task identity could be confounded with differences in participants, equipment, or recording conditions. We therefore used two versions of this benchmark. \emph{BROAD6} included all six tasks, whereas \emph{CLEAN4} was restricted to the four tasks that satisfied stricter criteria for selecting valid EEG intervals. The tasks and selection criteria are described in Table~S2, and their source references are provided in Table~S16.

Both of these benchmarks were used during model development. We compared different architectures, training datasets, and learning procedures on them, and used the results to guide subsequent experiments, including the selection of training data for Bridge3. 

To obtain an evaluation that was not used for these development decisions, we then assembled a separate benchmark, which we call \emph{D2}. It contained ten cognitive-task classes drawn from seven dataset identifiers. The final evaluation comprised 87,058 nonoverlapping eight-second EEG segments and 10,597 eligible 64-second groups. Participant counts reported across the seven sources sum to 290, although this should not be interpreted as 290 verified unique individuals, because identities cannot always be matched across independently distributed datasets.

Table~2 summarizes the datasets and task conditions included in this benchmark.

Across these benchmarks, we evaluated whether a model could identify a cognitive task without being calibrated on labeled EEG examples from the particular condition being tested. We refer to this as \emph{calibration-free} or \emph{zero-shot} (ZS) prediction. The model receives an EEG recording and descriptions of $K$ candidate tasks, and predicts which description best matches the recording. Labeled EEG examples from the target condition are not used to fit a new output classifier. However, zero-shot prediction can involve different degrees of novelty. We distinguish three cases: \emph{ZS-Study}, in which the recording comes from an unseen study, but the task is represented in training; \emph{ZS-Task}, in which the task is new but other tasks from the same study were included in the training; and \emph{ZS-Both}, in which neither the study nor an equivalent task was included in the training, according to our predefined task-equivalence rules. These categories describe the novelty of studies and tasks, not necessarily of individual participants. Indeed, two task classes in D2 include participants whose EEG was used during training on other tasks.

The evaluation procedure also differed between the development benchmarks and D2. For D2, we fixed the set of candidate models, task classes, and evaluation procedure before examining model scores. Some participant eligibility criteria were subsequently amended without access to those scores. After the D2 results became available, we conducted further analyses to understand the observed performance patterns. These included combining task classes across benchmarks, examining discrimination between tasks recorded within the same study, merging semantically related task labels, harmonizing evaluation metrics, and revisiting statistical comparisons. Because these analyses were formulated after D2 scoring, we regard them as exploratory rather than independent confirmation of model performance.

The recordings used in these experiments were obtained from public EEG repositories. Dataset
identifiers beginning with \texttt{ds} refer to OpenNeuro, while those beginning with \texttt{nm}
refer to native records distributed through the Neuroelectromagnetic Data Archive and Tools Resource (NEMAR). The prefix \texttt{on} identifies OpenNeuro datasets mirrored through NEMAR. Importantly, these identifiers indicate data distributions rather than necessarily independent studies. Some records are converted or mirrored versions of previously released datasets, and may contain recordings from the same participants. We therefore tracked the provenance of the dataset and source relationships when assembling the training and evaluation sets. Table~S15 lists the 73 source accessions represented in the later training and evaluation configurations, together with their roles.

\Needspace{105pt}
\par\medskip\noindent{\small\textbf{Table 2. Frozen D2 source and class accounting.}}
{\fontsize{8.8}{10.8}\selectfont\setlength{\tabcolsep}{3pt}\renewcommand{\arraystretch}{1.18}
\begin{longtable}{>{\raggedright\arraybackslash}p{54.30pt}>{\raggedright\arraybackslash}p{102.98pt}>{\raggedright\arraybackslash}p{37.45pt}>{\raggedright\arraybackslash}p{48.06pt}>{\raggedright\arraybackslash}p{48.06pt}>{\raggedright\arraybackslash}p{142.92pt}}
\toprule
\textbf{Source ID} & \textbf{Operational class} & \textbf{People / class} & \textbf{8-s windows} & \textbf{64-s groups} & \textbf{Novelty qualification} \\
\midrule\endfirsthead
\multicolumn{6}{l}{\textit{Table 2 (continued)}}\\
\toprule
\textbf{Source ID} & \textbf{Operational class} & \textbf{People / class} & \textbf{8-s windows} & \textbf{64-s groups} & \textbf{Novelty qualification} \\
\midrule\endhead
\midrule\multicolumn{6}{r}{\textit{Continued on next page}}\\\endfoot
\bottomrule\endlastfoot
ds002680 \citep{data_ds002680} & Go/no-go recognition & 14 & 5584 & 674 & Task holdout; known people \\
ds003061 \citep{data_ds003061} & Active auditory oddball & 13 & 3535 & 416 & Study holdout \\
ds003969 \citep{data_ds003969} & Active thinking & 98 & 14564 & 1753 & Strict common both \\
ds003969 & Breath-counting meditation & 98 & 7320 & 887 & Strict common both \\
ds004657 \citep{data_ds004657} & Manual driving & 18 & 1835 & 221 & Nominal both; model-dependent \\
ds004657 & Full assistance available & 18 & 1753 & 209 & Nominal both; model-dependent \\
ds004657 & Speed assistance available & 18 & 1748 & 210 & Nominal both; model-dependent \\
ds005189 \citep{data_ds005189} & Surprise recognition & 30 & 4771 & 573 & Task holdout; known people \\
ds006979 \citep{data_ds006979} & Visual working-memory task session & 53 & 33914 & 4182 & Study holdout \\
ds008711 \citep{data_ds008711} & RSVP language processing & 64 & 12034 & 1472 & Study holdout \\
\end{longtable}}
\vspace{-0.6em}\noindent{\footnotesize Total: 87,058 windows and 10,597 long groups. Repeated people in multiple classes are not additional participants. Source-scoped participant total: 290, not a verified cross-source unique-person count. RSVP: rapid serial visual presentation. ``Both'' denotes nominal study/task novelty, not uniform evidence of new cognitive mechanisms.}\par\medskip

\begin{figure}[!htbp]
\centering
\includegraphics[width=\linewidth,height=0.55\textheight,keepaspectratio]{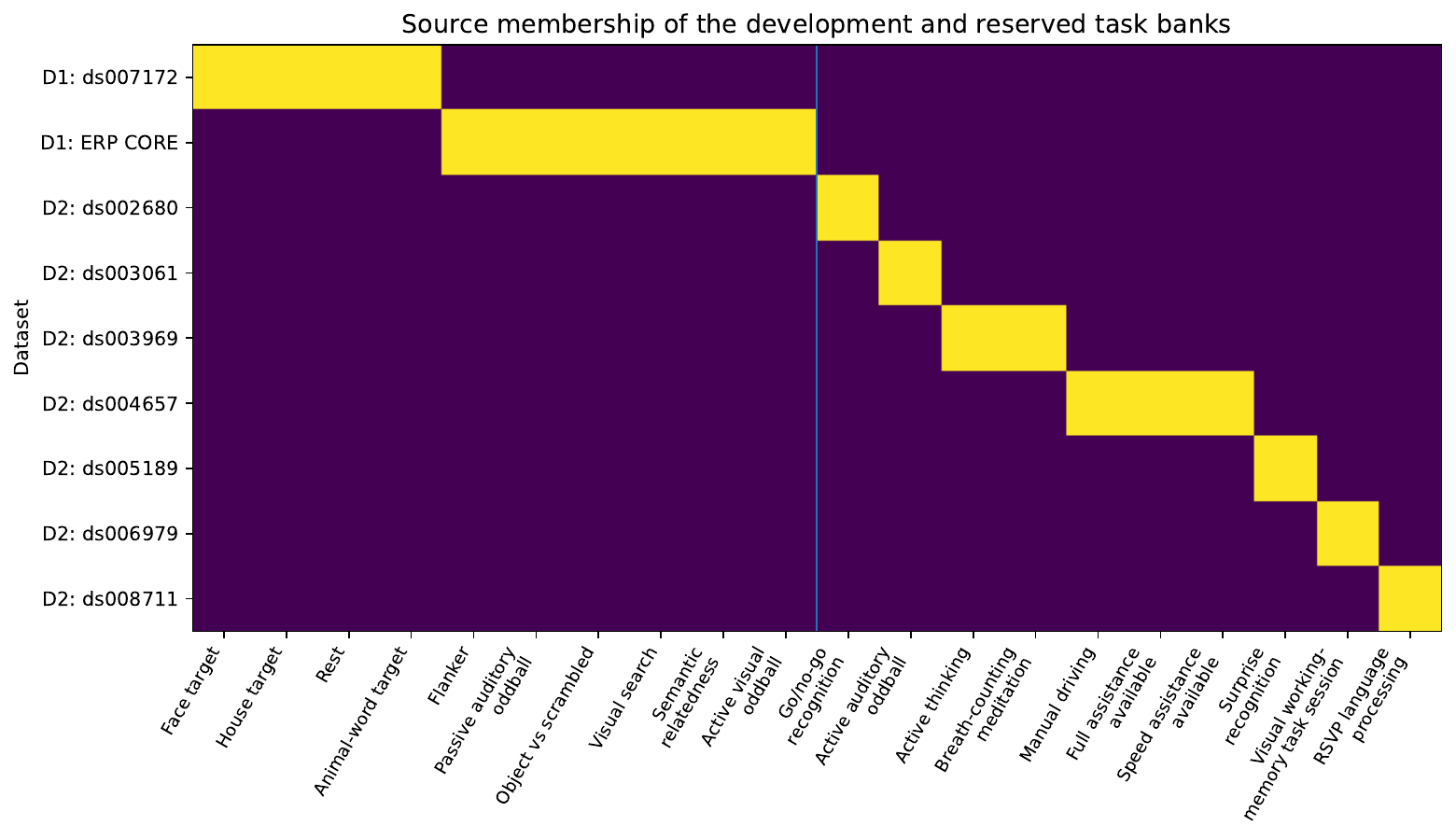}
\caption{\textbf{Task labels and their source datasets.} A filled cell indicates that a source contributes a task label. D1 has two multi-task datasets; D2 has five singleton-label datasets and two multi-condition datasets. The auditory-oddball and meditation/thinking datasets have shared investigator and facility provenance.}
\label{fig:task-source-matrix}
\end{figure}

\FloatBarrier
\section*{3. Results}\addcontentsline{toc}{section}{3. Results}

\subsection*{3.1 Models that performed well during development did not necessarily perform well on new evaluation data}

Our first question was whether models that performed well during development would also perform well when evaluated on different cognitive tasks and EEG datasets. To examine this, we compared nine model configurations on two benchmarks, each requiring the model to choose among ten possible cognitive tasks. The first benchmark, D1, had been used during model development, whereas the second, D2, was reserved for evaluating a predefined set of candidate models. The two benchmarks contained different tasks and recording sources. In both cases, the model was given an eight-second segment of single-channel EEG and had to identify the most likely task from ten candidate descriptions, without using labeled EEG from the evaluated condition to train a new classifier.

The nine configurations differed in three main ways. First, they were trained on different mixtures of EEG datasets, allowing us to examine whether including more studies and cognitive tasks improved generalization. Second, we compared models using only the raw EEG waveform (RAW) with models combining raw EEG and time--frequency representations (LATE). Third, we compared models that mapped EEG into a fixed semantic task space with models that also adapted the task representations during training (CLIP-X). These comparisons allowed us to examine whether the benefits of different training strategies were consistent across the development and reserved benchmarks.

\begin{figure}[!htbp]
\centering
\includegraphics[width=\linewidth,height=0.55\textheight,keepaspectratio]{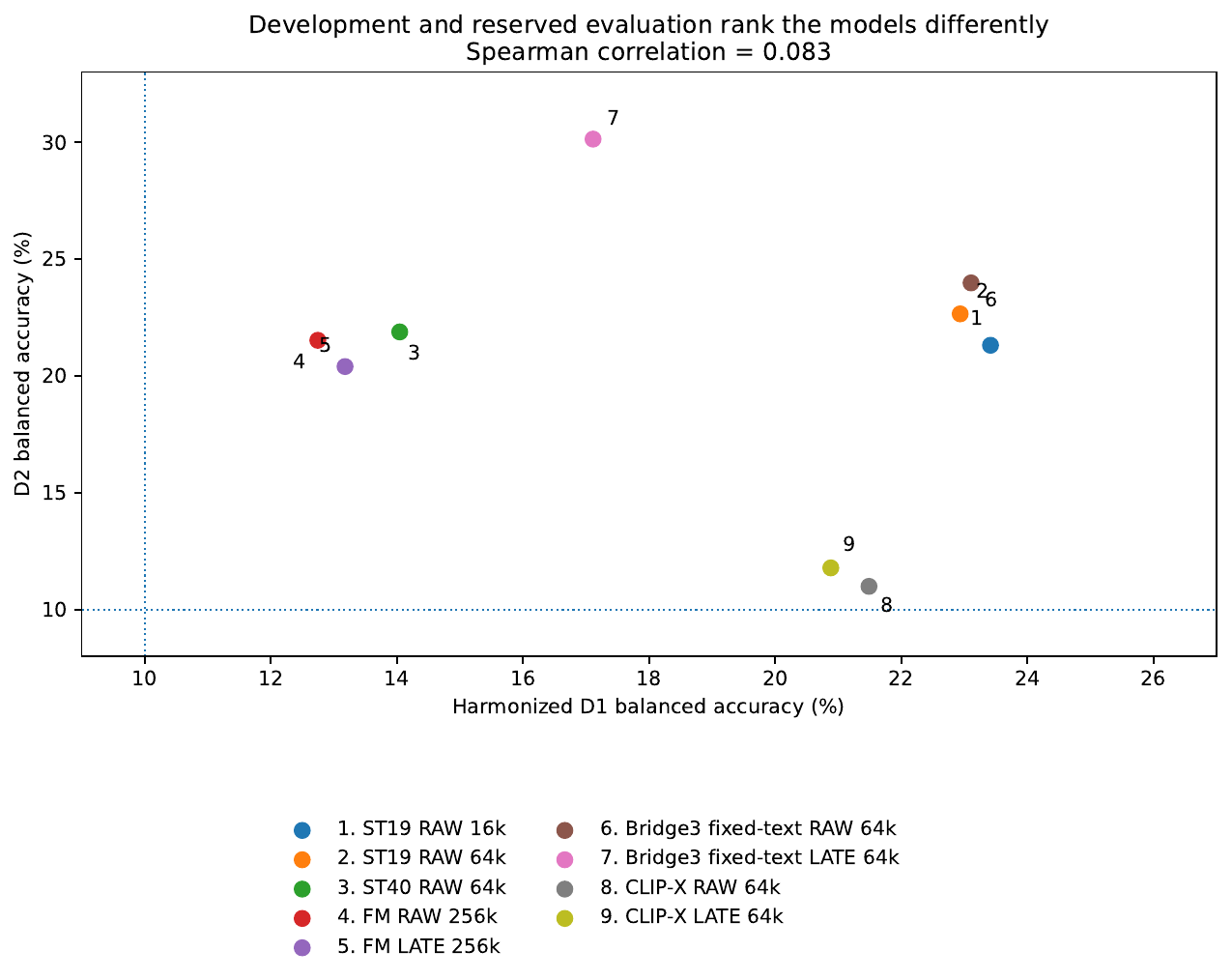}
\caption{\textbf{Model rankings differ between development and reserved evaluation.} Each point is a three-seed condition mean for eight-second inputs. D1 has been reaggregated using nonoverlapping windows and the same class--participant weighting as D2. Both candidate banks contain ten labels. The Spearman correlation across the nine condition means is 0.083.}
\label{fig:d1-d2-model-ranking}
\end{figure}

To compare performance on the same basis, we recomputed the D1 results after D2 scoring using
nonoverlapping EEG windows and the same weighting procedure used for D2. We measured performance using balanced accuracy (BA), which gives equal weight to each task class and, within each class, to each participant (Methods 5.6). Because both benchmarks involved ten task classes, chance-level balanced accuracy was 10\% in each case. Figure~\ref{fig:d1-d2-model-ranking} compares the mean performance of the nine model configurations on D1 and D2. Surprisingly, models that performed well on D1 were not necessarily the best models on D2. The correlation between their rankings was very weak (Spearman
$\rho=0.083$). This finding applies to the nine configurations examined here and does not establish
that development benchmarks are generally unreliable.

One example concerns the composition of the training data. The ST19 training mixture combined
multiple-task EEG datasets with additional sources, whereas ST40 expanded the collection to cover
a wider variety of studies and cognitive tasks, including more sources containing only one or a
few tasks. We expected that broader task coverage might improve generalization. However, on D2,
the RAW model trained with ST19 for 64,000 optimizer updates achieved 22.66\% balanced accuracy, compared with 21.88\% for the corresponding ST40 model (Table~3).

Thus, expanding the training collection did not improve performance in this comparison. Because
several aspects of the training mixture changed together, this result does not establish that adding single-task datasets was itself responsible for the difference.

A second example concerns how the EEG signal was represented. Within the Bridge3 configuration,
we compared a model using only raw EEG with one that additionally incorporated time--frequency
features (i.e. LATE model). Both used fixed semantic task vectors. In the development benchmark D1, the LATE model performed worse than RAW (17.11\% versus 23.11\% balanced accuracy). However, in D2, the result was reversed: LATE achieved 30.13\%, compared to 23.98\% for RAW. Thus, the apparent benefit of incorporating spectral information strongly depended on which tasks and datasets were used for evaluation.

Finally, we examined whether allowing the semantic task representations to adapt during
training improved performance. Compared with the fixed-text Bridge3 LATE model, CLIP-X LATE
improved balanced accuracy on D1 by 3.77 percentage points but reduced it on D2 by 18.34 points. The paired participant-bootstrap 95\% confidence interval for the difference in D2 was $-20.50$ to $-16.19$ percentage points, conditional on the sources evaluated and the trained models. The standard deviation of the difference across the three training seeds was 6.64 points. Thus, an approach that appeared beneficial on the development benchmark could substantially underperform on a different collection of tasks.

Together, these results show that the performance of the model was sensitive to the choice of evaluation data. Increasing the diversity of training datasets, adding spectral features, or adapting semantic task representations did not produce improvements that were consistent across D1 and D2. This motivated us to examine which tasks the models could recognize reliably and how much of their apparent task-prediction ability might reflect differences between studies and recording conditions.

\begin{figure}[!htbp]
\centering
\includegraphics[width=\linewidth,height=0.55\textheight,keepaspectratio]{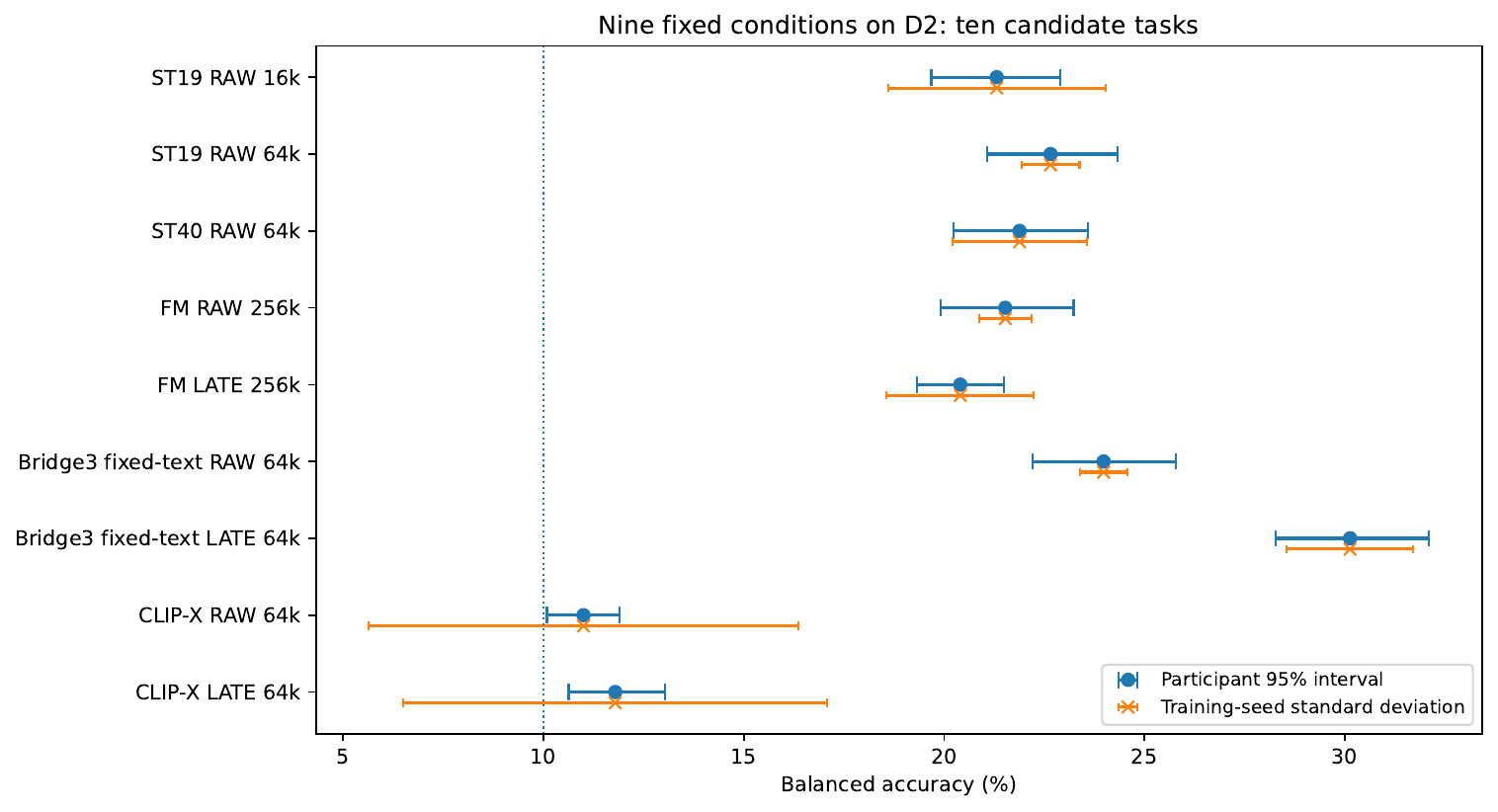}
\caption{\textbf{All nine conditions on the reserved ten-class bank (D2).} Circles show mean balanced accuracy with 95\% intervals from 5,000 joint participant resamples, conditional on the fixed sources and three trained seeds. Crosses show the same means with one training-seed standard deviation. The vertical line marks 10\% uniform chance. All inputs last eight seconds.}
\label{fig:d2-model-accuracy}
\end{figure}

Table 3 and Figure~\ref{fig:d2-model-accuracy} report all nine conditions. Fixed-text Bridge3 LATE had the highest observed mean: 30.13 $\pm$ 1.58\% BA, 25.65\% macro-F1 (the mean of per-class precision--recall F1 scores), and 76.57\% Top-5 accuracy. Its participant-bootstrap interval for BA was 28.28--32.11\%. The jointly adapted RAW and LATE models varied more across training seeds; in each architecture, two of three seeds were below the 10\% BA baseline.

This comparison changes a training package. In addition to text adaptation, the implementations differ in similarity scale, regularization and prototype-bank construction. It therefore does not isolate any single component or test the introduction of contrastive learning itself.

\Needspace{105pt}
\par\medskip\noindent{\small\textbf{Table 3. All nine frozen conditions on D2: 8 s and K = 10.}}
{\fontsize{8.8}{11.2}\selectfont\setlength{\tabcolsep}{3pt}\renewcommand{\arraystretch}{1.18}
\begin{longtable}{>{\raggedright\arraybackslash}p{127.94pt}>{\raggedright\arraybackslash}p{71.77pt}>{\raggedright\arraybackslash}p{71.77pt}>{\raggedright\arraybackslash}p{56.17pt}>{\raggedright\arraybackslash}p{53.05pt}>{\raggedright\arraybackslash}p{53.05pt}}
\toprule
\textbf{Condition} & \textbf{BA \ensuremath{\pm} seed SD (\%)} & \textbf{Participant CI (\%)} & \textbf{Macro-F1 (\%)} & \textbf{Top-5 (\%)} & \textbf{MRR} \\
\midrule\endfirsthead
\multicolumn{6}{l}{\textit{Table 3 (continued)}}\\
\toprule
\textbf{Condition} & \textbf{BA \ensuremath{\pm} seed SD (\%)} & \textbf{Participant CI (\%)} & \textbf{Macro-F1 (\%)} & \textbf{Top-5 (\%)} & \textbf{MRR} \\
\midrule\endhead
\midrule\multicolumn{6}{r}{\textit{Continued on next page}}\\\endfoot
\bottomrule\endlastfoot
ST19 RAW 16k & 21.31 \ensuremath{\pm} 2.71 & 19.68--22.90 & 17.30 & 71.10 & 0.420 \\
ST19 RAW 64k & 22.66 \ensuremath{\pm} 0.72 & 21.07--24.32 & 17.90 & 67.35 & 0.418 \\
ST40 RAW 64k & 21.88 \ensuremath{\pm} 1.67 & 20.23--23.60 & 19.75 & 70.54 & 0.424 \\
FM RAW 256k & 21.53 \ensuremath{\pm} 0.65 & 19.91--23.23 & 19.73 & 69.28 & 0.418 \\
FM LATE 256k & 20.40 \ensuremath{\pm} 1.84 & 19.32--21.50 & 18.76 & 67.06 & 0.401 \\
Bridge3 fixed-text RAW 64k & 23.98 \ensuremath{\pm} 0.59 & 22.21--25.79 & 20.61 & 70.40 & 0.438 \\
Bridge3 fixed-text LATE 64k & 30.13 \ensuremath{\pm} 1.58 & 28.28--32.11 & 25.65 & 76.57 & 0.492 \\
CLIP-X RAW 64k & 11.00 \ensuremath{\pm} 5.36 & 10.09--11.90 & 10.55 & 65.32 & 0.327 \\
CLIP-X LATE 64k & 11.79 \ensuremath{\pm} 5.29 & 10.63--13.02 & 9.60 & 70.59 & 0.351 \\
\end{longtable}}
\vspace{-0.6em}\noindent{\footnotesize The confidence intervals are a post-score statistical amendment: 5,000 participant resamples jointly across each person's classes, conditional on the fixed sources and three trained seeds. Seed standard deviation (SD) is reported separately; CI denotes a 95\% confidence interval and MRR denotes mean reciprocal rank. Top-5 uniform chance is 50\%; Top-1/BA chance is 10\%. }\par\medskip

\FloatBarrier
\subsection*{3.2 The leading model recognized some tasks much better than others}

The previous analysis showed that the relative performance of different models changed substantially between the development and reserved benchmarks. We next asked a different question: how well could the best-performing model on D2 recognize each of the ten cognitive tasks? This distinction matters because a model could achieve above-chance accuracy by recognizing a few tasks particularly well, even if it performed poorly on most others.

\begin{figure}[!htbp]
\centering
\includegraphics[
    width=\linewidth,
    height=0.55\textheight,
    keepaspectratio
]{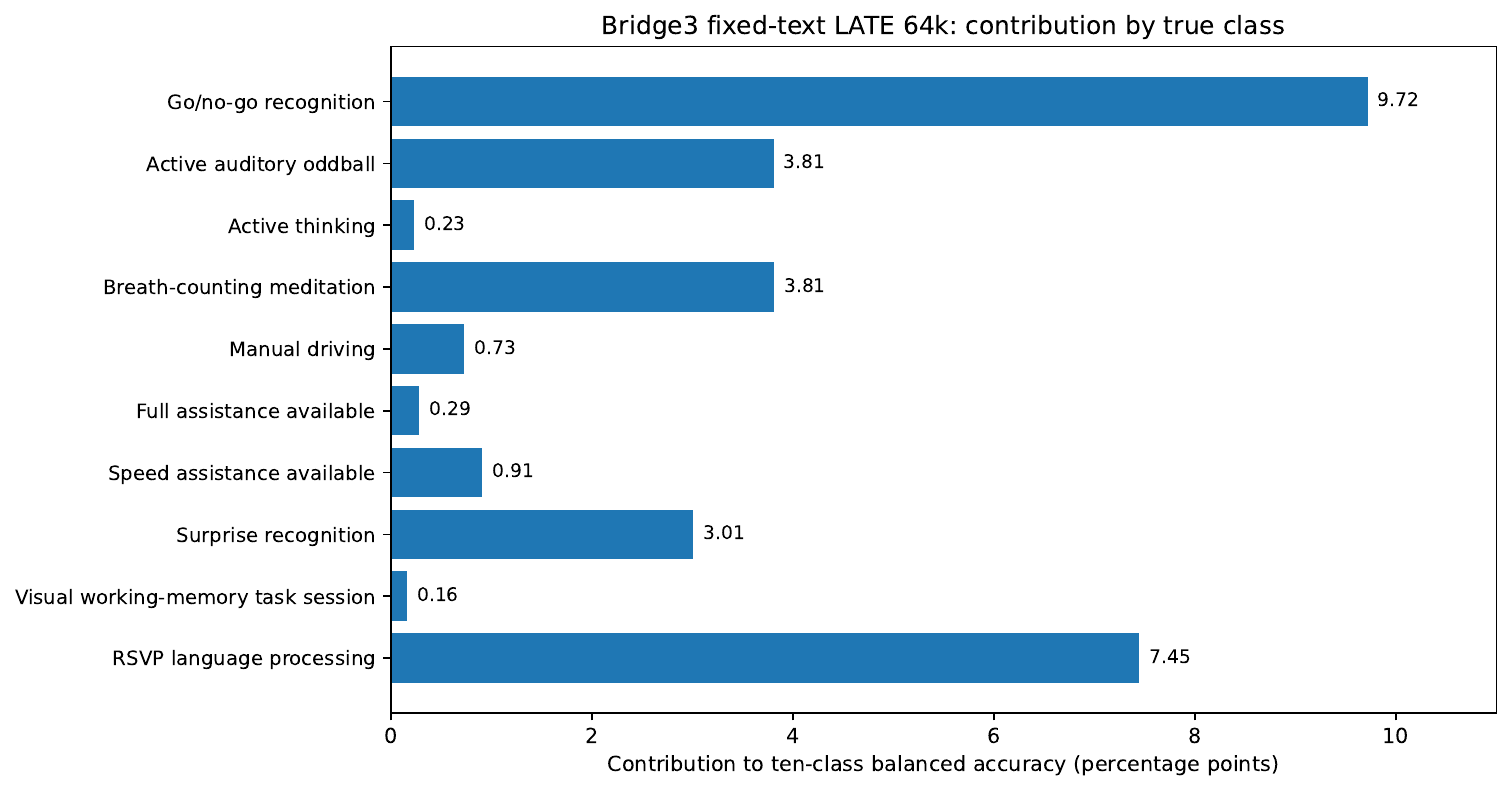}
\caption{
\textbf{A small number of tasks account for most of the leading model's balanced accuracy.}
Each bar shows the contribution of one task class to the ten-class balanced accuracy, calculated as its recall divided by ten. Go/no-go recognition, auditory oddball, and RSVP language processing together account for 69.63\% of the total score. Go/no-go recognition was excluded as a task during training, but its participants had contributed recordings from other tasks.
}
\label{fig:d2-class-contributions}
\end{figure}

We examined the fixed-text Bridge3 LATE model, which achieved the highest overall balanced accuracy
on D2 (30.13\%, compared with 10\% chance). Balanced accuracy gives equal weight to the ten
task classes, so each class contributes one-tenth of its recall to the final score. Figure~\ref{fig:d2-class-contributions} shows these contributions separately. The results were highly uneven. The model recognized the go/no-go recognition task with 97.22\% recall, which alone contributed 9.72 percentage points to the overall balanced accuracy. Together, go/no-go recognition, auditory oddball, and rapid serial visual presentation (RSVP) language processing contributed 20.98 of the 30.13 percentage points, accounting for 69.63\% of the total score. In contrast, the remaining
seven classes contributed only 9.15 percentage points, corresponding to an average recall of
approximately 13.1\%, only slightly above the 10\% chance level.

One important qualification concerns the go/no-go recognition task, which accounted for the largest contribution. Although this task itself had been excluded from training, its participants had previously contributed EEG recordings from other tasks. The same was true of the surprise-recognition class. Thus, these two classes tested generalization to new tasks, but not necessarily to entirely new participants. To examine how strongly the overall result depended on these classes, we recalculated mean recall after excluding them from the average. Performance across the remaining eight classes was 21.75\%. Importantly, the model still had to choose among the original ten candidate task descriptions for every prediction. This was therefore not an eight-class classification problem, and the chance-level baseline remained 10\%.

We also examined which task labels the model tended to predict, regardless of whether those
predictions were correct. This revealed another important pattern. When each true class was
given equal weight, the model assigned 32.88\% of predictions to auditory oddball, far more than the 10\% expected if predictions were distributed uniformly across the ten labels. However, only 11.59\% of predictions assigned to auditory oddball were correct. RSVP language processing was also selected frequently, accounting for 19.27\% of predictions, although its precision was higher, at 38.64\% (Table~S8). In other words, the model did not simply fail to recognize some tasks; it frequently classified them as a small set of preferred alternatives.

The confusion patterns make this more concrete. For example, EEG recorded during thinking and
meditation was classified as auditory oddball in 48.4\% and 50.1\% of cases, respectively. The driving tasks showed a similar tendency. These errors suggest that the model had difficulty distinguishing several cognitive activities from the EEG patterns associated with its frequently predicted labels. The auditory-oddball and meditation datasets also share investigator and recording-site provenance, which could contribute to the confusion. However, these results cannot determine whether the errors reflect shared recording characteristics, participant-related signals, or a more general tendency to overpredict auditory oddball.

These findings change how the overall 30.13\% balanced accuracy should be interpreted. The score demonstrates above-chance prediction across the ten-class benchmark, but it does not mean that the model can recognize all ten cognitive tasks equally well. The strong performance on a few classes, together with frequent misclassification into preferred labels, points to substantial differences in how well individual tasks are represented.

Importantly, predicting a single class repeatedly does not by itself increase balanced accuracy: a classifier that always chooses the same label achieves only $1/K$ balanced accuracy when there are $K$ equally weighted true classes. The concern here is therefore not an artificially inflated baseline, but the fact that an average score can conceal poor performance on most tasks. 

Taken together, these results suggest that achieving above-chance task prediction is only one step toward learning a representation that generalizes across cognitive activities. It is equally important to determine whether the model can distinguish tasks recorded under comparable conditions, rather than relying on differences associated with their sources. We examine this issue in the following analyses.

\FloatBarrier
\subsection*{3.3 Distinguishing cognitive tasks within the same study remains difficult}

The previous analysis showed that the model could recognize some tasks much better than others.
However, many of the evaluated tasks came from different studies, raising an important question of whether the model recognized differences in cognitive activity, or partly distinguished the studies in which the EEG had been recorded. Differences in recording equipment, participants, and experimental environments could provide clues to task identity when each task comes from a different source.

To investigate this possibility, we examined datasets in D2 that contained more than one task condition. We restricted the candidate labels to tasks recorded within the same dataset, so identifying the data source alone would no longer be sufficient to make a correct prediction. This analysis used the existing
trained models without further training or calibration. We examined two comparisons: thinking versus breath-counting meditation, and three driving conditions that differed in their experimental demands. Unlike the original ten-class evaluation, these were two-class and three-class classification problems, with chance levels of 50\% and 33.33\%, respectively. Because these comparisons were formulated after examining the D2 results, they should be regarded as exploratory.

\begin{figure}[!htbp]
\centering
\includegraphics[
    width=\linewidth,
    height=0.55\textheight,
    keepaspectratio
]{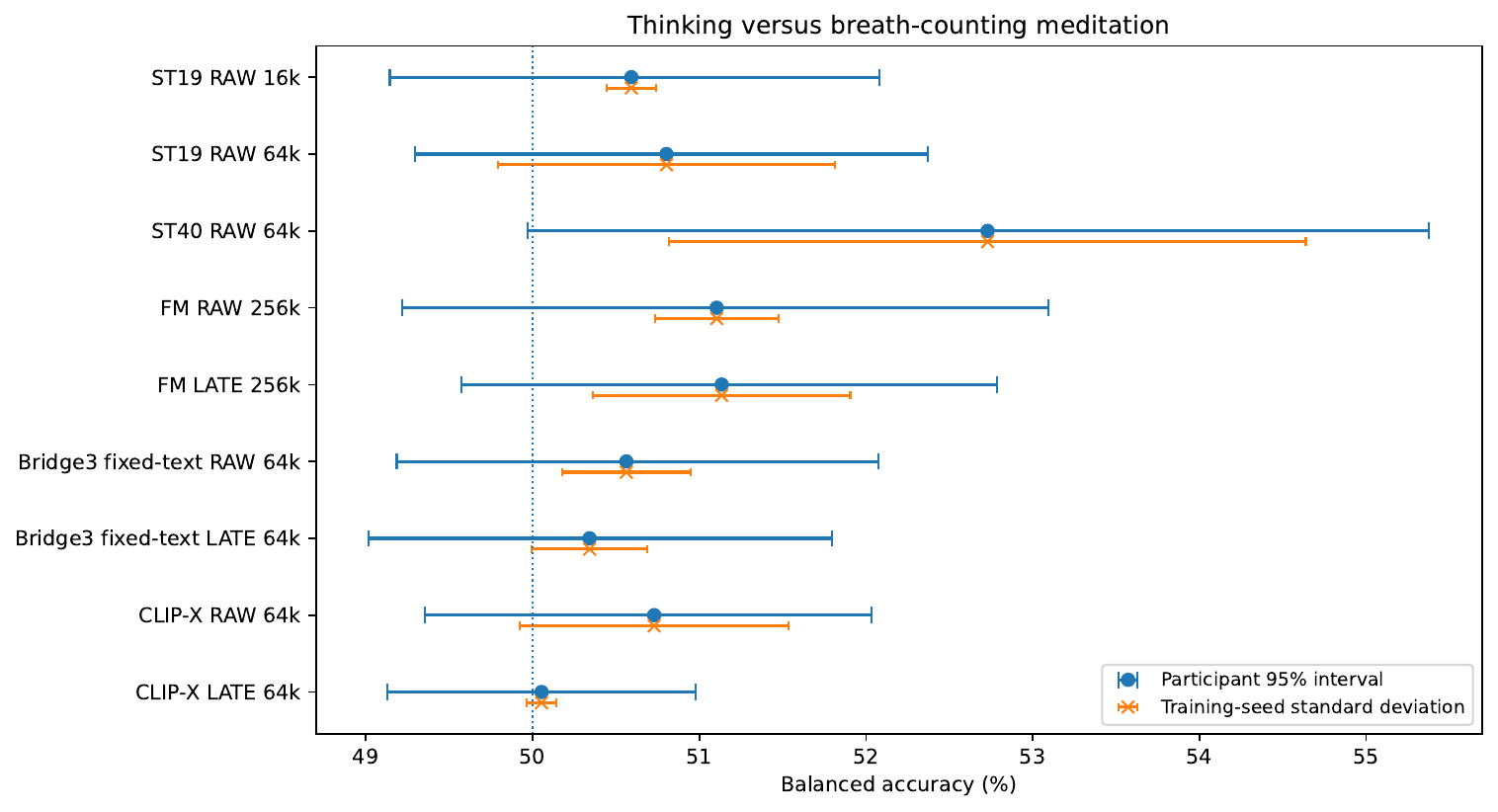}
\caption{
\textbf{Distinguishing thinking from breath-counting meditation within the same EEG dataset.}
Exploratory two-class evaluation involving 98 participants. The model selected between thinking and meditation rather than between the ten tasks in the full D2 benchmark. Chance balanced accuracy was 50\%.
Circles and crosses indicate the participant-bootstrap intervals and training-seed standard deviations, respectively, as defined in Figure~\ref{fig:d2-model-accuracy}.}
\label{fig:within-meditation}
\end{figure}

\begin{figure}[!htbp]
\centering
\includegraphics[
    width=\linewidth,
    height=0.55\textheight,
    keepaspectratio
]{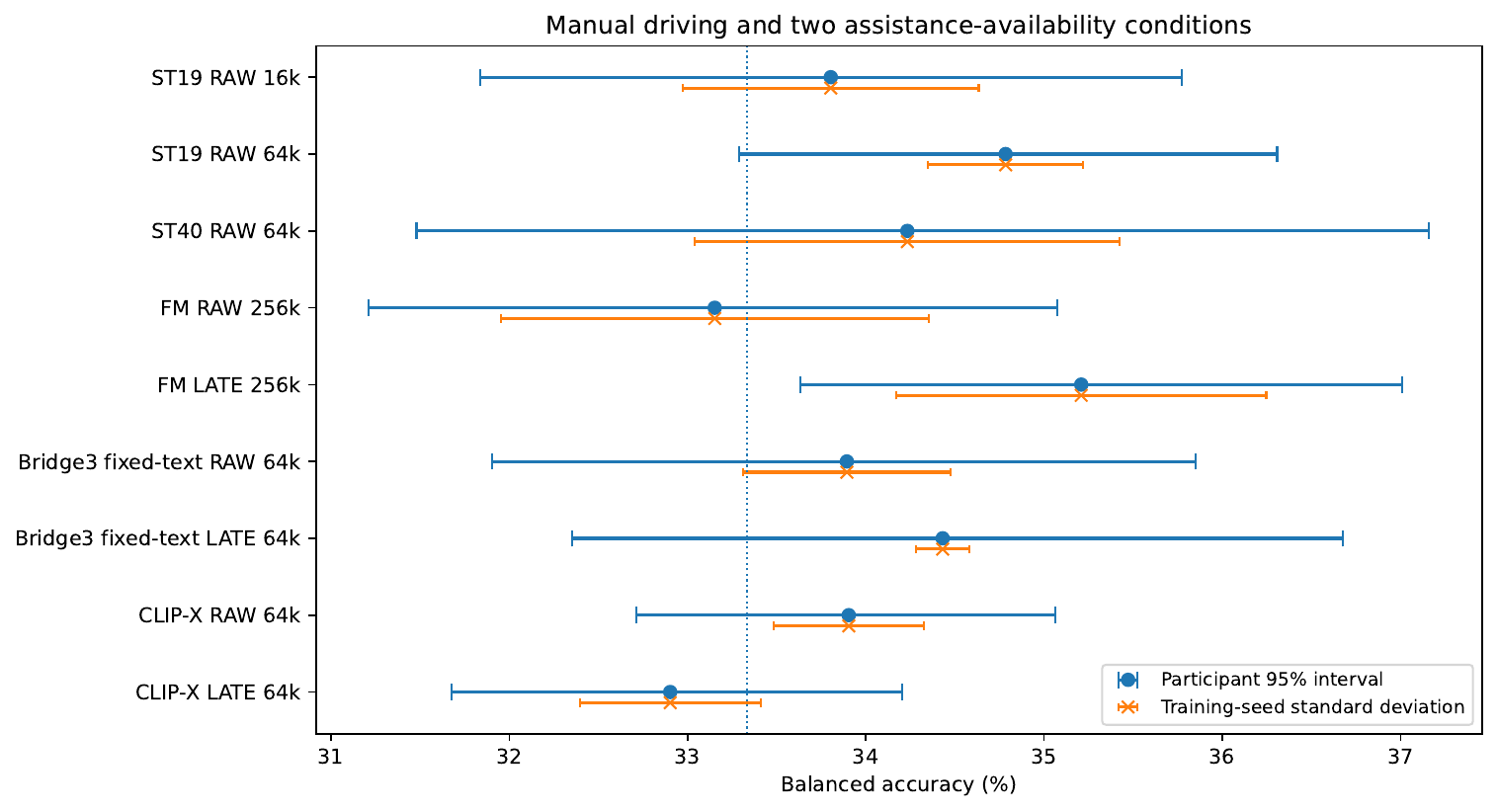}
\caption{
\textbf{Distinguishing three driving conditions within the same EEG dataset.}
Exploratory three-class evaluation involving 18 participants. Candidate predictions were restricted to the three driving conditions, giving a chance balanced accuracy of 33.33\%. Circles and crosses
indicate the participant-bootstrap intervals and training-seed standard deviations, respectively, as defined in Figure~\ref{fig:d2-model-accuracy}.
}
\label{fig:within-driving}
\end{figure}

We first evaluated the fixed-text Bridge3 LATE model, which had achieved the highest overall balanced accuracy in the ten-class D2 evaluation. For thinking versus meditation, its balanced accuracy was 50.34\%, almost identical to the 50\% chance level (Figure~\ref{fig:within-meditation}). For the three driving conditions, balanced accuracy was 34.43\%, compared with 33.33\% chance (Figure~\ref{fig:within-driving}). Thus, the model's above-chance performance in the broader ten-class evaluation did not translate into reliable discrimination between these particular conditions recorded within the same source.

Across all nine models, balanced accuracy ranged from 50.05\% to 52.73\% for thinking versus meditation,
and from 32.90\% to 35.21\% for the driving conditions. One model, FM LATE, showed a small above-chance effect for driving, with a participant-bootstrap interval of 33.63--37.01\%. However, these comparisons were exploratory, were not corrected for multiple comparisons, and involved differences in driving-related
training exposure across models. We therefore do not regard this isolated result as strong evidence of generalizable within-study task discrimination.

There are at least two possible explanations for these near-chance results. One is that the models relied partly on differences between recording sources when predicting tasks in the broader benchmark. Restricting the candidates to conditions from the same source removes this particular source-level cue, making it harder to use such differences to identify the correct task.

A second possibility is that the task conditions examined here are themselves difficult to distinguish from short single-channel EEG recordings. Thinking and breath-counting meditation differ in their intended cognitive processes, but both involve sustained internal mental activity and may share many
aspects of their sensory and behavioral context. Their differences in attention or mental state may therefore produce EEG changes that are subtle, variable across individuals, or difficult to detect from a single anterior--posterior differential channel. Similarly, the three driving conditions may differ in the level of driver engagement or assistance while sharing much of the same visual, motor, and attentional
context. Such differences could be meaningful at the behavioral level without producing consistently distinguishable EEG patterns in eight-second windows.

These explanations are not mutually exclusive, and the present analysis cannot determine their relative contributions. In particular, performance near chance does not demonstrate that the cognitive states are identical or that their neural correlates are absent. Longer recordings, additional electrodes, or measurements targeting specific task-related processes might reveal differences that this evaluation could not detect.

It is also important not to generalize these findings to all within-study task comparisons. Developmental analyses using D1 showed better discrimination between some conditions recorded within the same source, and related analyses of shared tasks also yielded positive results (Supplement~S5). The few-shot calibration results presented below provide another example of improved discrimination under different conditions. However, such performance may partly reflect differences in stimuli, responses, eye movements, or experimental blocks, rather than cognitive processes alone. For example, ERP CORE paradigms were recorded in separate blocks \citep{li2021perils,kappenman2021}.

Together, these results illustrate two challenges for learning cognitive representations from heterogeneous EEG datasets. When different studies contribute different tasks, a model may exploit source-specific information. When several tasks are recorded within the same study, their neural signatures may nevertheless be too similar or too weakly measured to distinguish reliably. Building models that generalize across studies will require training data that address both problems: recordings of multiple well-characterized tasks within participants, together with repeated measurements of comparable tasks across independent studies.

\FloatBarrier
\subsection*{3.4 Adding more training studies broadens task coverage, but improvements are uneven}

A central motivation of this study was to take advantage of the large amount of EEG data already available from different research groups. Most of these datasets were collected for different scientific purposes and contain different cognitive tasks. Our task-semantic alignment approach was designed to make it possible to learn from these heterogeneous recordings within a common framework. We therefore asked a simple question: does adding more studies and a wider variety of tasks to the training data improve the model's ability to recognize cognitive tasks in recordings not used for training?

We first investigated this question in an early experiment conducted during model development. Starting with a fixed collection of EEG datasets, we progressively added 5, 10, 15, and finally all 19 additional
datasets that had passed our data-quality and eligibility checks. We refer to these five training mixtures as \emph{SC0}, \emph{SC5}, \emph{SC10}, \emph{SC15}, and \emph{SCALL}, respectively. These names
indicate the number of datasets added to the original collection, rather than the total number of independent studies used for training.

To make the comparison interpretable, we evaluated every mixture on the same development benchmark, D1, using its ten cognitive-task classes and nonoverlapping eight-second EEG segments. We repeated training with three random seeds and examined performance after 16,000, 32,000, and 64,000 optimization updates. This allowed us to examine whether adding training datasets was beneficial across different stages of learning.

\begin{figure}[!htbp]
\centering
\includegraphics[
    width=\linewidth,
    height=0.55\textheight,
    keepaspectratio
]{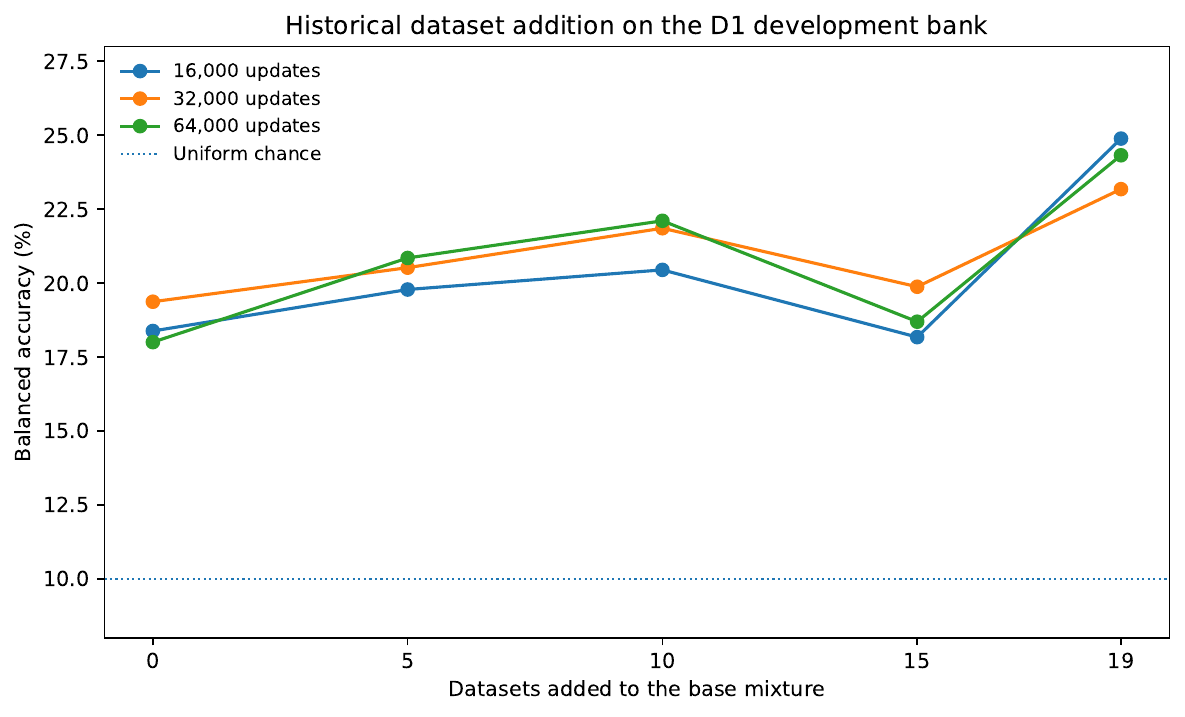}
\caption{
\textbf{Adding EEG datasets improves development performance unevenly.}
Balanced accuracy on the ten-task D1 development benchmark using eight-second EEG segments. Each point is the mean of three training seeds. The five training mixtures contain 0, 5, 10, 15, or 19 additional datasets, denoted SC0, SC5, SC10, SC15, and SCALL. Lines connect results obtained after the same number of optimizer updates. Although the fully expanded mixture outperformed the original mixture, intermediate additions did not produce monotonic improvements. The historical SCALL training recipe is distinct from the later ST19-MATCHED recipe.}
\label{fig:study-addition}
\end{figure}

The overall result was encouraging, but not straightforward (Figure~\ref{fig:study-addition}). After 16,000 updates, mean balanced accuracy increased from 18.38\% with the original training mixture to 24.89\% with all 19 additional datasets, an improvement of 6.50 percentage points calculated from the unrounded values. The largest mixture also outperformed the original mixture at later checkpoints: 23.18\% versus 19.37\% after 32,000 updates, and 24.32\% versus 18.01\% after 64,000 updates. Thus, the expanded training collection produced consistently higher development scores than the original collection at these three checkpoints.

However, adding datasets did not produce a steady improvement at every stage. In particular, increasing the number of added datasets from 10 to 15 reduced balanced accuracy by 2.27, 1.98, and 3.41 percentage points at the three checkpoints, respectively. Performance then recovered when all 19 datasets
were included. These fluctuations suggest that the effect of adding data depends not only on how much is added, but potentially also on which studies and tasks are included. Importantly, this experiment changed several properties of the training data simultaneously, including task coverage, participants, and sampling exposure. It therefore does not establish which of these factors caused the observed changes.

These results raised a more fundamental question about what it means to increase the diversity of EEG training data. Adding a new dataset can introduce a cognitive task that the model has never encountered before. This expands the range of activities represented in training. However, if that task appears in only one study, the model may also learn characteristics specific to that study, its participants, or its
recording equipment. In contrast, when comparable tasks are recorded in multiple studies, the model
has an opportunity to learn which EEG features are shared across different recording conditions. Similarly, datasets containing multiple tasks provide information about how EEG changes between
activities while some aspects of the recording context are held relatively constant.

Therefore, we examined the structure of the later training collections, rather than considering only their
total number of datasets (Table~4). For this analysis, we distinguished dataset identifiers, independent study families, and registered cognitive-task categories (VFTs). These counts capture different aspects of the available training information.

The ST19-MATCHED mixture contained 38 dataset identifiers representing 32 study families and 41 VFTs.
The larger ST40 mixture contained 59 dataset identifiers, 53 families, and 59 VFTs. Thus, expanding from
ST19-MATCHED to ST40 substantially increased both the number of represented studies and the range of cognitive tasks.

Surprisingly, however, two measures of task repetition remained unchanged. In both mixtures, only 16 VFTs were represented in at least two study families, and only 19 families contained at least two VFTs. In other words, the larger training collection covered more distinct tasks, but it did not increase these counts of tasks repeated across studies or families containing multiple task types. The selectively expanded Bridge3 mixture occupied an intermediate position, covering 44 VFTs across 35 study families.

\Needspace{245pt}
\noindent\begin{minipage}{\linewidth}
\par\medskip\noindent
{\small\textbf{Table 4. Training-data composition:
task diversity versus repetition across studies.}}
\par\smallskip

{\fontsize{9}{11.5}\selectfont
\setlength{\tabcolsep}{4pt}
\renewcommand{\arraystretch}{1.20}

\begin{tabular}{
@{}
>{\raggedright\arraybackslash}p{0.40\linewidth}
*{3}{>{\centering\arraybackslash}p{0.17\linewidth}}
@{}
}
\toprule
\textbf{Measure}
& \textbf{ST19-MATCHED}
& \textbf{Bridge3}
& \textbf{ST40} \\
\midrule

\multicolumn{4}{@{}l}{
\textit{Amount and diversity of training data}
}\\

Dataset identifiers
& 38 & 41 & 59 \\

Registered study families
& 32 & 35 & 53 \\

Original task identifiers
& 89 & 92 & 110 \\

Task categories (VFTs)
& 41 & 44 & 59 \\

Eight-second window entries
& 1,608,919 & 1,663,914 & 2,704,726 \\

\addlinespace[5pt]
\midrule

\multicolumn{4}{@{}l}{
\textit{Task repetition within and across studies}
}\\

Task categories found in
$\geq 2$ study families
& 16 & 16 & 16 \\

Study families containing
$\geq 2$ task categories
& 19 & 19 & 19 \\

\bottomrule
\end{tabular}
}

\par\smallskip
\noindent{\scriptsize
Dataset IDs refer to data distributions; study families are registered source
 groups, not independently verified cohorts. Native IDs are original task labels;
 VFTs are harmonized task categories. Repetition counts describe VFTs in multiple
 families and families with multiple VFTs, not necessarily within-person contrasts.
 Four-second auxiliary data are excluded; window entries may overlap. ST19/ST40
 are historical condition names, not total dataset counts.
}\par
\end{minipage}
\par\medskip

The key observation in Table~4 is that expanding the training collection increased the number of studies and task categories, but did not increase the number of task categories represented across multiple study families. Likewise, the number of study families containing multiple task categories remained unchanged.

This distinction matters because adding a new task from a new study provides more examples of what that task looks like in a particular recording environment, but does not necessarily help the model identify which EEG features are specific to the task rather than to that environment. Learning such distinctions may require either recording multiple tasks under comparable conditions or observing the same task across different studies.

These counts describe the structure of the training data rather than demonstrating a causal effect on model performance. Nevertheless, they help explain why greater task coverage did not consistently translate into better generalization.

We carried out two further developmental analyses to better understand these effects. First, we examined whether the order in which studies were added influenced the results. This randomized-addition experiment
completed 240 training runs across six different addition orders, using one optimization seed. However, only 29 external evaluation reports were available in the inspected archive. Figure~\ref{fig:randomized-addition} therefore shows the available common evaluation slices rather than a complete comparison of all 240 runs.

Second, we investigated whether performance could improve simply by training longer without changing the composition of the data. This fixed-composition experiment separated additional optimization from dataset
expansion, with results shown in Figures~\ref{fig:fixed-mixture-raw} and~\ref{fig:fixed-mixture-late}.

Across these follow-up analyses, improvement was not uniform across evaluation benchmarks.
Some configurations improved on D1-ST, which contains more diverse tasks drawn from
different sources, while performance on the original D1 benchmark declined.
This pattern is consistent with a possible trade-off between broader task coverage and the ability to distinguish tasks already represented in the original development data. However, because each
D1-ST class comes from a different source, improved performance on D1-ST alone cannot establish better recognition of cognitive processes independently of study-specific information.

Together, these experiments suggest that the value of additional EEG data depends on how the new recordings complement the existing training collection. Expanding the number of tasks is useful, but increasing the number of independent observations of comparable tasks may be equally important. This motivates training datasets that combine a broad range of cognitive activities with repeated task measurements across participants and independent recording studies.

\FloatBarrier
\subsection*{3.5 Few-shot calibration improves task prediction, particularly with examples from
the same participant}

So far, we have evaluated how well the models could identify cognitive tasks without using labeled EEG examples from the target condition. Although some tasks could be recognized above chance, performance was often limited and varied considerably across tasks. However, poor zero-shot prediction does not necessarily mean that the learned EEG representations contain no useful information about cognitive activity. Some of that information might become easier to extract if a small number of labeled EEG examples were available.

This possibility is also important for practical applications. For example, a person using a simple EEG device might be willing to perform a few known activities to calibrate the system. We therefore investigated a \emph{few-shot} setting, in which the model was given a small number of labeled EEG examples to help identify subsequent recordings. We asked two questions: could examples from another participant improve prediction, and would examples from the person being evaluated provide an additional benefit?

We examined these questions using an earlier model trained on the fully expanded dataset mixture (SCALL), after 16,000 optimizer updates. The analysis used one training seed and the ten-task D1 development benchmark, rather than the reserved D2 benchmark. Each prediction was made from a 64-second EEG recording, while calibration examples consisted of labeled eight-second EEG windows. We varied the number of calibration examples, referred to as \emph{support windows}, available for each task class.

We compared two ways of selecting these examples. In the \emph{cross-participant} condition, the support examples came from one participant in each target study, and predictions were evaluated on
different participants. This tested whether a small amount of labeled EEG from the relevant study could improve prediction for other people. In the \emph{same-participant} condition, support examples came from the person being evaluated. Support and query recordings were selected to avoid any overlap in their raw EEG samples, although they still came from the same task recording.

\begin{figure}[!htbp]
\centering
\includegraphics[
    width=\linewidth,
    height=0.55\textheight,
    keepaspectratio
]{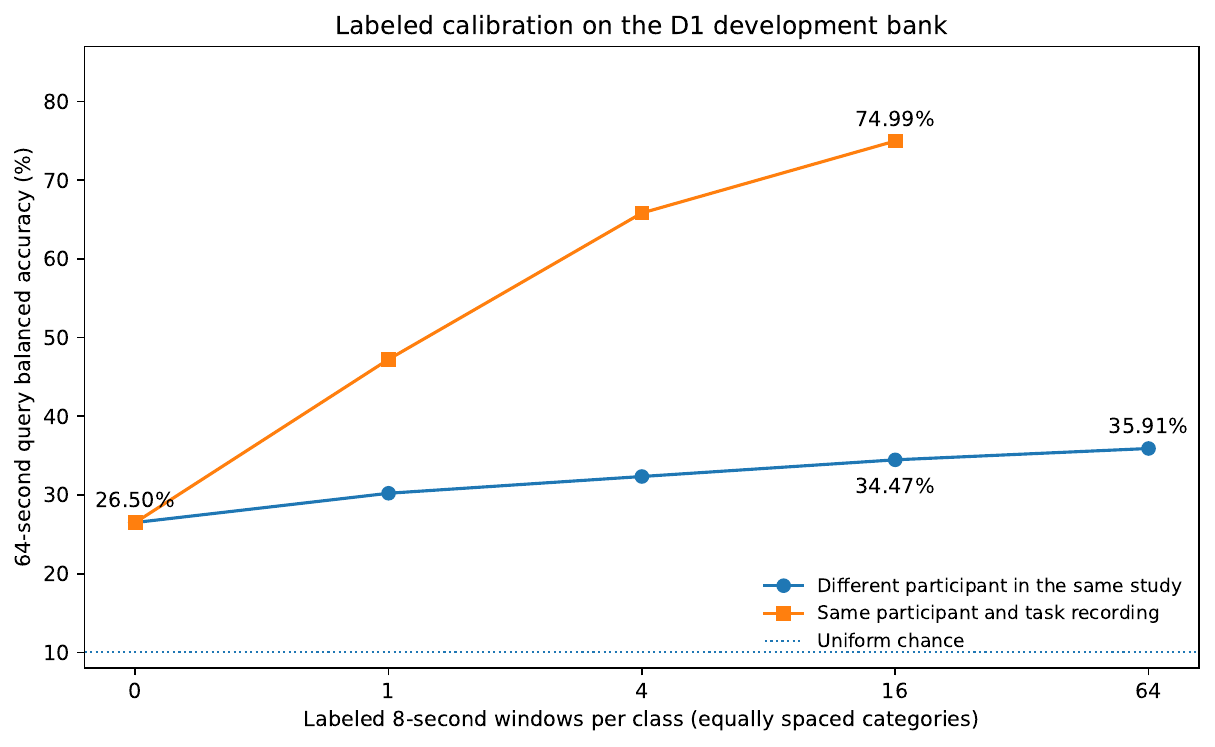}
\caption{
\textbf{Few-shot calibration substantially improves developmental task prediction, especially with examples from the same participant.}
Results for the historical full-addition model (SCALL), trained for 16,000 updates with seed 42, on the ten-task D1 development benchmark. Predictions use 64-second EEG queries, while support examples are labeled eight-second windows per available task class. Cross-participant support comes from one participant per target study, with other participants used for queries. Same-participant support comes from the queried participant's own task recording, excluding any overlap in raw EEG samples. The zero-support reference is 26.50\% for this seed; the corresponding three-seed mean is 28.98\%. Chance balanced accuracy is 10\%. Horizontal positions indicate support categories rather than a linear sample-count scale. The same-participant 64-window result is omitted because too few eligible queries remained.
}
\label{fig:fewshot-calibration}
\end{figure}

The results showed a clear difference between these two settings (Figure~\ref{fig:fewshot-calibration}). Without labeled support examples, the model achieved 26.50\% balanced accuracy on the ten-task benchmark, compared with a 10\% chance level. With cross-participant support, balanced accuracy was 30.22\%, 32.35\%, 34.47\%, and 35.91\% with 1, 4, 16, and 64 labeled windows per available class, respectively. Thus, examples from another participant were associated with a modest improvement.

The results were substantially higher when the support examples came from the same participant. Balanced accuracy reached 47.24\% with just one labeled window per available class, increased to 65.84\% with four windows, and reached 74.99\% with sixteen. These results suggest that a small amount of person-specific calibration can make task distinctions much easier to read out from the learned EEG representation.

The large benefit associated with same-participant support is encouraging, but its interpretation requires care. These calibration and evaluation examples came from the same task recording, even though overlapping EEG samples were excluded. The model could therefore benefit from participant-specific patterns, recording conditions, or other features shared between support and query data, rather than solely from cognitive-task information. Furthermore, increasing the number of support windows reduced the pool of eligible queries, so the reported differences do not constitute a
controlled comparison on an identical set of test recordings.

This was also a developmental analysis using only one trained seed. Its 26.50\% zero-support result differs from the corresponding three-seed mean of 28.98\% for 64-second queries. The findings should therefore not be interpreted as independent validation of a personalized model, or as evidence that similar performance would be obtained across recording sessions or days.

Nevertheless, these results suggest a promising direction for practical EEG applications. A model trained across many studies may provide a shared representation of cognitive activity, while a small amount of labeled EEG from an individual could help adapt predictions to
that person's signals. This possibility is especially relevant to low-burden wearable EEG,
where perfect zero-shot decoding may be difficult to achieve. Establishing whether such personalization truly generalizes will require experiments with fixed evaluation sets, independent
recording sessions, and repeated measurements across days.

\FloatBarrier
\subsection*{3.6 Grouping related cognitive tasks reveals information beyond exact task labels}

The previous analyses evaluated whether the models could identify the exact cognitive task associated with an EEG recording. However, tasks used in different experiments are not always completely distinct in terms of the cognitive processes they involve. Two studies may use different
instructions, stimuli, or response procedures while engaging similar forms of attention, memory, or decision-making. A model might therefore confuse two closely related tasks while still capturing something meaningful about their shared cognitive demands.

We asked whether this possibility could be examined by grouping semantically related tasks into broader categories. Rather than requiring the model to distinguish every individual task, we evaluated whether it could identify the broader type of cognitive activity. We refer to this process as \emph{semantic coarsening}, because it replaces fine-grained task labels with coarser categories based on their meaning.

For this exploratory analysis, we combined the task labels from D1 and D2 into a single evaluation bank containing 20 candidate tasks. The best model on this combined bank, fixed-text Bridge3 LATE, achieved 14.59\% balanced accuracy using eight-second EEG segments, compared with 5\% chance. We then constructed three alternative task groupings containing 15, 13, or 8 broader categories, denoted F15, F13, and F8, respectively. When predictions were made using the corresponding merged semantic task prototypes, the leading balanced accuracies were 15.57\%, 19.49\%, and 29.30\%.

These results suggest that the model can more readily distinguish broader task categories than individual tasks. However, this improvement must be interpreted carefully. Reducing the number of categories also makes the classification problem easier: uniform chance increases from 5\% for
20 classes to 6.67\%, 7.69\%, and 12.5\% for F15, F13, and F8, respectively. Furthermore, some distinctions that were difficult in the earlier analyses disappear when tasks are grouped together. In particular, F8 no longer requires the model to distinguish thinking from meditation or the individual driving conditions. Higher accuracy on these broader categories therefore does not necessarily indicate improved recognition of fine-grained cognitive states.

We next asked a more specific question: do the improvements depend on grouping tasks according to their cognitive meaning, or could similar improvements arise from combining labels arbitrarily? This distinction is especially important because task identity is strongly associated with recording source in our datasets. If a grouping simply combines tasks from particular studies, its apparent benefit might reflect study-specific signals rather than cognitive similarity.

To address this issue, we compared the semantic groupings with randomized groupings that preserved the same relationship to recording sources. We kept each model's original 20-class predictions unchanged and reassigned task labels only within their source datasets. This procedure preserved both the number of original task labels in each group and the exact number contributed by each source. It therefore allowed us to ask whether the chosen semantic groupings agreed with the model's existing predictions better than alternative groupings with the same source composition. We generated 10,000 randomized groupings for each partition and evaluated this control across all nine models (Table~S12).

\begin{figure}[!htbp]
\centering
\includegraphics[
    width=\linewidth,
    height=0.55\textheight,
    keepaspectratio
]{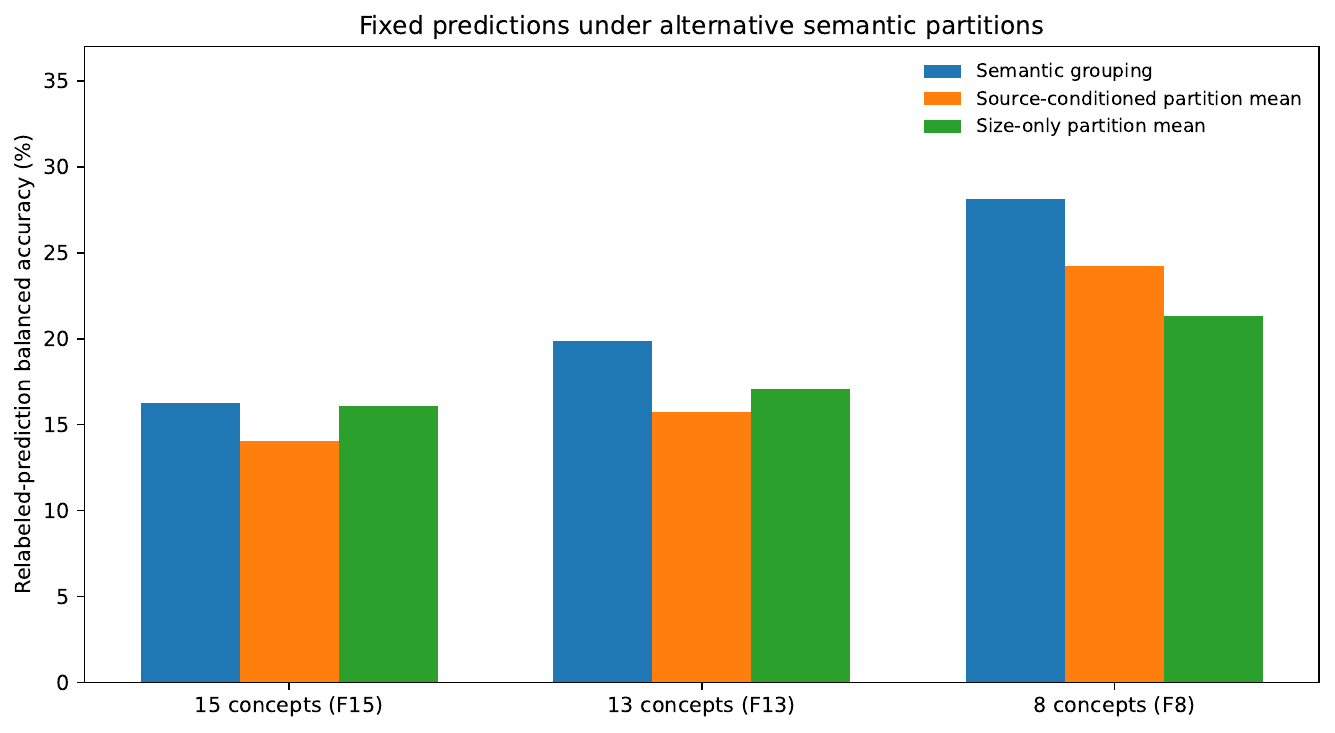}
\caption{
\textbf{Semantically related tasks are grouped more consistently with model predictions than source-matched random task groupings.}
Exploratory analysis of fixed-text Bridge3 LATE 64k on the combined 20-task D1+D2 benchmark. The original 20-class predictions are held fixed and evaluated after relabeling tasks into 15, 13, or 8 broader categories (F15, F13, and F8). Bars compare the observed semantic groupings with the mean accuracy across 10,000 randomized partitions. Source-conditioned partitions preserve both group size and the exact contribution of each recording source; size-only partitions preserve group size alone. The observed advantages over source-conditioned random groupings are 2.19, 4.18, and 3.91 percentage points for F15, F13, and F8, respectively.}
\label{fig:semantic-coarsening}
\end{figure}

For fixed-text Bridge3 LATE, the results favored the semantic groupings (Figure~\ref{fig:semantic-coarsening}). With the 13-category grouping (F13), assigning the original predictions to their broader semantic categories yielded 19.89\% balanced accuracy. The mean across source-conditioned random groupings was 15.71\%, a difference of 4.18 percentage points. Only 0.25\% of the randomized groupings reached or exceeded the observed score, under the counting convention
described in Methods~5.7.

The same pattern appeared at the other levels of grouping. For F15, the observed accuracy was 16.26\%, compared with 14.07\% for the source-conditioned random groupings, a difference of 2.19 points. For F8, the corresponding values were 28.14\% and 24.23\%, a difference of 3.91 points. Thus, grouping tasks according to their semantic relationships provided a modest advantage over randomized groupings that preserved the same source composition.

These results should be distinguished from the merged-prototype evaluation described above. In that evaluation, the model selected among newly constructed representations of the broader task categories. In the randomization analysis, the original 20-class predictions were held fixed and only their assigned category labels were changed. The two procedures therefore answer different questions and need not produce identical accuracy values.

These findings provide tentative evidence that some of the model's errors are structured by relationships between task descriptions, rather than being entirely arbitrary. The advantage over source-conditioned random groupings is particularly interesting because it cannot be explained simply by the number of merged categories or their distribution across the recorded sources.

Nevertheless, this analysis does not establish that the model learned an invariant neural representation of cognitive meaning. Tasks grouped by their descriptions may also share sensory stimuli, motor responses, or experimental procedures. Moreover, tasks coming from sources with only one evaluated label cannot provide informative within-source permutations. Both the semantic groupings and this analysis were developed after examining the evaluation results, so the randomization frequencies should be regarded as exploratory descriptions rather than confirmatory statistical evidence.

Taken together, these results suggest that evaluating EEG models only on exact experimental task labels may overlook some of the structure shared between related cognitive activities. At the same time, semantic grouping cannot substitute for demonstrating that such structure transfers across independent studies. A stronger test would require comparable cognitive operations recorded under different experimental conditions, with the semantic categories and evaluation procedure defined before model testing.

\FloatBarrier
\FloatBarrier
\section*{4. Discussion}\addcontentsline{toc}{section}{4. Discussion}

\subsection*{4.1 What task-semantic alignment achieved, and what remains unresolved}

The main goal of this study was to find a way to learn from EEG recordings collected for different cognitive experiments. These recordings are abundant, but their task labels are difficult to combine because each study uses different instructions, stimuli, and response requirements. We approached this problem by representing the tasks themselves in a shared semantic space and training an EEG encoder to predict those representations. This provides a common prediction framework in which recordings from different studies can contribute to learning without requiring every experiment to use the same task labels.

The results show both the promise and the limitations of this approach. The best model achieved 30.13\% balanced accuracy when choosing among ten cognitive tasks in the reserved evaluation, compared with 10\% chance. This demonstrates above-chance task prediction from a single anterior--posterior differential EEG channel without fitting a new classifier using labeled examples from the evaluated condition. However, the result should not be interpreted as reliable recognition of ten cognitive states. Performance differed substantially across tasks, nearly 70\% of the total balanced-accuracy score came from three classes, and two evaluated classes included participants whose EEG had been used during training on other tasks. The models also struggled to distinguish the thinking and meditation conditions and the three driving conditions when predictions were restricted to tasks recorded within the same source.

A second important finding was that the model's apparent quality depended strongly on the evaluation data. Models that performed well on the development benchmark were not necessarily the best on the reserved benchmark. In addition, grouping related task labels revealed that some prediction errors were consistent with semantic relationships between tasks, even after controlling for the distribution of those labels across recording sources. This suggests that the representation may contain information about broader task similarities, but it does not establish that the model has learned neural representations of cognitive operations that are invariant across people and studies.

The few-shot results offer a complementary perspective. When labeled examples were provided from the person being evaluated, task prediction improved markedly. This suggests that the learned representation may contain information that is difficult to use through a fixed, description-based readout but easier to access with participant-specific calibration. Nevertheless, the strongest calibration results used support and query examples from the same task recording and therefore cannot establish generalization to a new session or day.

Taken together, these findings distinguish three different achievements: learning an EEG representation from heterogeneous task-labeled data, predicting tasks without target-condition calibration, and adapting predictions using a small amount of labeled EEG. The first provides a useful way to integrate existing recordings; the second remains limited and uneven; and the third is promising but requires more independent validation. Understanding how to improve each of them is a more useful goal than interpreting any single accuracy score as evidence of general cognitive-state decoding.

\subsection*{4.2 Why multi-task contrasts and broader task coverage are complementary}

One of the central questions raised by our results is what kind of EEG data should be added when building a larger model. More recordings are not necessarily more informative in the same way. A new dataset may provide additional participants, new recording equipment, or an entirely new cognitive task. Alternatively, it may provide repeated measurements of tasks already represented in the training data. These additions create different opportunities for learning, and they may also introduce different sources of bias.

Studies in which the same participants perform multiple tasks are particularly valuable for learning task distinctions. Because several activities are recorded from the same individuals, often using the same device and experimental environment, the model can compare EEG patterns across tasks while some participant- and study-specific factors remain relatively stable. This creates an opportunity to learn what changes with the cognitive task rather than simply learning who was recorded or where the data came from. Such comparisons are not perfectly controlled: different tasks may still involve different stimuli, eye movements, motor responses, or recording blocks. Moreover, a model trained only on a small number of multi-task studies may encounter too narrow a range of cognitive operations and recording contexts.

Datasets containing only one or a few tasks offer a different benefit. They can introduce previously unseen activities and broaden the range of task descriptions with which the EEG encoder is trained to align. They also increase the diversity of participants and recording conditions. This could be important for zero-shot prediction of tasks that were not represented in the original training collection. The difficulty is that, when a particular task appears in only one study, its identity is closely associated with the characteristics of that study. The model may learn to distinguish recording sources instead of the cognitive operations they contain. At a fixed training budget, adding many such recordings can also reduce the relative exposure to the within-participant task contrasts already present in the training mixture.

Our experiments illustrate this tension. In the historical dataset-addition experiment, the largest training mixture outperformed the original mixture on the development benchmark, showing that additional heterogeneous recordings could be useful (Figure~\ref{fig:study-addition}). However, the gains were not monotonic: some intermediate additions reduced performance. Similarly, the broader ST40 training mixture did not improve reserved-benchmark accuracy over ST19 in the corresponding RAW comparison (Table~3). These results do not demonstrate that single-task datasets are harmful, because task composition, participant coverage, sampling exposure, and other factors changed together. Rather, they show that expanding the training collection does not guarantee better transfer.

The structure of the training data provides a clue to why this may happen. From ST19-MATCHED to ST40, the number of represented task categories increased from 41 to 59. Yet the number of task categories appearing in at least two registered study families remained at 16, and the number of families containing multiple task categories remained at 19. The model therefore gained access to more kinds of tasks without an increase in these two counts of repeated task observations. This does not mean that the amount of data available for each repeated task was unchanged. It does mean that broader coverage did not automatically produce broader replication of comparable tasks across studies (Table~4).

The ideal strategy may therefore combine the strengths of both types of data. We need recordings in which the same people perform several tasks, allowing meaningful comparisons within individuals, as well as recordings in which comparable tasks are repeated across independent groups, devices, and environments. In such a design, the model would encounter both differences between tasks under relatively similar conditions and similarities between tasks across different conditions. Table~5 summarizes these complementary roles and the remaining uncertainties.

\Needspace{236pt}
\par\medskip\noindent{\small\textbf{Table 5. Complementary contributions of multi-task and task-sparse EEG datasets.}}
{\fontsize{9.0}{11.3}\selectfont\setlength{\tabcolsep}{3.6pt}\renewcommand{\arraystretch}{1.12}
\begin{longtable}{>{\raggedright\arraybackslash}p{77pt}>{\raggedright\arraybackslash}p{160pt}>{\raggedright\arraybackslash}p{205pt}}
\toprule
\textbf{Training data} & \textbf{Potential gain} & \textbf{Main ambiguity or limitation} \\
\midrule\endfirsthead
\multicolumn{3}{l}{\textit{Table 5 (continued)}}\\
\toprule
\textbf{Training data} & \textbf{Potential gain} & \textbf{Main ambiguity or limitation} \\
\midrule\endhead
\midrule\multicolumn{3}{r}{\textit{Continued on next page}}\\\endfoot
\bottomrule\endlastfoot
Multiple tasks per person & Task contrasts with relatively shared participant, device and study characteristics; supports discriminative representations. & Limited task and context coverage; task order, stimuli and responses remain possible shortcuts. \\
One or few tasks per study & Wider task vocabulary, additional people and hardware conditions; possible support for unseen-task transfer. & Task identity can identify the source; new exposure can dilute contrasts in a finite sampling budget. \\
Crossed hybrid design & Multiple tasks within people and repeated operations across independent studies and days. & Requires coordinated collection and held-out tests; its benefit has yet to be measured under matched conditions. \\
\end{longtable}}
\vspace{-0.6em}\noindent{\footnotesize Mechanistic expectations and future tests, rather than a causal conclusion from the present mixtures. ``Single-task'' describes the study's task diversity, not necessarily one label per participant.}\par\medskip

The importance of separating task information from recording-source information is also illustrated by two additional analyses. Under the label structure of D2, a hypothetical classifier that knew only the source dataset could achieve as much as 70\% class-balanced accuracy, despite having no ability to distinguish conditions within a source. This is a property of the evaluation set, not the performance of our EEG models. Separately, a supervised classifier trained to identify recording sources from simple EEG features achieved 78.83\% balanced accuracy across nine sources, compared with 11.11\% chance (Methods~5.10). This probe was trained after the reserved benchmark had been examined, and it does not show that the task models used the same features. Nevertheless, it demonstrates that source-related information can be strongly expressed in the recordings.

A decisive future comparison would vary the amount of within-person task contrast and the degree of task repetition across studies while holding total EEG duration, model training budget, and other relevant factors as constant as possible. Evaluation should include both unseen studies and discrimination between tasks recorded within the same source. The goal would be to determine whether repeated evidence about the same cognitive operation across contexts improves generalization more reliably than adding new task labels alone.

\subsection*{4.3 What the model learns depends on its task descriptions and EEG inputs}

Our approach assumes that descriptions of experimental tasks provide useful targets for learning EEG representations. While this is a practical assumption, the meaning assigned to a task depends on how the task was described. Two experiments may use similar instructions while differing in sensory stimulation or response demands. Conversely, two experiments engaging similar cognitive operations may be described using different terminology. The semantic vectors are therefore a structured form of supervision, not direct measurements of the cognitive processes taking place in the brain.

This limitation seems to be indeed present in our study. For example, one resting-state task specifies a manual response in its canonical description but no response in its richer description. We retained the descriptions and vectors used in the original experiments so that the reported results would remain faithful to the evaluated models. Revising the annotations after observing the reserved-benchmark results and then selecting models on that basis would require a new development and evaluation cycle.

The semantic-grouping analysis provides a useful but limited indication that the chosen task descriptions capture relationships relevant to EEG prediction. Some broader semantic groupings agreed with the model's errors more than randomized groupings with the same recording-source composition (Section~3.6). However, similar task descriptions can also reflect similar stimuli, response rules, or experimental procedures. We therefore cannot conclude from this analysis alone that the model has learned a shared neural representation of abstract cognitive functions. A stronger test would use tasks with similar intended cognitive demands but different sensory and motor implementations, evaluated in independent studies.

The choice of learning architecture introduces a related uncertainty. In our experiments, both the fixed-text model and CLIP-X already used bidirectional contrastive alignment between EEG and task representations. CLIP-X additionally changed several components, including the adaptation of text vectors, the similarity scale, regularization, and candidate-task-bank construction. Its lower performance on the reserved benchmark cannot therefore be attributed solely to making the task representations trainable. Controlled comparisons that change one component at a time would be needed to identify why these approaches generalized differently. Related language-conditioned EEG models have also reported sensitivity to the supplied linguistic context \citep{jiang2026leaf,zhang2026prime}.

The EEG representation itself may impose further limits. The RAW encoder summarizes locally processed waveform features over an eight-second window using temporal averages and variation, rather than explicitly modeling the full sequence of events. Adding spectral representations provides complementary information, but may also expose differences in recording bandwidth, noise, or preprocessing. Resampling recordings to a common frequency does not remove these differences. Because the original source-level processing histories are not fully reconstructed (Table~S3), we cannot completely separate task-related signals from acquisition-related effects or claim exact end-to-end reproduction of every training run.

Finally, the motivation for using task-semantic supervision should not be mistaken for a conclusion that self-supervised learning is ineffective for EEG. Our earlier signal-only pretraining experiments produced only small and inconsistent benefits under the protocols we tested (Supplement~S0). Other architectures, larger training collections, different adaptation methods, or combinations of self-supervised and task-supervised objectives may behave differently \citep{kostas2021,jiang2024labram,wang2025cbramod,elouahidi2025reve,yang2026worth,liu2026compass,sukhbaatar2026singlem,an2026stem}. The present work does not include a protocol-matched comparison with public EEG foundation models or compact supervised baselines such as EEGNet \citep{lawhern2018}. Those comparisons are necessary to establish how much of the observed performance is attributable specifically to the proposed semantic-alignment approach.

\subsection*{4.4 From laboratory EEG to everyday cognitive-state prediction}

The long-term motivation for using a single EEG channel is practical. If a useful representation of cognitive activity could be obtained from a small number of electrodes, it might support devices that are easier to wear and use outside the laboratory. Our experiments therefore used one anterior--posterior differential signal formed from two physical electrode locations. This deliberately restricted input is different from attempting to maximize accuracy using all available laboratory electrodes.

However, the present results do not establish that a consumer wearable can already recognize everyday cognitive states. The underlying recordings came from research datasets, and the evaluated labels referred to experimentally defined tasks. We did not test a new wearable device during normal daily activities. Furthermore, using one channel may discard neural information useful for distinguishing subtle mental states. A direct comparison between the same anterior--posterior channel, sparse multi-electrode configurations, and conventional higher-density EEG would help measure that trade-off. Such comparisons should use the same participants, task periods, and evaluation splits so that the benefit of additional electrodes can be assessed alongside the burden of wearing them.

The distinction between zero-shot and few-shot prediction is especially relevant for practical deployment. A zero-shot model could, in principle, recognize activities without requiring each user to perform an initial calibration procedure. Our present zero-shot results show that this remains difficult. Few-shot calibration offers a more realistic alternative if a person is willing to provide a small number of labeled examples. The strong same-participant developmental result is encouraging in that respect, but its support and query examples came from the same recording. We do not yet know whether similar gains would persist after the device is removed and reapplied, or when the user performs the activity on another day. A practical personalized system would need to be evaluated on later, independent sessions using the same held-out query recordings across different calibration budgets.

Everyday recordings also introduce problems that are largely absent from our present evaluation. People move between activities, perform several things at once, change their goals, and encounter situations outside the model's predefined task vocabulary. Electrode contact, motion artifacts, and recording conditions change throughout the day. A useful system would therefore need to recognize transitions, represent uncertainty, and sometimes decline to assign one of the known task labels. Its predictions would also need to be interpreted as uncertain estimates of broad activities or task demands, not direct access to a person's moment-to-moment thoughts or clinical state.

Several limitations of the current evidence reinforce this distinction between a laboratory result and a practical device. The few-shot analysis used one training seed and a changing set of eligible queries; the within-source comparisons involved only two particular sources; and the semantic-partition and source-identification analyses were exploratory. The reported participant-bootstrap intervals are conditional on the available studies and trained models, and intervals that include chance do not establish that neural information is absent. These limitations define the next questions to test rather than ruling out the possibility of useful low-burden cognitive-state prediction.

\subsection*{4.5 Designing data that separate cognitive activity from recording context}

The most important implication of these findings may concern future EEG data collection. If the aim is to predict cognitive activity across people and situations, simply increasing the total number of EEG windows is unlikely to be sufficient. The training data must also make it possible to distinguish an activity from the participant, recording environment, and experimental procedure associated with it. The results on source confounding motivate repeating comparable tasks across independent studies, while the few-shot results motivate collecting repeated sessions from the same individuals.

Several existing resources already contain useful elements of this design. ERP CORE provides multiple experimental paradigms \citep{kappenman2021}, and the Healthy Brain Network includes a common EEG task battery \citep{alexander2017hbn}. The Zurich Cognitive Language Processing Corpus combines reading-related EEG with eye tracking \citep{hollenstein2018zuco}. The Brain, Body, and Behavior Dataset studies learning and attention during educational videos \citep{madsen2026bbbd}, while the Everyday Activity Science and Engineering Table Setting Dataset combines EEG with synchronized behavioral annotations \citep{meier2026ease}. These datasets show that multi-task and naturalistic EEG collection is already feasible. What is still needed for the present goal is a benchmark that combines repeated tasks within individuals, comparable operations across independent recording environments, and evaluations designed specifically to separate cognitive information from recording context.

A practical collection could begin with a small set of everyday activities performed by every participant: checking messages, reading, watching videos, writing, planning, and taking quiet breaks. Each activity would be repeated on different days, with the device removed and reapplied between sessions. The same core activities would be collected at multiple sites, while optional modules could provide additional task diversity. Controlled computer-based sessions could establish reliable task boundaries and experimental contrasts, followed by more naturalistic recordings at home or at work. Stimuli, task order, response schedules, and time of day should be varied so that the model cannot identify activities simply from a particular block or recording routine.

Importantly, an everyday activity is not equivalent to a single cognitive operation. Checking messages might involve finding a fact, prioritizing requests, remembering information, or composing a reply. Watching a video might involve passive viewing, learning material for a later test, or monitoring for an event. We therefore propose varying the goal while keeping some aspects of the activity or stimulus comparable. Conversely, the same cognitive goal should be tested across different applications and sensory modalities. For example, retaining information could be studied while reading a message, reading an article, or watching a video. This would allow us to ask whether the model recognizes the cognitive demand rather than the application or stimulus format. Table~6 provides examples of these activity--goal contrasts.

\Needspace{340pt}
\par\medskip\noindent{\small\textbf{Table 6. Proposed activity--goal contrasts for a future dataset.}}
{\fontsize{9.1}{11.6}\selectfont\setlength{\tabcolsep}{4pt}\renewcommand{\arraystretch}{1.15}
\begin{longtable}{>{\raggedright\arraybackslash}p{82pt}>{\raggedright\arraybackslash}p{196pt}>{\raggedright\arraybackslash}p{164pt}}
\toprule
\textbf{Everyday activity} & \textbf{Within-person contrast} & \textbf{Replication and held-out test} \\
\midrule\endfirsthead
\multicolumn{3}{l}{\textit{Table 6 (continued)}}\\
\toprule
\textbf{Everyday activity} & \textbf{Within-person contrast} & \textbf{Replication and held-out test} \\
\midrule\endhead
\midrule\multicolumn{3}{r}{\textit{Continued on next page}}\\\endfoot
\bottomrule\endlastfoot
Checking messages & Find a named fact, prioritize requests, or retain a request for a later reply. Record reading, deciding and typing separately. & New messages, interface layouts and days; repeat information retention in documents and videos. \\
Viewing videos & View freely, learn for a later test, or monitor a specified event in matched clips. Annotate pauses and concurrent activity. & New clips and days; test learning goals in reading or listening contexts. \\
Reading and writing & Read for gist or locate a fact; copy supplied text or compose an answer. Match response format where feasible. & New topics and applications; separate language demands from typing and gaze. \\
Planning & Look up a stated schedule or integrate constraints to construct a plan using comparable calendar material. & New schedules, interfaces and independently collected cohorts. \\
Switching and breaks & Continue an activity or interrupt and resume it; sample quiet rest and unprompted pauses as well as instructed tasks. & New transition sequences and days; recognize boundaries and permit unknown or overlapping labels. \\
\end{longtable}}
\vspace{-0.6em}\noindent{\footnotesize Proposed collection, not an experiment reported here. The contrasts concern instructed goals and observable behavior, not direct ground truth for latent cognitive processes. Shared goals should recur across activities and sites, with repeated sessions for the same people.}\par\medskip

Collecting the EEG signal is only part of the challenge. The labels must also specify what was observed and what was inferred. We would distinguish the activity being performed, the instruction or goal given to the participant, the sensory events and responses occurring during it, and any evidence that the intended cognitive operation was actually engaged. A task instruction is not proof of a mental state. Brief reports, comprehension tests, or delayed memory tests could provide additional evidence, but their timing and uncertainty should be retained.

The temporal structure of everyday cognition also requires more detailed annotations than a single label assigned to every eight-second EEG window. Opening a message or pressing a button is an event with a particular onset, whereas maintaining information or sustaining attention may extend over an interval. Different activities can overlap, and their boundaries are often uncertain. The dataset should therefore preserve continuous EEG alongside accurately synchronized event markers and interval annotations. Our eight-second prediction task can remain a common benchmark, but shorter events and longer-lasting states should be evaluated at appropriate timescales. When testing EEG-only decoding from continuous recordings, the model must not be given the true task boundaries or behavioral logs as hidden inputs.

The intended wearable signal should be recorded using the device that would actually be deployed. A subset of sessions could include simultaneous higher-density EEG to determine what information is lost when using fewer electrodes. Eye movements, muscle activity, and motion measurements would be useful for understanding alternative explanations for model predictions. However, these measurements should be used for annotation or separately reported control analyses unless the experiment explicitly evaluates a multimodal system. They must not silently become additional inputs to a model claimed to operate on EEG alone.

Privacy is another central consideration for everyday activity recordings. Real messages, documents, and screen contents may reveal sensitive information about participants or third parties. Controlled sessions with study-provided materials would create a shareable reference dataset, while naturalistic sessions would require explicit consent, careful minimization of retained content, and participant review. In many cases, event timing and audited semantic annotations may be scientifically useful without releasing private message content or identifying details. The derived annotations themselves would still require privacy assessment.

Finally, the evaluation must be designed to answer the questions that the present study could not resolve. Separate tests should assess unfamiliar participants, later sessions from known participants, independent study sites, novel stimuli or applications, and new combinations of activities and goals. Task and participant exposure during all forms of training, including self-supervised pretraining, should be documented. Few-shot calibration should use examples from earlier sessions or other people while keeping the future-session test set fixed. Continuous evaluation should include transitions, overlapping activities, and unknown states, not only balanced accuracy for clearly separated task blocks. The final test sets should be fixed before model selection rather than repeatedly reused during development.

The decisive experiment would compare training collections that differ in the amount of within-person task diversity and cross-study task repetition while matching total EEG duration, sampling, and optimization as closely as possible. Greater diversity may improve transfer, but it may also introduce noise or new shortcuts. The relevant measure of progress is therefore performance on prespecified independent tests, not simply the number of tasks represented during training or the accuracy of distinguishing one familiar application from another. This would test the hypothesis suggested by our exploratory results: for cognitive EEG foundation models, the most useful form of scale may be repeated evidence about the same cognitive operations across people and contexts, rather than additional recordings from isolated task--study combinations.

\FloatBarrier

\Needspace{190pt}
\section*{5. Methods}\addcontentsline{toc}{section}{5. Methods}

The methods below describe how recordings collected for different experiments were converted into a common prediction problem. We first explain how we defined the EEG input and tracked which studies and tasks had been used in training. We then describe how task descriptions and EEG segments were represented, how the two types of representations were aligned during training, and how predictions were evaluated. Finally, we detail the additional analyses used to investigate possible recording-source effects and the limits of the learned task representations.

\subsection*{5.1 Defining the EEG input and tracking study and task exposure}

To compare recordings from studies using different EEG systems, we restricted the primary input to one physically measured anterior--posterior differential signal, referred to as AP2E. We formed this signal by subtracting a recorded posterior channel from a recorded anterior channel, or by using an admitted pair close to these electrode locations. The model received 4,000 samples per primary example, corresponding to eight seconds at 500~Hz. Importantly, the signal was derived from two actual electrodes; a virtual channel reconstructed by interpolation was not treated as equivalent. Source-specific preprocessing and interval selection followed the archived preprocessing records. The processing details recoverable for this release are summarized in Table~S3, and should not be taken as a complete reconstruction of every source pipeline.

Because the central question concerns generalization to new studies and tasks, we also tracked what each model might already have encountered during training. The composition of the reserved D2 evaluation set was revised during quality assurance without consulting model scores, and was then fixed before scoring. Two D2 task holdouts, from \texttt{ds002680} and \texttt{ds005189}, included participants who had contributed training EEG for other tasks. The driving conditions have different novelty classifications across models because ST40 and Full-Mixture had been exposed to driving-related tasks. Under the strict study-and-task novelty definition shared across the evaluated models, the common ZS-Both subset consists of the thinking and meditation conditions from one source. Unless we explicitly state that candidate labels were restricted, subgroup recall was computed while retaining the original ten possible task predictions.

We checked the source identifiers used in the D1 and D1-ST development evaluations against the primary training indices of ST19-MATCHED, ST40 and Bridge3, as well as the common auxiliary source M3CV (\texttt{nm000166}) \citep{data_nm000166}. None of these exact evaluation identifiers appeared in those indices (Table~S2). This is a check of recorded identifiers, not proof that all participants, files or related study families were independent: such overlaps may be hidden by aliases or different processing branches. We therefore defined study and task novelty with respect to each model's frozen training register, while recording participant exposure separately.

We also used those training registers to describe how well tasks were represented within and across studies. A \emph{study-family--task edge} was counted whenever an accepted eight-second training-index entry linked a registered study family to a verified functional task (VFT) category. We counted each such relationship once, regardless of how many EEG windows it contained. The VFT categories group source-native tasks according to registered functional equivalence; they may still combine conditions with different stimuli, responses or task phases. Native four-second auxiliary examples were recorded separately and were not included in these primary-graph counts.

\subsection*{5.2 Representing cognitive tasks and making predictions}

The model needs a common set of possible outputs even when different studies use different task names. We supplied this through textual task descriptions. For each candidate task, we used two complementary descriptions: a more detailed \emph{rich} description and a field-structured \emph{canonical} description. Both were embedded using the same all-MiniLM-L6-v2 sentence encoder, fixed at revision \nolinkurl{1110a243fdf4706b3f48f1d95db1a4f5529b4d41}. The sentence-embedding framework, MiniLM model architecture and pretrained model card are documented separately \citep{reimers2019,wang2020minilm,allminilmcard}. The intended model-facing descriptions omitted explicit dataset and device names, although details of task procedures and some inconsistencies between descriptions remained possible sources of information.

Each description produces a vector in the same semantic coordinate system. For a candidate task $c$, let $p_{c1}$ and $p_{c2}$ denote its rich and canonical vectors, and define $\operatorname{norm}(u)=u/\|u\|_2$. In the fixed-text models, we first normalize each vector, add them, and normalize again to obtain a single task prototype, $p_c=\operatorname{norm}(\operatorname{norm}(p_{c1})+\operatorname{norm}(p_{c2}))$. In jointly adapted models, each text view passes through the learned adapter before the two views are combined. Separately, the EEG encoder produces a representation $z(x)$, which is projected by $W$ into the task space and normalized as $e(x)$. Prediction then requires no newly trained task-specific output classifier: it chooses the candidate description with the greatest cosine similarity to the EEG representation,

\begin{equation}\widehat y(x)=\underset{c\in\mathcal C}{\operatorname{argmax}}\;e(x)^\top p_c,\qquad e(x)=\operatorname{norm}(Wz(x))\label{eq:1}\end{equation}

where $\mathcal C$ is the candidate set supplied for that evaluation. Changing the candidate set changes the prediction problem, even when the EEG encoder is unchanged.

We also tested a distinct, lower-dimensional task representation during development, called B128-LIT. Its purpose was to examine whether a more compact text space could provide a useful target for EEG alignment. We fitted principal component analysis (PCA) to 15,304 text strings represented by normalized MiniLM vectors: 857 Cognitive Atlas task names, 918 Cognitive Atlas concepts, 13,459 NeuroQuery article titles, and 70 training-task descriptions in that build \citep{poldrack2011atlas,dockes2020neuroquery}. The 11 D1 task descriptions were excluded from PCA fitting. For prediction, we subtracted the fitted corpus mean, projected vectors into 128 principal components, and normalized them. Those components explained 82.25\% of the text-vector variance in the fitting corpus. The shared-task and matched-training analyses used their respective registered transformations; B128-LIT was not silently substituted for the final fixed-text task interface. More generally, replacing the sentence encoder requires aligning the EEG projection to the replacement text space: equal vector dimensions do not imply that two embedding spaces share the same coordinates.

\subsection*{5.3 Extracting EEG features and constructing the training mixtures}

We compared two ways of representing exactly the same one-channel EEG input. The \emph{RAW} model learns features directly from the voltage waveform, whereas \emph{LATE} adds a parallel representation of how signal power changes across frequencies and time. The purpose of LATE was to test whether spectral information could complement waveform features; it did not receive additional EEG electrodes.

The RAW encoder consists of three one-dimensional convolutional layers with 32, 64 and 128 output channels, kernel lengths of 51, 15 and 9 samples, and strides of 5, 2 and 2. The layers use Gaussian error linear unit (GELU) nonlinearities, and the input amplitude is divided by a fixed factor of 20. After convolution, we summarize the feature sequence by its mean and standard deviation over time, producing 256 features. A learned 256-to-256 projection followed by normalization creates the shared EEG representation. A separate bias-free 256-to-384 projection $W$ maps it to the semantic task space. The convolutional stack contains 106,304 parameters, the 256-to-256 projection 65,792, and $W$ 98,304, for 270,400 parameters along the common RAW inference path. Before the global temporal pooling, the nominal receptive field is $51+14\times5+8\times10=201$ input samples.

For LATE, we compute three short-time Fourier representations of the same EEG segment. We use periodic Hann windows of 256, 512 and 1,024 samples, with respective hops of 64, 128 and 256 samples; transforms are neither centered nor padded. Their configured frequency range is 0.5--100~Hz. The one-sided power spectral density (PSD) is transformed to $\log_{10}(\mathrm{PSD}+10^{-12})$, without fitting a separate spectral normalization to each dataset. Each resolution is processed by its own two-layer two-dimensional convolutional branch: first 1-to-8 channels with a $(5,3)$ frequency-by-time kernel and $(2,1)$ padding, then 8-to-16 channels with a $(3,3)$ kernel and $(1,1)$ padding. Both layers have stride $(1,1)$, include biases, and are followed by GELU. Averaging over frequency and then pooling the temporal mean and population standard deviation produces 32 features per resolution; the variance calculation adds $10^{-6}$ before the square root. The three spectral branches therefore supply 96 features and contain 3,888 parameters in total.

LATE combines these 96 features with the 256 waveform features, giving a 352-dimensional input to a 352-to-256-to-256 fully connected fusion network. This network includes biases and GELU after each layer and contains 156,160 parameters. Together with the waveform encoder (106,304), spectral branches (3,888) and text projection (98,304), the LATE inference path contains 364,656 parameters. These parameter totals exclude source-native classification heads and instantiated auxiliary heads, whose numbers depend on the training configuration (Supplement~S3). Keeping these counts separate avoids mistaking a training-only head for an input to the final task prediction.

All final EEG models documented in the inspected training specifications were initialized separately from scratch for each training seed; they did not continue training from one common signal-only self-supervised checkpoint. As a concrete example of the optimization protocol, the Full-Mixture configuration used AdamW with an encoder learning rate of $10^{-4}$ and a projection/head learning rate of $3\times10^{-4}$. Its optimizer parameters were $\beta=(0.9,0.95)$, weight decay 0.01 and gradient-norm clipping at 1. An effective batch contained 1,536 examples: 1,426 primary eight-second windows and 110 native four-second auxiliary examples. Learning-rate warm-up covered 5\% of its 256,000-update training horizon, followed by cosine decay to one tenth of the initial learning rate. Bridge3 and CLIP-X used 64,000-update horizons in the final comparison. Checkpoints with the same update count should not be assumed to have identical learning-rate histories; the actual schedules are specified in their training records.

Dataset composition was controlled separately from model architecture and training duration. For the primary data stream, sampling first selected a VFT category and then a namespaced participant, a task unit and an accepted EEG window. Bridge3 preserved approximately 95\% of its primary sampling mass from ST19-MATCHED and allocated the remaining 5\% to three added datasets: \texttt{ds003458}, \texttt{ds004200} and \texttt{ds004917}. CLIP-X used the same primary source and interval indices, the same 95:5 primary allocation, and the common native four-second auxiliary pool in its final training mode. Its internal pilot used separate held-out splits to choose the training objective. The dataset-index audit counts windows eligible for sampling, not the exact number of times each window was drawn. Moreover, models or seeds with the same nominal mixture can still encounter different sequences of sampled examples.

\subsection*{5.4 Training the EEG encoder against fixed task representations}

The fixed-text model learned to identify tasks in two complementary ways. First, it used the original task labels within individual studies, which can help distinguish conditions recorded in a similar environment. Second, it aligned EEG representations with shared semantic descriptions, so that tasks from different sources could participate in one prediction space. The total objective combined these two components with equal weight.

For the first component, let $z_i$ be the normalized 256-dimensional EEG representation of example $i$, $t_i$ its source-native task label, $f_i$ its registered study family, and $v_i$ its VFT category. Each represented family $f$ has its own learned native-task classifier $h_f$. We compute cross-entropy within that family's examples and then average across families. Families whose native head has only one possible output are omitted because such a head cannot distinguish task conditions. This is therefore not one global classifier trained to recognize study sources:

\begin{equation}\mathcal L_{\mathrm{native}}=\frac{1}{|\mathcal F_B|}\sum_{f\in\mathcal F_B}\frac{1}{|I_f|}\sum_{i\in I_f}\operatorname{CE}(h_f(z_i),t_i)\label{eq:2}\end{equation}

The second component brings an EEG example close to the descriptions of its corresponding VFT while contrasting it against other task descriptions. Each VFT has two frozen text views, $p_{vr}$ with $r\in\{1,2\}$, and both are treated as positive targets for an EEG example of that task. We calculate cosine-similarity logits $a_{ivr}=e_i^\top p_{vr}/\tau$, using $e_i=\operatorname{norm}(Wz_i)$ and a fixed temperature $\tau=0.1$. The EEG-to-text loss compares the total similarity assigned to the two correct views with the total across the full task-prototype bank:

\begin{equation}\mathcal L_{E\to T}=-\frac{1}{B}\sum_{i=1}^{B}\log\frac{\sum_{r=1}^{2}\exp(a_{i,v_i,r})}{\sum_{v\in\mathcal V}\sum_{r=1}^{2}\exp(a_{i,v,r})}\label{eq:3}\end{equation}

We also learn in the reverse direction. For each task view, all EEG examples in the batch assigned to that VFT count as positives, while the full EEG batch forms the comparison set. This text-to-EEG loss is averaged over both text views and the VFT categories present in the batch:

\begin{equation}\mathcal L_{T\to E}=-\frac{1}{2|\mathcal V_B|}\sum_{v\in\mathcal V_B}\sum_{r=1}^{2}\log\frac{\sum_{i:v_i=v}\exp(a_{i,v,r})}{\sum_{i=1}^{B}\exp(a_{i,v,r})}\label{eq:4}\end{equation}

Finally, the two directions contribute equally to the alignment loss, which is added to native-task supervision:

\begin{equation}\mathcal L_{\mathrm{align}}=\tfrac{1}{2}(\mathcal L_{E\to T}+\mathcal L_{T\to E}),\qquad\mathcal L_{\mathrm{fixed}}=\mathcal L_{\mathrm{native}}+\mathcal L_{\mathrm{align}}\label{eq:5}\end{equation}

Here $\mathcal F_B$ denotes the represented families with multi-output native classifiers, $I_f$ their examples in the batch, $\mathcal V$ the task-prototype bank, and $\mathcal V_B$ the VFTs represented in that batch. Instantiated facet heads were inactive in these fixed-text baseline runs. The union prototype bank could contain descriptions of tasks that were not sampled in a particular smaller training condition; the released inventory does not consolidate every run's exact denominator membership and execution bindings. Thus, the equations specify the archived objective, but the documented differences in bank membership remain relevant when comparing historical runs.

\subsection*{5.5 Adapting task vectors jointly with the EEG representation}

The fixed-text approach assumes that the pretrained task-description space is already suitable for EEG decoding. We also tested whether allowing task representations to change during training would improve alignment. The resulting CLIP-X model used the same bidirectional, multi-positive EEG--text contrastive structure as the fixed-text model, but it added text adaptation and two regularization terms. Thus, the comparison is between learning packages, not between contrastive and noncontrastive training.

CLIP-X applies a small residual adapter to each original 384-dimensional task vector $p$. It first normalizes the vector, passes it through LayerNorm, a 384-to-192 linear transformation, GELU and a 192-to-384 output transformation, and adds the resulting correction to the normalized original vector before normalizing again. The final layer's weights and bias start at zero so that the adapter initially leaves the original task vector unchanged. The model also learns its similarity scale, initialized to 10 and capped at 100; the initial effective temperature is therefore 0.1. Its full-mode task bank contains the union of eligible primary and four-second auxiliary VFT categories. A software correction that added auxiliary-only entries to this bank was documented before final D2 scoring.

\begin{equation}\begin{aligned} A(p)&=\operatorname{norm}\big(\operatorname{norm}(p)+r(p)\big),\\ r(p)&=W_2\operatorname{GELU}\big(W_1\operatorname{LayerNorm}(p)+b_1\big)+b_2 .\end{aligned}\label{eq:6}\end{equation}

CLIP-X uses the same EEG-to-text and text-to-EEG positive-pair definitions as Equations~\ref{eq:3}--\ref{eq:5}, replacing the frozen text vectors with their adapted versions and using the learned similarity scale. We added two constraints to discourage unhelpful changes. The first, a \emph{text-geometry penalty}, encourages the relative similarities among task descriptions to remain close to their pretrained structure. We form normalized rich/canonical centroids for each task and penalize the squared difference between the adapted and original pairwise cosine-similarity matrices:

\begin{equation}\mathcal L_{\mathrm{geometry}}=\frac{1}{|\mathcal V|^2}\sum_{u,v\in\mathcal V}\left(\bar p_u^{\top}\bar p_v-\bar p_{u,0}^{\top}\bar p_{v,0}\right)^2\label{eq:7}\end{equation}

The second, a \emph{cross-study consistency penalty}, addresses the possibility that EEG samples of the same registered task may occupy different parts of the representation space depending on their source. It operates on the normalized 256-dimensional EEG vector $z$, before the semantic projection $W$. For each VFT present in at least two sources in a batch, we compute the mean representation within each source, normalize these means, and penalize their pairwise cosine distance. The term averages across eligible tasks and is defined as zero if no task appears in at least two sources in that batch:

\begin{equation}\mathcal L_{\mathrm{cross}}=\frac{1}{|\mathcal V_{\ge2}|}\sum_{v\in\mathcal V_{\ge2}}\frac{1}{\binom{|\mathcal S_v|}{2}}\sum_{s<t\in\mathcal S_v}\left(1-\mu_{vs}^{\top}\mu_{vt}\right)\label{eq:8}\end{equation}

In Equations~\ref{eq:7}--\ref{eq:8}, $\bar p_v$ and $\bar p_{v,0}$ are the adapted and original text centroids, $\mathcal S_v$ is the set of sources contributing task $v$ in the batch, and $\mu_{vs}$ is the normalized mean EEG representation for that task and source. Both additional terms have weight 0.05 in the final training objective:

\begin{equation}\mathcal L_{\mathrm{X}}=\mathcal L_{\mathrm{native}}+\mathcal L_{\mathrm{align}}+0.05\mathcal L_{\mathrm{geometry}}+0.05\mathcal L_{\mathrm{cross}}\label{eq:9}\end{equation}

An internal development pilot selected the combination of geometry preservation and cross-study consistency over the geometry-only variant. We then evaluated the resulting compound CLIP-X model on D2. Because the model also differed in text adaptation, similarity scaling and bank construction, that evaluation does not isolate the effect of any single addition.

\subsection*{5.6 Scoring predictions and estimating uncertainty}

A ten-task classification score can be dominated by studies or participants contributing the most EEG windows. We therefore used a balanced-accuracy (BA) estimator that gives the same weight to each true task class and, within that class, to each participant. Specifically, we first calculate the fraction of correct windows for each participant and class, then average those fractions across participants within the class, and finally average across the $K$ classes. This was the primary D2 metric, and D1 was recomputed using the same weighting when model rankings were compared. Macro-F1 and the share of predictions assigned to each label were derived from the resulting class--participant-equal confusion matrix. Table~S1 distinguishes these estimates from earlier window-pooled and person-first summaries.

\begin{equation}\operatorname{BA}=\frac{1}{K}\sum_{c=1}^{K}\frac{1}{N_c}\sum_{p\in P_c}\frac{1}{n_{pc}}\sum_{i=1}^{n_{pc}}\mathbf{1}[\widehat y_{pci}=c]\label{eq:10}\end{equation}

In Equation~\ref{eq:10}, $P_c$ is the set of participants who supplied windows for class $c$, $N_c=|P_c|$, and $n_{pc}$ is the number of accepted windows from participant $p$ in that class. Under uniform guessing among $K$ candidates, the expected BA is $1/K$.

We reported predictions from both eight-second windows and some longer query intervals, but the 64-second protocols must be distinguished. In D2, a 64-second query consists of exactly eight consecutive, nonoverlapping eight-second windows from one task recording. We sum each candidate's similarity scores over those eight windows before choosing a label; using the mean would give the same ranking. In historical B2 and shared-task analyses, the procedure instead averages 29 overlapping eight-second scores with two-second spacing across a 64-second interval. We kept these protocols separate rather than combining windows across recording gaps, tasks or unrelated four-second auxiliary epochs.

To describe uncertainty from the evaluated participants, we used 5,000 bootstrap resamples with random seed 20261010, introduced after D2 scoring. Resampling was stratified within the fixed source datasets, and a participant sampled for one class was carried jointly across all classes they contributed. The same resamples were used for all models, enabling paired differences. For the reported three-seed model mean, we averaged a participant's recalls across the three fixed trained seeds before bootstrap resampling. As a result, the intervals quantify variability among the observed participants under the selected sources and trained models; standard deviations across training seeds are reported separately. Neither measure accounts for newly sampled studies, selection of the model panel, or new optimization seeds.

Some supplementary comparisons were also specified after seeing D2 results. These include restricting candidate labels to conditions recorded within one source and correlating the rankings of nine model configurations between development and reserved benchmarks. Because those nine conditions were related and were not independent repetitions, we did not assign a p-value to their rank correlation. We likewise did not introduce an equivalence margin or claim multiplicity-adjusted superiority retrospectively. For a source-family sensitivity check, \texttt{ds003061} and \texttt{ds003969} were removed together because of their shared provenance, while other sources were removed one at a time. These analyses retained the original ten task candidates, but removing a source changes the composition of true classes; the resulting differences should not be read as a confidence interval for one unchanged population quantity.

\subsection*{5.7 Testing source-only prediction and semantic task groupings}

Two exploratory diagnostics addressed a common concern: a model might benefit from knowing which study recorded an EEG segment, without recognizing its underlying cognitive task. First, we calculated the strongest balanced accuracy available to a hypothetical classifier that knows only the source identity $s$. This \emph{source-only oracle} is a theoretical property of the evaluation label structure, not a trained EEG model. For each source, the oracle assigns the task label that maximizes class-balanced accuracy under the observed class-to-source distribution. For a randomized source-only decision rule $q(c\mid s)$, the optimum is

\begin{equation}\operatorname{BA}_{\mathrm{source\ oracle}}=\frac{1}{K}\sum_s\max_c P_w(S=s\mid Y=c)\label{eq:11}\end{equation}

where $P_w$ denotes the conditional source distribution under the weighting used for the benchmark. When each class belongs to exactly one source and the sets of labels associated with the sources do not overlap, the expression reduces to the number of occupied sources divided by the number of classes. For different weighting schemes or broader merged categories, the conditional probabilities must be recalculated. The oracle gives an upper bound for source-only classification under this benchmark structure, not a measurement of whether the EEG task decoder exploits those source cues.

Second, we asked whether semantically merging related task labels agrees with the errors made by the existing task decoder more than alternative mergers that share the same source structure. This is different from asking the model to predict newly averaged semantic prototypes. We kept its original fine-label predictions fixed and evaluated them after assigning both the true labels and predicted labels to a proposed broader category. Given the original class--participant-equal confusion matrix $C_{ij}$, we sum each original true class's prediction mass across labels $j$ belonging to its broader group, average across original labels within that group, and then average across the merged groups to obtain their balanced accuracy.

Simply comparing semantic groupings with arbitrary label partitions would be insufficient, because an arbitrary grouping could combine labels from different studies in ways the semantic grouping did not. We therefore generated a \emph{source-conditioned null}: within each original dataset, its task labels were randomly permuted before forming the broader categories. This preserved each merged group's number of labels and exact composition by recording source. A second, size-only null permuted labels without the source restriction. Both leave the model's EEG predictions and text parameters unchanged.

For each of the F15, F13 and F8 semantic partitions, we evaluated 10,000 Monte Carlo draws with random seed 20261011. The draws encountered 720, 3,848 and 2,145 distinct partitions, respectively, so some partitions were sampled more than once. We reported the fraction of draws at least as accurate as the observed semantic grouping using $(1+\text{number of draws at least as large as observed})/(1+\text{number of draws})$. This is an exploratory tail frequency: neither the semantic partitions nor the diagnostic was selected without looking at evaluation results, and the calculation is not corrected for that selection.

\subsection*{5.8 Measuring few-shot calibration with labeled EEG examples}

Our main task-prediction tests required no labeled examples from the evaluated target condition. A separate historical developmental analysis asked whether the same representation would be easier to use after seeing a few labeled EEG windows. It used the full-addition model after 16,000 training updates, with training seed 42, and evaluated the original ten-class D1 task bank using 64-second queries. The labeled \emph{support} sets contained 1, 4, 16 or 64 eight-second windows for each hierarchy class available in the relevant support study. Task classes absent from a support study retained their frozen text prototypes. The archived report describes adaptation using EEG centroids computed for the target study; it was not an additional self-supervised pretraining experiment.

We distinguished who supplied the support recordings. In the \emph{cross-participant} setting, one participant from each D1 study provided support, while the other participants supplied queries. In the \emph{same-participant} setting, support examples came from the participant being evaluated. We excluded every eight-second or 64-second query sharing any raw EEG sample with a support window, although same-participant examples could still come from the same task recording. This protects against exact sample overlap but does not make the test independent of participant- or session-specific recording features.

Figure~\ref{fig:fewshot-calibration} presents the archived fine-10 results. The 64-support same-participant point was omitted because too few eligible queries remained. Increasing support can also change which queries are eligible, so the different support budgets are not necessarily compared on identical query sets. Only one model seed was evaluated; uncertainty across training seeds was not estimated for this calibration analysis. The supplied analysis materials document the reported values and support definitions.

\subsection*{5.9 Visualizing the task-description space without retraining the model}

We visualized the text representations to help readers understand what kind of task relationships the EEG encoder was asked to learn. This analysis uses text vectors only; it cannot show whether the model successfully decodes their corresponding neural states. It was added after the EEG decoding experiments and involved no additional model selection or performance evaluation. The frozen prototype file has shape $68\times2\times384$, indicating 68 registered functional-task entries, two text views per entry and 384 embedding coordinates. We selected the 44 entries documented as observed primary-training VFTs in the archived Bridge3 specification. Auxiliary-only or unexposed entries were not included in the display.

We verified the SHA-256 checksums of both the float32 prototype array and its task index against the frozen training specification. Display names expanded shorthand but did not alter the archived task identities or vectors. We neither imputed missing task-description fields nor reran the sentence encoder. For each task, we unit-normalized the two stored text views using float64 arithmetic, summed them and normalized the result to produce a centroid $c_i$. Pairwise similarities were computed directly in the original text space as $S_{ij}=c_i^\top c_j$.

For the two-dimensional plot in Figure~\ref{fig:taskspace}, we supplied the corresponding nonnegative cosine distances $1-S_{ij}$ to t-distributed stochastic neighbor embedding (t-SNE) \citep{vandermaaten2008}. The primary display used perplexity 10, random initialization, seed 42, automatic learning rate, the exact solver and 2,000 iterations; these settings were fixed before inspecting the plot. We also evaluated every combination of perplexities 5, 10 and 20 with random seeds 42, 43 and 44, without choosing a new primary plot based on visual appearance. We assessed local neighborhood preservation using five-neighbor trustworthiness, computed from the original-space distances and the two-dimensional coordinates \citep{sklearnTrustworthiness}. The primary score was 0.903, with a sensitivity range of 0.538--0.952 across the tested settings (Table~S14). The plot is therefore a descriptive visualization, not stable evidence about global geometry; reported nearest neighbors and cosine similarities come from $S$ rather than positions on the plot. Software versions, coordinates, similarity values and source fingerprints accompany the analysis code.

\subsection*{5.10 Measuring how easily recording sources can be identified}

Finally, we asked a diagnostic question complementary to the source-only oracle: is information about the recording source readily available in simple features of these EEG recordings? We trained a separate supervised classifier to identify the source dataset, rather than the cognitive task. This analysis was performed after D2 scores were examined and does not reveal which features the frozen task-prediction models actually used. Its aim was to measure the strength of a possible recording-source signal under an explicitly source-supervised probe.

The probe distinguished nine datasets from the combined D1/D2 bank. We extracted 22 prespecified signal features from 16,960 eight-second windows belonging to 429 source-namespaced participant groups. We used at most 16 uniformly spaced eligible windows per participant and task class. Participant groups were defined within sources; matching participant numbers in different datasets were not assumed to indicate the same individual.

For each EEG window, we subtracted its mean and estimated spectral power using Welch's method at 500~Hz, with 1,000-sample segments, 500-sample overlap and a 1,024-point transform. We integrated power in nine bands: 0.5--2, 2--4, 4--8, 8--13, 13--20, 20--30, 30--45, 45--70 and 70--100~Hz. The feature vector contained nine base-10 log absolute band powers, nine log powers relative to the sum over these bands, and four base-10 log amplitude statistics: standard deviation, median absolute deviation, the 99.5th percentile of absolute amplitude and standard deviation of the first differences. Before taking logarithms, band powers were floored at $10^{-10}$ and amplitude statistics at $10^{-5}$. The archived D1 feature extractor restored its normalized stored signal by multiplying by 20, whereas D2 features used its processed signal directly. These differences, along with the original source-specific processing histories, are relevant when interpreting the source classifier.

To avoid testing on windows from a participant group seen during classifier fitting, we used five-fold stratified group cross-validation with random seed 20261009. Each group belonged entirely to one fold. Within each training fold, we fitted a StandardScaler and then logistic regression with $C=1$, class-balanced weights and a maximum of 1,200 iterations. The scaler was applied to the held-out fold without refitting. Source-class balanced accuracy was calculated from all out-of-fold predictions by giving equal weight to the nine source recalls. It reached 78.83\%, compared with 11.11\% chance for nine sources. Because this analysis explicitly used source labels in training and tested held-out participants \emph{within} sources rather than held-out sources, the result is evidence that source information is detectable, not evidence that the EEG task decoder relies on the same features.

\subsection*{6.1 Data and code availability}
The EEG recordings analyzed in this study were obtained
from existing publicly distributed research datasets.
The original data remain with their respective providers,
and the dataset accessions and associated publications
are documented in Tables S15--S17 and the bibliography.
No new participants were recruited, and no wearable-device
recordings were collected for this study.

The code and version-matched supporting materials will be
made publicly available upon acceptance of this paper.
The planned release includes analysis and figure-generation
code, reference model and evaluation-metric implementations,
source-accession and citation inventories, aggregate results,
and the task-space similarity matrix. 

\subsection*{6.2 Competing interests}
Ryota Kanai is the founder and Chief Executive Officer of Araya, Inc. The company's work includes artificial intelligence and neurotechnology, and wearable EEG is an intended application of this research. This company leadership role is disclosed as a potential competing interest.

\subsection*{6.3 Funding}

Japan Science and Technology Agency (JST) Moonshot R\&D Program, Grant Number JPMJMS2012.

\subsection*{6.4 Acknowledgments}

This work was supported by the Japan Science and Technology Agency (JST) Moonshot R\&D Program, Grant Number JPMJMS2012. The author thanks the researchers who made the cited EEG datasets available.

\subsection*{6.5 AI assistance}
ChatGPT (OpenAI) assisted with research planning, code development and execution support, aggregate
analyses, figure preparation, literature searches, and manuscript drafting and revision. Claude Opus 5.5
(Anthropic) was used as a simulated peer reviewer to provide critical feedback on manuscript drafts
during preparation and revision. The author retains full responsibility for the scientific analyses, source verification, claims, and final manuscript.

\section*{7. Conclusion}\addcontentsline{toc}{section}{7. Conclusion}
Task-semantic alignment provides a common prediction interface for heterogeneous single-channel EEG recordings, supporting both zero-shot candidate ranking and few-shot calibration. Multi-task recordings offer useful within-person contrasts, while task-sparse studies can expand the range of candidate cognitive operations. Their balance remains unresolved: historical additions sometimes helped and sometimes harmed transfer, and greater task coverage did not ensure discrimination within a new study. The next test is whether independent task replication and multi-task contrasts improve generalization at a fixed data and training budget. Longitudinal recordings of everyday activities would provide a practical setting for that test.

\clearpage

{\small
\setlength{\bibsep}{3pt}
\bibliographystyle{unsrtnat}
\bibliography{references}
}
\clearpage\setcounter{figure}{0}\renewcommand{\thefigure}{S\arabic{figure}}\renewcommand{\theHfigure}{suppl.\arabic{figure}}\section*{Supplementary material}\addcontentsline{toc}{section}{Supplementary material}

\FloatBarrier
\subsection*{S0. Developmental SSL pilots that motivated the semantic strategy}

The historical pilot figures below are transcribed from the dated project research summary of 28 September 2026, which records completed downstream evaluations. They are included to explain the research trajectory, not as a new external-model benchmark. The original pretraining pool comprised five named source families: PEERS/Courier \citep{data_ds004395,data_ds004706,katerman2022,dougherty2022}, SparrKULee \citep{data_sparrkulee,accou2024}, THINGS-EEG1 \citep{data_ds003825,grootswagers2022}, EEG Moments \citep{data_ds008257,gifford2026moments} and Language ERP \citep{data_language_erp,stone2023}, with 219 person groups and an audited union of 584.7 exposed recording hours. This exposure measure is different from a later task-semantic training pool or from cumulative repeated sample-hours.

The pilot used a width-128, four-layer, four-head causal encoder, eight-second inputs and batch size 128. A masked-latent objective trained a student to predict an exponential-moving-average teacher representation using Smooth L1, approximately 50\% span masking, 5\% warm-up and cosine learning-rate decay. A fixed sequence of 4,096 batch draws was repeated four times to reach 16,384 updates. Its cumulative presentation was 4,660.34 sample-hours. The 8,192- and 16,384-update evaluations below refer to checkpoints of that run.

\Needspace{105pt}
\par\medskip\noindent{\small\textbf{Table S0. Archived SSL pilot downstream results (developmental; percent balanced accuracy).}}
{\fontsize{8.8}{11.2}\selectfont\setlength{\tabcolsep}{3pt}\renewcommand{\arraystretch}{1.18}
\begin{longtable}{>{\raggedright\arraybackslash}p{189.11pt}>{\raggedright\arraybackslash}p{67.54pt}>{\raggedright\arraybackslash}p{94.55pt}>{\raggedright\arraybackslash}p{94.55pt}}
\toprule
\textbf{Downstream problem} & \textbf{Scratch} & \textbf{SSL: 8,192 updates} & \textbf{SSL: 16,384 updates} \\
\midrule\endfirsthead
\multicolumn{4}{l}{\textit{Table S0 (continued)}}\\
\toprule
\textbf{Downstream problem} & \textbf{Scratch} & \textbf{SSL: 8,192 updates} & \textbf{SSL: 16,384 updates} \\
\midrule\endhead
\midrule\multicolumn{4}{r}{\textit{Continued on next page}}\\\endfoot
\bottomrule\endlastfoot
COG-BCI, 10 tasks & 34.41 & 34.33 & 35.41 \\
CMx7-MM, 6 tasks (motor imagery excluded) & 20.86 & 20.60 & 20.09 \\
\end{longtable}}
\vspace{-0.6em}\noindent{\footnotesize Within-study supervised downstream evaluation; five folds and three seeds. Values come from the 28 September archival summary, not an independent rerun. They are not directly comparable to the final calibration-free task banks. SSL checkpoint steps describe pretraining, not downstream labeled exposure.}\par\medskip

The COG-BCI and CMx7-MM results are archived within-study, person-group-equal balanced accuracies from five-fold, three-seed downstream comparisons. Their task banks and supervised adaptation differ from the final calibration-free D2 evaluation. The approximately one-point COG-BCI gain coexists with a decline in the six-task comparison. The archive also reports a separate four-task pilot on ds004148 with 16 training, eight inner-development and 12 outer-development participants: scratch 28.56\%, SSL-initialized task training 30.80\%, and a frozen SSL encoder plus classifier 28.79\%. Its reported paired improvement has a participant-bootstrap interval of \ensuremath{-}0.80 to +5.20 points. Those outer participants had already been involved in exploration.

These observations motivated supplying semantic task information explicitly, but they do not demonstrate that semantic training is superior to SSL under equal input, labels, budget or model capacity. The complete fold-level pilot files were not independently re-executed for this preprint; their archived summary and scope are retained in the release provenance. The final models were initialized afresh, so these pilots are predecessors in the decision history rather than a shared pretrained ancestor of the final panel.

\clearpage

\FloatBarrier
\subsection*{S1. Endpoint and novelty records}

\Needspace{105pt}
\par\medskip\noindent{\small\textbf{Table S1. Endpoint dictionary and source-only oracle assumptions.}}
{\fontsize{8.8}{11.2}\selectfont\setlength{\tabcolsep}{3pt}\renewcommand{\arraystretch}{1.18}
\begin{longtable}{>{\raggedright\arraybackslash}p{80.17pt}>{\raggedright\arraybackslash}p{73.76pt}>{\raggedright\arraybackslash}p{163.55pt}>{\raggedright\arraybackslash}p{128.27pt}}
\toprule
\textbf{Endpoint} & \textbf{Candidates / chance} & \textbf{Weights and windows} & \textbf{Source-only oracle} \\
\midrule\endfirsthead
\multicolumn{4}{l}{\textit{Table S1 (continued)}}\\
\toprule
\textbf{Endpoint} & \textbf{Candidates / chance} & \textbf{Weights and windows} & \textbf{Source-only oracle} \\
\midrule\endhead
\midrule\multicolumn{4}{r}{\textit{Continued on next page}}\\\endfoot
\bottomrule\endlastfoot
B2 native local & 5 and 6; equal-source chance 18.33\% & Per-person local BA; then equal source means. 8-s windows overlap; 64-s bags average 29 windows at 2-s spacing. & Source supplied; local chance differs by source. \\
B2 native pooled-11 & 11 / 9.09\% & Original evaluator forms a pooled window-count confusion matrix. Not the later person-equal endpoint. & 2/11 = 18.18\%, if classes are source-exclusive. \\
D1 fine-10 historical & 10 / 10\% & Later 4+6 functional classes; person-equal BA, overlapping 8-s windows in original summaries. & 20\% under class-equal weights; 22.60\% under 99/40 person weights with all classes present. \\
D1 fine-10 harmonized & 10 / 10\% & Nonoverlapping 8 s; equal classes, then equal people within class. & 20\%. \\
D1 local fine-10 & 4 or 6 & Candidate restriction; source-equal chance 20.83\%; 99/40 person-weighted chance 22.60\%. & Local source identity already supplied. \\
D1-ST CLEAN4 / BROAD6 & 4 / 25\%; 6 / 16.67\% & One task per source; per-person summaries. Accepted 8-s phase windows; 64 s disabled. & 100\% for both banks. \\
Integrated16 & 16 / 6.25\% & D1 fine-10 plus BROAD6; eight distinct dataset IDs. Historical estimator retained. & 50\% under class-equal weights and disjoint source-exclusive classes; recompute for other weights. \\
D2 pooled & 10 / 10\% & Nonoverlapping 8 s; class \ensuremath{\to} person \ensuremath{\to} windows. Original64: eight contiguous nonoverlapping windows. & 70\% for the seven dataset-ID groups. \\
D2 subgroup recall & Still 10 / 10\% & Average selected true classes; predictions retain all ten candidates. & Depends on the selected true classes; not a two- or eight-way classifier. \\
D2 within-source & 2 / 50\%; 3 / 33.33\% & Restricted candidate bank. Equal weight to classes, then to people. Uncertainty from the joint participant bootstrap (Methods 5.6). & Local source identity supplied. \\
D1+D2 pooled-20 & 20 / 5\% & Nonoverlapping 8 s; class \ensuremath{\to} person \ensuremath{\to} windows. & 45\%, nine dataset IDs with source-exclusive classes. \\
F15 / F13 / F8 & 15 / 6.67\%; 13 / 7.69\%; 8 / 12.5\% & Concept \ensuremath{\to} original task \ensuremath{\to} person; separate merged-prototype and fixed-argmax relabeling analyses. & Archived weighted oracle: 46.67\% / 52.56\% / 75.00\%. Not simply S/K. \\
\end{longtable}}
\vspace{-0.6em}\noindent{\footnotesize For source-exclusive classes and person-first averaging, the oracle is the mean of $1/K_{\mathrm{source}}$ under the actual person weights. The often-quoted 20\% D1 oracle applies to class-equal evaluation, not every historical D1 estimator. All oracle values are structural capacities, not trained EEG baselines.}\par\medskip

The B2 local candidate count was checked directly against the five ds007172 and six ERP CORE task cards and evaluator. Its equal-source uniform baseline is 18.33\%, not the 20.83\% obtained from the later 4/6 functional bank. This correction changes interpretation of the endpoint label, not the archived prediction values. The historical person-first D1 source-only oracle also differs from a class-equal S/K calculation; using 99 and 40 people with four and six classes gives 22.60\%.

\Needspace{105pt}
\par\medskip\noindent{\small\textbf{Table S2. D1/D1-ST novelty record: audited scope and remaining uncertainty.}}
{\fontsize{8.8}{11.2}\selectfont\setlength{\tabcolsep}{3pt}\renewcommand{\arraystretch}{1.18}
\begin{longtable}{>{\raggedright\arraybackslash}p{73.76pt}>{\raggedright\arraybackslash}p{89.79pt}>{\raggedright\arraybackslash}p{131.48pt}>{\raggedright\arraybackslash}p{150.72pt}}
\toprule
\textbf{Evaluation source} & \textbf{Operational target} & \textbf{Direct training-source check} & \textbf{Task-level status} \\
\midrule\endfirsthead
\multicolumn{4}{l}{\textit{Table S2 (continued)}}\\
\toprule
\textbf{Evaluation source} & \textbf{Operational target} & \textbf{Direct training-source check} & \textbf{Task-level status} \\
\midrule\endhead
\midrule\multicolumn{4}{r}{\textit{Continued on next page}}\\\endfoot
\bottomrule\endlastfoot
D1 ds007172 & Native 5 \ensuremath{\to} functional 4; visual targets, rest, words & Absent from audited ST19/ST40/Bridge3 primary indices and auxiliary nm000166. & Broad operation relatives exist; exact model \ensuremath{\times} task support and raw-family overlap remain to be certified. \\
D1 ERP CORE & Six ERP-task protocols & Absent under its dataset identifier from those indices. & Related flanker, oddball, visual search and semantic tasks exist in training; no inference of exact equivalence from names alone. \\
ds004117 & Selective verbal WM encoding & Absent from the same audited training indices. & Historical task-equivalence rule: ZS-Study; verbal Sternberg VFT present, evaluated phase not identical. \\
ds004860 & Moral narrative comprehension & Absent from the same audited training indices. & Historical task-equivalence rule: nominal ZS-Both; related speech/reading/semantic tasks. \\
ds006142 & Movie recognition & Absent from the same audited training indices. & Historical task-equivalence rule: nominal ZS-Both; related emotion viewing, encoding and recall. \\
nm000219 & Covert spatial attention & Absent from the same audited training indices. & Historical task-equivalence rule: nominal ZS-Both; related search/oddball/flanker. \\
ds006446 & Intertemporal choice & Absent from the same audited training indices. & Historical task-equivalence rule: nominal ZS-Both; related valuation and autobiographical recall. \\
ds004579 & Interval timing & Absent from the same audited training indices. & Historical task-equivalence rule: nominal ZS-Both; related reaction-time and motor tasks. \\
\end{longtable}}
\vspace{-0.6em}\noindent{\footnotesize This is a dataset-identifier check against three primary execution indices and their shared auxiliary data stream. It is not a complete per-checkpoint provenance, duplicate-person, raw-hash or public-model-pretraining audit. These novelty annotations follow an earlier task-equivalence rule. They should not be applied to later mixtures without rechecking. D1 and D1-ST were repeatedly inspected during development.}\par\medskip

For task overlap, a related training description is not enough to certify exact operational equivalence. VFT-level novelty and stimulus/response/phase relations should therefore be released independently. The table preserves the historical task-equivalence annotations for D1-ST while recording the narrower direct-source audit completed for this release.

\clearpage

\FloatBarrier
\subsection*{S2. Documented physical-input coverage}

\Needspace{105pt}
\par\medskip\noindent{\small\textbf{Table S3. Physical-input metadata retained in the inspected registry.}}
{\fontsize{8.8}{11.2}\selectfont\setlength{\tabcolsep}{3pt}\renewcommand{\arraystretch}{1.18}
\begin{longtable}{>{\raggedright\arraybackslash}p{73.76pt}>{\raggedright\arraybackslash}p{109.03pt}>{\raggedright\arraybackslash}p{150.72pt}>{\raggedright\arraybackslash}p{112.24pt}}
\toprule
\textbf{Source} & \textbf{Pair in inspected registry} & \textbf{Native reference field} & \textbf{Headers with pair / checked} \\
\midrule\endfirsthead
\multicolumn{4}{l}{\textit{Table S3 (continued)}}\\
\toprule
\textbf{Source} & \textbf{Pair in inspected registry} & \textbf{Native reference field} & \textbf{Headers with pair / checked} \\
\midrule\endhead
\midrule\multicolumn{4}{r}{\textit{Continued on next page}}\\\endfoot
\bottomrule\endlastfoot
ds002680 & FP1 \ensuremath{-} OZ & Common (registry) & 350/350 \\
ds003061 & Fp1 \ensuremath{-} Oz & CMS/DRL (registry) & 39/39 \\
ds005189 & Fp1 \ensuremath{-} Oz & Contralateral mastoids TP9/TP10 & 0/0 recorded; not coverage proof \\
ds004117 & FP1 \ensuremath{-} OZ & Common (registry) & 85/85 \\
ds004860 & Fp1 \ensuremath{-} Oz & Mean left/right mastoids & 0/0 recorded; not coverage proof \\
ds006446 & Fp1 \ensuremath{-} Oz & CMS (registry) & 29/29 \\
\end{longtable}}
\vspace{-0.6em}\noindent{\footnotesize These are partial registry records, not a complete preprocessing specification. Native rates, anti-alias and notch settings, and final signal-hash bindings are not consolidated in this release. A common reference cancels under differencing, but unequal channel filters and prior processing do not.}\par\medskip

The registry records no header-coverage count for the two 0/0 entries. Corresponding source-level records for ds007172, ERP CORE, ds003969, ds004657, ds006979, ds008711, ds006142, nm000219, ds004579 are not consolidated here. Later quality-assurance verification records are not fully linked to the evaluated file hashes in this release. Source-specific preprocessing therefore remains a limitation of exact reproduction; the common 500-Hz model rate does not fill these gaps.

\FloatBarrier
\subsection*{S3. Architecture, initialization and historical lineage}

The common RAW and LATE prediction-path parameter counts are 270,400 and 364,656. In the inspected union schema there are 125 native-head outputs and 65 instantiated facet outputs, adding 32,125 and 16,705 parameters, respectively. These schema-level totals should not be mistaken for the active loss-bearing parameters on every condition. The residual task adapter has 148,800 parameters, with one additional learned logit-scale scalar. All full-model totals require the corresponding checkpoint schema to be named.

The archived training specification for B2 used an earlier waveform encoder, trained from scratch. It took 8-s primary and 4-s auxiliary inputs. Later ST19, ST40, Bridge3, Full-Mixture and CLIP-X training specifications also use fresh initialization. The original SSL work belongs to the project's history, not a causal explanation for every final checkpoint. Source-specific physical input in the early branch has not been fully re-certified for this release, so early calibration and shared-task findings are not presented as a perfectly input-matched comparison with the final AP2E panel.

The SCALL 16k mean of 24.89\% and the Full-Mixture RAW 16k mean of 15.70\% come from different mixtures, training schedules and run generations. A checkpoint with the same update count is not a matched exposure or optimizer state. Later ST19-MATCHED 16k also has a distinct historical D1 mean of 25.93\%. We report these run generations separately.

\Needspace{105pt}
\par\medskip\noindent{\small\textbf{Table S4. Historical observation-to-decision record (evidence status retained).}}
{\fontsize{8.8}{11.2}\selectfont\setlength{\tabcolsep}{3pt}\renewcommand{\arraystretch}{1.18}
\begin{longtable}{>{\raggedright\arraybackslash}p{97.50pt}>{\raggedright\arraybackslash}p{149.50pt}>{\raggedright\arraybackslash}p{204.75pt}}
\toprule
\textbf{Stage and evidence status} & \textbf{Observation available at that point} & \textbf{Design consequence and bounded outcome} \\
\midrule\endfirsthead
\multicolumn{3}{l}{\textit{Table S4 (continued)}}\\
\toprule
\textbf{Stage and evidence status} & \textbf{Observation available at that point} & \textbf{Design consequence and bounded outcome} \\
\midrule\endhead
\midrule\multicolumn{3}{r}{\textit{Continued on next page}}\\\endfoot
\bottomrule\endlastfoot
Initial SSL / legacy screening
Developmental & Weak semantic retrieval did not identify whether the bottleneck was the EEG representation or its interface. Input and endpoint definitions evolved. & Moved to physically constrained task alignment and separate probes. No clean final SSL-versus-scratch effect is claimed. \\
Task-coverage and sampler/loss studies, by 3 Oct
Developmental & Task/subject/source information coexisted; task exposure was strongly imbalanced. Source confounding was explicitly recognized. & Global task balancing; task-card correction; local/global readout separation. Relation-based and adversarial losses were not universal improvements. \\
Semantic embedding, 4 Oct
Developmental; matched comparisons & Some pooled gains failed shared-class controls; text prototypes could collide. & B128-LIT improved specific shared-task contrasts. The gain was limited to developmental shared-task tests. \\
Study addition / random orders
Developmental; design frozen locally & Nested mixtures were nonmonotonic; study count also changes composition. & Six randomized addition orders; 240 training checkpoints. The common evaluations of subsets with 20 or 30 added families show between-subset point-estimate variation. \\
Full-Mixture scaling
Fixed composition; developmental endpoints & More studies and more training could not be interpreted as the same intervention. & RAW/LATE trajectories to 256k. Different benchmarks peak at different horizons. \\
Bridge3 / CLIP-X
D1-informed mixture; internal objective selection & An anchor-preserving mixture was hypothesized to avoid loss of earlier transfer. Alignment gains depended on architecture and readout. & Bridge3 matched controls and CLIP-X advanced together. Internal selection does not erase the mixture's developmental origin. \\
D2 amended reserve, 9 Oct
Frozen before scoring & Four-second reserve data did not meet the eight-second requirement; novelty and source audits required qualification. & Source replacements and nine-condition, three-seed model panel frozen before prediction. CLIP-X gains failed to transfer to D2. \\
D2 diagnostics and fusion, 9 Oct
Explicitly post-score & Pooled gains concentrated in a subset of labels; related cross-bank tasks competed. & Within-study restriction, metric harmonization, joint K20, source probing and F15/F13/F8 fusion. These post-score diagnoses motivated independent validation. \\
\end{longtable}}
\vspace{-0.6em}\noindent{\footnotesize Dates identify documented stages, not a claim of exhaustive preregistration. The main text follows scientific dependencies; this record preserves the actual order of knowledge.}\par\medskip

\begin{figure}[!htbp]
\centering
\includegraphics[width=\linewidth,height=0.55\textheight,keepaspectratio]{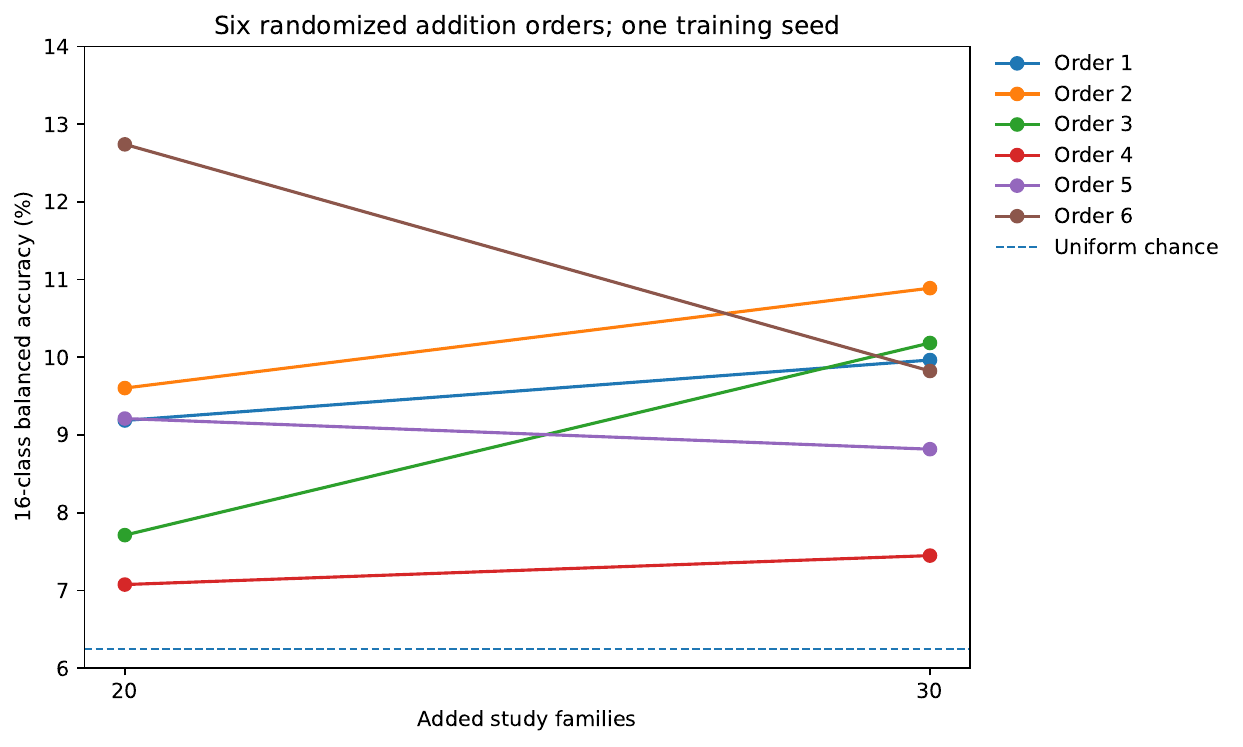}
\caption{\textbf{Six randomized dataset-addition orders.} The displayed subsets contain 20 or 30 added study families, denoted R20 and R30 in the original records. All orders use training seed 42. The points are the common archived 16-class evaluations available for all six orders. The dashed line is 6.25\% uniform chance.}
\label{fig:randomized-addition}
\end{figure}

\begin{figure}[!htbp]
\centering
\includegraphics[width=\linewidth,height=0.55\textheight,keepaspectratio]{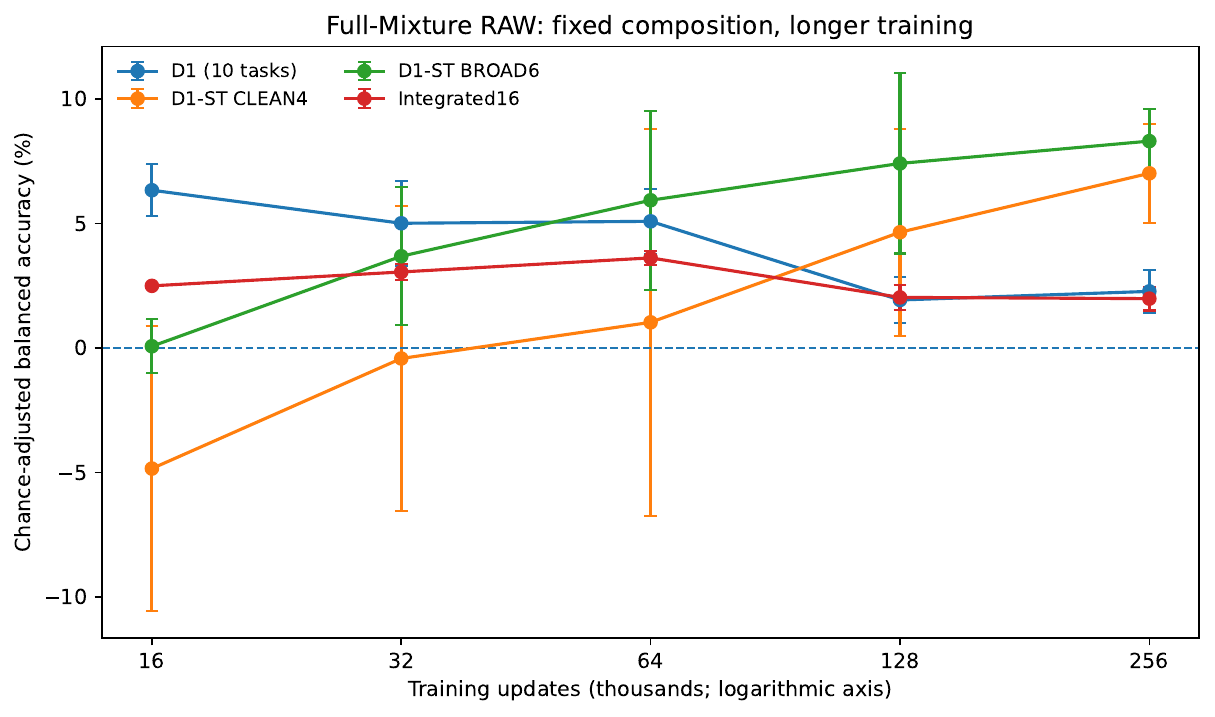}
\caption{\textbf{Longer training with a fixed mixture: waveform-only encoder.} Points are three-seed means and error bars are one training-seed standard deviation. The plotted chance-adjusted score is $100(\mathrm{BA}-1/K)/(1-1/K)$, with BA as a proportion and $K$ the candidate count. Error bars are scaled by the same factor $100/(1-1/K)$. The training-update axis is logarithmic.}
\label{fig:fixed-mixture-raw}
\end{figure}

\begin{figure}[!htbp]
\centering
\includegraphics[width=\linewidth,height=0.55\textheight,keepaspectratio]{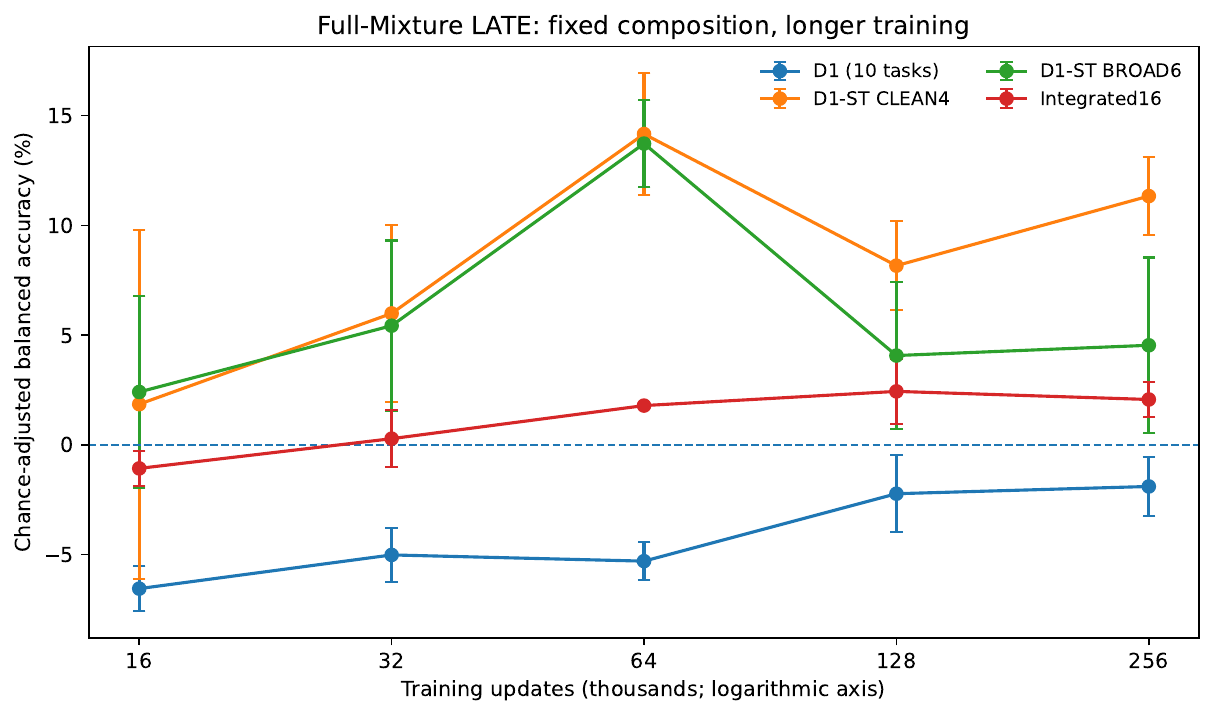}
\caption{\textbf{Longer training with a fixed mixture: waveform--spectral encoder.} Points and error bars show three-seed means and one training-seed standard deviation. The score is $100(\mathrm{BA}-1/K)/(1-1/K)$; error bars use the same linear scaling. The training-update axis is logarithmic. Negative values denote below-uniform-chance performance for that candidate bank.}
\label{fig:fixed-mixture-late}
\end{figure}

\FloatBarrier
\subsection*{S4. Complete D2 uncertainty, concentration and weighting diagnostics}

\Needspace{105pt}
\par\medskip\noindent{\small\textbf{Table S5. Restricted-candidate D2 diagnostics with both uncertainty summaries.}}
{\fontsize{8.8}{11.2}\selectfont\setlength{\tabcolsep}{3pt}\renewcommand{\arraystretch}{1.18}
\begin{longtable}{>{\raggedright\arraybackslash}p{129.71pt}>{\raggedright\arraybackslash}p{82.26pt}>{\raggedright\arraybackslash}p{69.60pt}>{\raggedright\arraybackslash}p{82.26pt}>{\raggedright\arraybackslash}p{75.93pt}}
\toprule
\textbf{Condition} & \textbf{Thinking / meditation BA \ensuremath{\pm} SD} & \textbf{Participant CI} & \textbf{Driving BA \ensuremath{\pm} SD} & \textbf{Participant CI} \\
\midrule\endfirsthead
\multicolumn{5}{l}{\textit{Table S5 (continued)}}\\
\toprule
\textbf{Condition} & \textbf{Thinking / meditation BA \ensuremath{\pm} SD} & \textbf{Participant CI} & \textbf{Driving BA \ensuremath{\pm} SD} & \textbf{Participant CI} \\
\midrule\endhead
\midrule\multicolumn{5}{r}{\textit{Continued on next page}}\\\endfoot
\bottomrule\endlastfoot
ST19 RAW 16k & 50.59 \ensuremath{\pm} 0.15 & 49.14--52.08 & 33.80 \ensuremath{\pm} 0.83 & 31.84--35.77 \\
ST19 RAW 64k & 50.80 \ensuremath{\pm} 1.01 & 49.29--52.37 & 34.78 \ensuremath{\pm} 0.44 & 33.29--36.31 \\
ST40 RAW 64k & 52.73 \ensuremath{\pm} 1.91 & 49.97--55.38 & 34.23 \ensuremath{\pm} 1.19 & 31.48--37.16 \\
FM RAW 256k & 51.10 \ensuremath{\pm} 0.37 & 49.22--53.09 & 33.15 \ensuremath{\pm} 1.20 & 31.21--35.08 \\
FM LATE 256k & 51.13 \ensuremath{\pm} 0.77 & 49.58--52.78 & 35.21 \ensuremath{\pm} 1.04 & 33.63--37.01 \\
Bridge3 fixed-text RAW 64k & 50.56 \ensuremath{\pm} 0.39 & 49.19--52.08 & 33.89 \ensuremath{\pm} 0.58 & 31.91--35.85 \\
Bridge3 fixed-text LATE 64k & 50.34 \ensuremath{\pm} 0.35 & 49.02--51.80 & 34.43 \ensuremath{\pm} 0.15 & 32.35--36.68 \\
CLIP-X RAW 64k & 50.73 \ensuremath{\pm} 0.80 & 49.35--52.03 & 33.90 \ensuremath{\pm} 0.42 & 32.71--35.06 \\
CLIP-X LATE 64k & 50.05 \ensuremath{\pm} 0.09 & 49.13--50.98 & 32.90 \ensuremath{\pm} 0.51 & 31.68--34.20 \\
\end{longtable}}
\vspace{-0.6em}\noindent{\footnotesize All numbers are percentages. K = 2 (chance 50\%) in ds003969, n = 98; K = 3 (chance 33.33\%) in ds004657, n = 18. CIs are unadjusted, conditional participant intervals. FM LATE driving has a small interval above chance; this is a post-score, one-source observation, not a multiplicity-adjusted novel-task confirmation.}\par\medskip

\Needspace{105pt}
\par\medskip\noindent{\small\textbf{Table S6. Prediction concentration and standardized agreement on D2.}}
{\fontsize{8.8}{11.2}\selectfont\setlength{\tabcolsep}{3pt}\renewcommand{\arraystretch}{1.18}
\begin{longtable}{>{\raggedright\arraybackslash}p{132.88pt}>{\raggedright\arraybackslash}p{66.44pt}>{\raggedright\arraybackslash}p{69.60pt}>{\raggedright\arraybackslash}p{53.78pt}>{\raggedright\arraybackslash}p{117.06pt}}
\toprule
\textbf{Condition} & \textbf{Max share (\%)} & \textbf{Entropy / log 10} & \textbf{Class-equal \ensuremath{\kappa}} & \textbf{Recall excluding known people (\%)} \\
\midrule\endfirsthead
\multicolumn{5}{l}{\textit{Table S6 (continued)}}\\
\toprule
\textbf{Condition} & \textbf{Max share (\%)} & \textbf{Entropy / log 10} & \textbf{Class-equal \ensuremath{\kappa}} & \textbf{Recall excluding known people (\%)} \\
\midrule\endhead
\midrule\multicolumn{5}{r}{\textit{Continued on next page}}\\\endfoot
\bottomrule\endlastfoot
ST19 RAW 16k & 39.58 & 0.735 & 0.126 & 12.85 \\
ST19 RAW 64k & 31.59 & 0.763 & 0.141 & 13.58 \\
ST40 RAW 64k & 23.87 & 0.876 & 0.132 & 12.45 \\
FM RAW 256k & 25.75 & 0.879 & 0.128 & 11.43 \\
FM LATE 256k & 25.25 & 0.854 & 0.116 & 9.00 \\
Bridge3 fixed-text RAW 64k & 33.04 & 0.816 & 0.155 & 15.22 \\
Bridge3 fixed-text LATE 64k & 32.88 & 0.821 & 0.224 & 21.75 \\
CLIP-X RAW 64k & 25.09 & 0.815 & 0.011 & 8.30 \\
CLIP-X LATE 64k & 34.92 & 0.709 & 0.020 & 12.61 \\
\end{longtable}}
\vspace{-0.6em}\noindent{\footnotesize Maximum share and entropy are means across seed-specific summaries. \ensuremath{\kappa} is derived from the class-equal confusion; it equals (BA \ensuremath{-} 0.1)/0.9 and adds no independent evidence beyond BA. Last column omits true classes 0 and 7 but retains K = 10 for every prediction; its chance remains 10\%. It is not an eight-way score.}\par\medskip

\Needspace{105pt}
\par\medskip\noindent{\small\textbf{Table S7. Separating window spacing from metric weighting.}}
{\fontsize{8.8}{11.2}\selectfont\setlength{\tabcolsep}{3pt}\renewcommand{\arraystretch}{1.18}
\begin{longtable}{>{\raggedright\arraybackslash}p{126.55pt}>{\raggedright\arraybackslash}p{72.77pt}>{\raggedright\arraybackslash}p{88.58pt}>{\raggedright\arraybackslash}p{79.09pt}>{\raggedright\arraybackslash}p{72.77pt}}
\toprule
\textbf{Condition} & \textbf{D1 historical (\%)} & \textbf{D1 nonoverlap, person (\%)} & \textbf{D1 class--person (\%)} & \textbf{D2 class--person (\%)} \\
\midrule\endfirsthead
\multicolumn{5}{l}{\textit{Table S7 (continued)}}\\
\toprule
\textbf{Condition} & \textbf{D1 historical (\%)} & \textbf{D1 nonoverlap, person (\%)} & \textbf{D1 class--person (\%)} & \textbf{D2 class--person (\%)} \\
\midrule\endhead
\midrule\multicolumn{5}{r}{\textit{Continued on next page}}\\\endfoot
\bottomrule\endlastfoot
ST19 RAW 16k & 25.93 & 25.79 & 23.41 & 21.31 \\
ST19 RAW 64k & 24.67 & 24.73 & 22.93 & 22.66 \\
ST40 RAW 64k & 13.71 & 13.67 & 14.04 & 21.88 \\
FM RAW 256k & 12.05 & 11.97 & 12.75 & 21.53 \\
FM LATE 256k & 8.29 & 8.21 & 13.18 & 20.40 \\
Bridge3 fixed-text RAW 64k & 25.01 & 24.99 & 23.11 & 23.98 \\
Bridge3 fixed-text LATE 64k & 14.13 & 14.16 & 17.11 & 30.13 \\
CLIP-X RAW 64k & 23.06 & 23.04 & 21.49 & 11.00 \\
CLIP-X LATE 64k & 22.37 & 22.39 & 20.88 & 11.79 \\
\end{longtable}}
\vspace{-0.6em}\noindent{\footnotesize The checkpoints are fixed. D1 reaggregation is post-score. The candidate count is ten in each benchmark, but the task sets differ. For FM LATE, nonoverlap changes BA by \ensuremath{-}0.08 percentage points and reweighting by +4.96 points (computed before rounding). Pooled person-first weighting gives the 99-person visual source more weight than class-first weighting of its four classes.}\par\medskip

For FM LATE, the historical person-first measure gives more weight to the 99-person visual source than the class-first measure, which assigns it four of ten classes. The model's different performance across source/task groups therefore changes the pooled value substantially when weights change. It is not evidence that removing overlapping windows created nearly five points of new discrimination.

\Needspace{105pt}
\par\medskip\noindent{\small\textbf{Table S8. Where the leading D2 aggregate comes from.}}
{\fontsize{8.8}{11.2}\selectfont\setlength{\tabcolsep}{3pt}\renewcommand{\arraystretch}{1.18}
\begin{longtable}{>{\raggedright\arraybackslash}p{132.88pt}>{\raggedright\arraybackslash}p{63.27pt}>{\raggedright\arraybackslash}p{75.93pt}>{\raggedright\arraybackslash}p{75.93pt}>{\raggedright\arraybackslash}p{91.75pt}}
\toprule
\textbf{True class} & \textbf{Recall (\%)} & \textbf{BA contribution (pp)} & \textbf{Predicted share (\%)} & \textbf{Balanced precision (\%)} \\
\midrule\endfirsthead
\multicolumn{5}{l}{\textit{Table S8 (continued)}}\\
\toprule
\textbf{True class} & \textbf{Recall (\%)} & \textbf{BA contribution (pp)} & \textbf{Predicted share (\%)} & \textbf{Balanced precision (\%)} \\
\midrule\endhead
\midrule\multicolumn{5}{r}{\textit{Continued on next page}}\\\endfoot
\bottomrule\endlastfoot
Go/no-go recognition & 97.22 & 9.72 & 15.53 & 62.61 \\
Auditory oddball & 38.11 & 3.81 & 32.88 & 11.59 \\
Active thinking & 2.33 & 0.23 & 0.67 & 34.79 \\
Breath meditation & 38.14 & 3.81 & 11.04 & 34.55 \\
Manual driving & 7.33 & 0.73 & 4.64 & 15.81 \\
Full assistance & 2.87 & 0.29 & 2.14 & 13.36 \\
Speed assistance & 9.10 & 0.91 & 3.70 & 24.62 \\
Surprise recognition & 30.11 & 3.01 & 4.63 & 64.98 \\
Visual working memory & 1.64 & 0.16 & 5.50 & 2.97 \\
RSVP language & 74.47 & 7.45 & 19.27 & 38.64 \\
\end{longtable}}
\vspace{-0.6em}\noindent{\footnotesize Bridge3 fixed-text LATE, three-seed mean confusion. Precision is computed after class/person balancing and seed averaging; it is not prevalence-calibrated deployment precision. Contributions sum to 30.13 percentage points and are arithmetic, not causal attributions.}\par\medskip

\begin{figure}[!htbp]
\centering
\includegraphics[width=\linewidth,height=0.55\textheight,keepaspectratio]{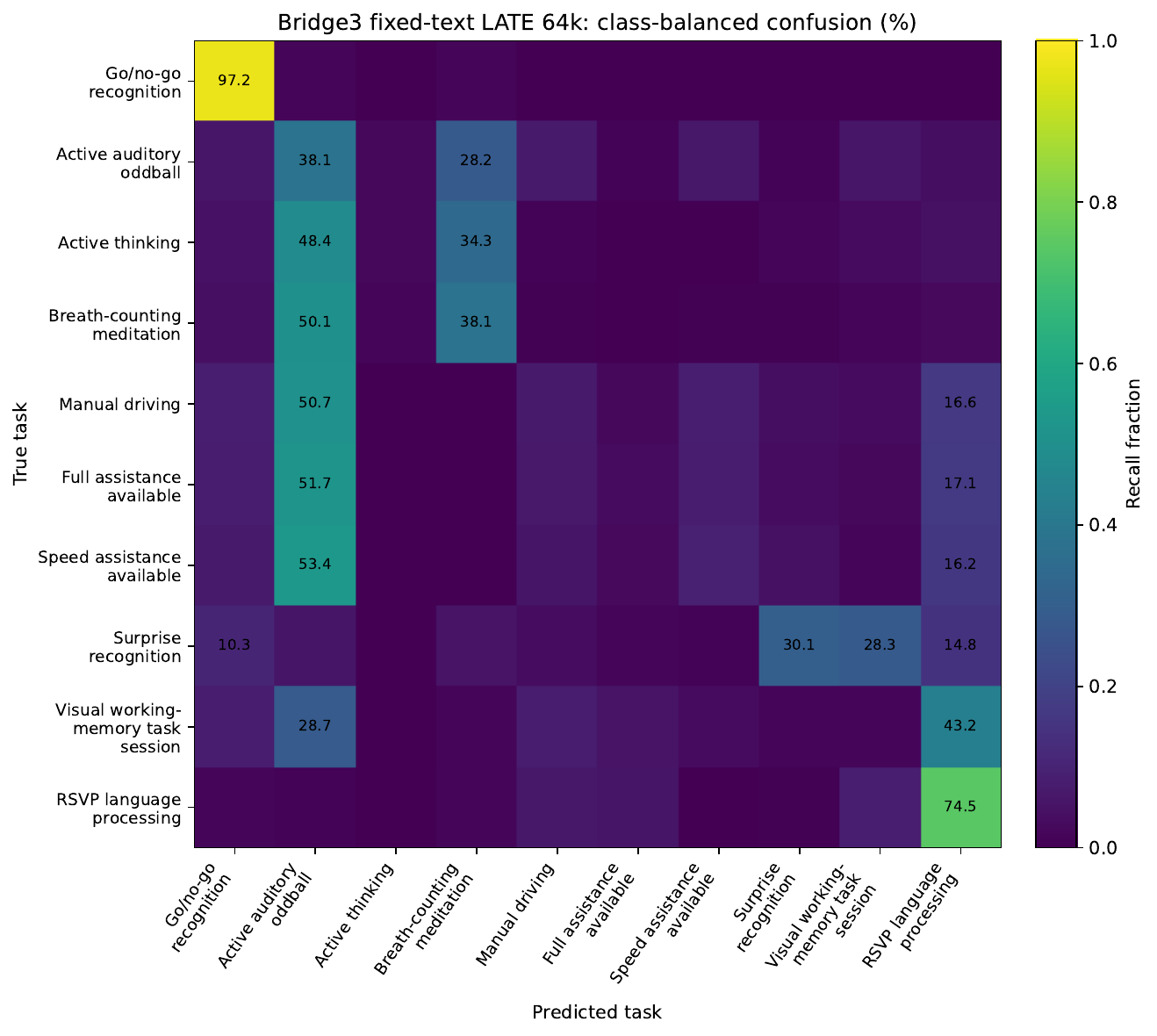}
\caption{\textbf{Ten-class confusion matrix for Bridge3 fixed-text LATE 64k.} Rows are true labels and columns are predicted labels. Participant recalls are averaged within each class and then across three training seeds. Precision and predicted-share summaries appear in Table S8.}
\label{fig:d2-confusion}
\end{figure}

\begin{figure}[!htbp]
\centering
\includegraphics[width=\linewidth,height=0.55\textheight,keepaspectratio]{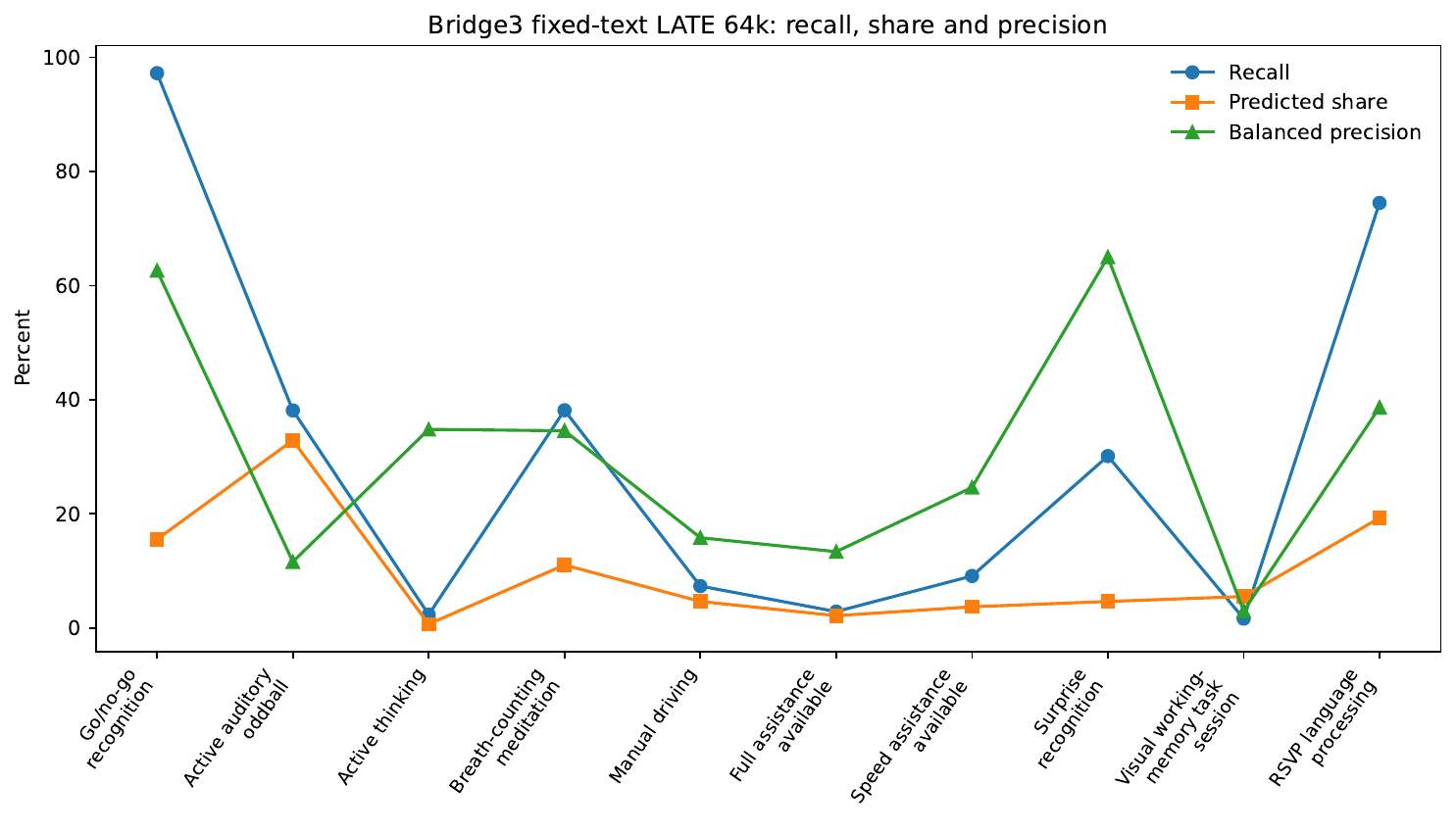}
\caption{\textbf{Class recall, predicted share and precision.} Values are calculated from the three-seed mean class-balanced confusion matrix of Bridge3 fixed-text LATE 64k. Precision therefore refers to this balanced evaluation distribution. All ten candidates remain in the prediction bank.}
\label{fig:d2-class-profile}
\end{figure}

The paired D2 CLIP-X-minus-fixed-text effects are \ensuremath{-}12.98 points for RAW (participant interval \ensuremath{-}14.67 to \ensuremath{-}11.36) and \ensuremath{-}18.34 for LATE (\ensuremath{-}20.50 to \ensuremath{-}16.19). The fixed-text LATE-minus-RAW effect is +6.15 points (4.63 to 7.72). These conditional comparisons use identical participant draws; their uncertainty does not generalize over new study families.

Omitting one conservative source family at a time changes the leading model's remaining-class K10 recall from 22.68\% to 40.29\%. The family group containing ds003061 and ds003969 is treated jointly. The large range reflects different omitted labels as well as source coverage. It is reported as a sensitivity analysis, not as a sampling distribution for a fixed benchmark score.

Below-chance cases must also be interpreted under the specified weights and candidate bank. The historical FM LATE D1 curve is below the 10\% uniform baseline, whereas the class-balanced harmonized 256k score is 13.18\%. CLIP-X has individual below-chance seeds in D2. The archived confusion matrices retain systematic off-diagonal predictions; a low mean should not be described as harmless random guessing. Target-label remapping chosen after inspecting these matrices would be a new exploratory decoder, not a correction to the original primary result.

\Needspace{105pt}
\par\medskip\noindent{\small\textbf{Table S9. Frozen D2 novelty-group recall against all ten candidates.}}
{\fontsize{8.8}{11.2}\selectfont\setlength{\tabcolsep}{3pt}\renewcommand{\arraystretch}{1.18}
\begin{longtable}{>{\raggedright\arraybackslash}p{129.71pt}>{\raggedright\arraybackslash}p{66.44pt}>{\raggedright\arraybackslash}p{82.26pt}>{\raggedright\arraybackslash}p{75.93pt}>{\raggedright\arraybackslash}p{85.42pt}}
\toprule
\textbf{Model} & \textbf{ZS-Study
3 classes} & \textbf{ZS-Task
known people} & \textbf{Nominal both
5 classes} & \textbf{Strict common both
2 classes} \\
\midrule\endfirsthead
\multicolumn{5}{l}{\textit{Table S9 (continued)}}\\
\toprule
\textbf{Model} & \textbf{ZS-Study
3 classes} & \textbf{ZS-Task
known people} & \textbf{Nominal both
5 classes} & \textbf{Strict common both
2 classes} \\
\midrule\endhead
\midrule\multicolumn{5}{r}{\textit{Continued on next page}}\\\endfoot
\bottomrule\endlastfoot
ST19 RAW 16k & 25.65 & 55.15 & 5.18 & 8.59 \\
ST19 RAW 64k & 26.34 & 58.97 & 5.92 & 11.56 \\
ST40 RAW 64k & 12.91 & 59.63 & 12.17 & 13.74 \\
FM RAW 256k & 10.50 & 61.92 & 11.98 & 14.88 \\
FM LATE 256k & 5.21 & 66.00 & 11.28 & 18.55 \\
Bridge3 fixed-text RAW 64k & 28.15 & 59.03 & 7.46 & 8.38 \\
Bridge3 fixed-text LATE 64k & 38.07 & 63.66 & 11.95 & 20.24 \\
CLIP-X RAW 64k & 6.38 & 21.81 & 9.45 & 18.64 \\
CLIP-X LATE 64k & 15.60 & 8.54 & 10.81 & 20.23 \\
\end{longtable}}
\vspace{-0.6em}\noindent{\footnotesize Participant-/class-equal recall (\%). All predictions retain K = 10. The strict common subset contains thinking and meditation from one study, not two independent study replications. The nominal driving-task novelty depends on each model's training exposure.}\par\medskip

\FloatBarrier
\subsection*{S5. Additional developmental evidence}

The principal nested-addition and calibration results appear in the main text (Sections 3.4--3.5; Figures~\ref{fig:study-addition} and~\ref{fig:fewshot-calibration}). This section retains the complementary sampler and shared-task analyses.

The earlier sampling/readout program compared semantic--study balancing, within-cell balancing and global task balancing. Its selected global-task-balanced recipe (historical identifier B2) had a local endpoint of 25.25 \ensuremath{\pm} 0.68\% at 8 s and 28.24 \ensuremath{\pm} 1.40\% at 64 s under an equal-source summary of native 5/6-way problems. These values describe developmental selection. The global pooled-11 path had different averaging and candidate competition, and should not be compared numerically as though it were the same endpoint.

\begin{figure}[!htbp]
\centering
\includegraphics[width=\linewidth,height=0.55\textheight,keepaspectratio]{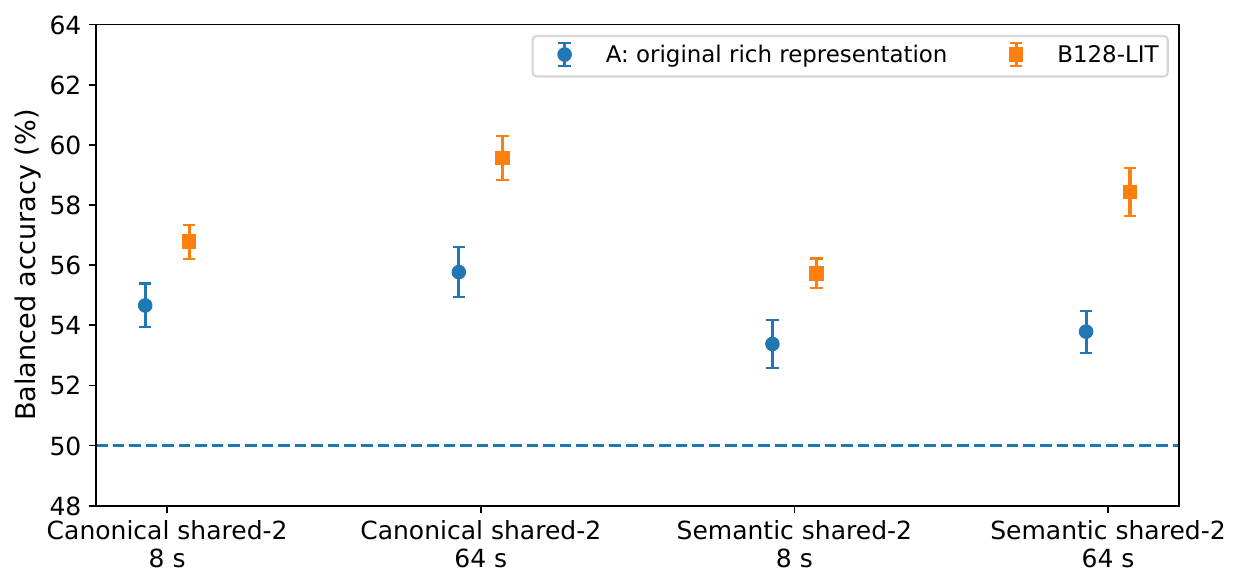}
\caption{\textbf{Shared-task development comparison.} The original rich-text representation (A) and the 128-dimensional literature-derived representation (B128-LIT) are evaluated on two task categories represented in both D1 sources. Pooled scores average participants rather than equal source means. These historical 64-second scores aggregate 29 overlapping eight-second windows.}
\label{fig:shared-task-development}
\end{figure}

B128-LIT's matched shared-task tests restricted both D1 sources to two task categories present in each source. Canonical-readout means increased from 54.66\% to 56.78\% at 8 s and from 55.77\% to 59.56\% in the historical 64-s protocol. The pooled estimator averages participant-level scores, not the two source means equally; eligible counts can differ by duration. The grouping and D1 data were inspected during development (Figure~\ref{fig:shared-task-development}).

\clearpage

\FloatBarrier
\subsection*{S6. Semantic partitions and their separate estimators}

\Needspace{105pt}
\par\medskip\noindent{\small\textbf{Table S10. Preserved semantic partitions of the original twenty labels.}}
{\fontsize{8.8}{11.2}\selectfont\setlength{\tabcolsep}{3pt}\renewcommand{\arraystretch}{1.18}
\begin{longtable}{>{\raggedright\arraybackslash}p{32.07pt}>{\raggedright\arraybackslash}p{137.90pt}>{\raggedright\arraybackslash}p{137.90pt}>{\raggedright\arraybackslash}p{137.90pt}}
\toprule
\textbf{Group} & \textbf{F15} & \textbf{F13} & \textbf{F8} \\
\midrule\endfirsthead
\multicolumn{4}{l}{\textit{Table S10 (continued)}}\\
\toprule
\textbf{Group} & \textbf{F15} & \textbf{F13} & \textbf{F8} \\
\midrule\endhead
\midrule\multicolumn{4}{r}{\textit{Continued on next page}}\\\endfoot
\bottomrule\endlastfoot
1 & [0,1,6] Visual object categorization & [0,1,6] Visual object categorization & [0,1,6] Visual object categorization \\
2 & [2] Rest & [2] Rest & [2,12,13] Internal-state domain \\
3 & [3] Animal-word lexical detection & [3,8,19] Visual linguistic/semantic processing & [3,8,19] Visual linguistic processing \\
4 & [4] Flanker & [4,7] Visuospatial selection & [4,7] Visuospatial selection \\
5 & [5,11] Auditory standard/deviant processing & [5,11] Auditory standard/deviant processing & [5,9,11] Oddball domain \\
6 & [7] Visual search & [9] Active visual oddball & [10,17] Visual episodic recognition \\
7 & [8,19] Linguistic semantics & [10,17] Visual episodic recognition & [14,15,16] Driving context \\
8 & [9] Active visual oddball & [12] Active thinking & [18] Visual working memory \\
9 & [10,17] Visual episodic recognition & [13] Breath meditation & --- \\
10 & [12] Active thinking & [14] Manual driving & --- \\
11 & [13] Breath meditation & [15] Full assistance available & --- \\
12 & [14] Manual driving & [16] Speed assistance available & --- \\
13 & [15] Full assistance available & [18] Visual working memory & --- \\
14 & [16] Speed assistance available & --- & --- \\
15 & [18] Visual working memory & --- & --- \\
\end{longtable}}
\vspace{-0.6em}\noindent{\footnotesize The original index order is the left-to-right column order in Figure~\ref{fig:task-source-matrix}, beginning at zero. Rows align group positions for compact display, not one-to-one equivalence between partitions. Auditory activity/passivity, modality and response distinctions are intentionally lost in some groupings.}\par\medskip

\Needspace{105pt}
\par\medskip\noindent{\small\textbf{Table S11. Archived merged-prototype scores, retained separately from the relabeling diagnostic.}}
{\fontsize{8.8}{11.2}\selectfont\setlength{\tabcolsep}{3pt}\renewcommand{\arraystretch}{1.18}
\begin{longtable}{>{\raggedright\arraybackslash}p{161.35pt}>{\raggedright\arraybackslash}p{69.60pt}>{\raggedright\arraybackslash}p{69.60pt}>{\raggedright\arraybackslash}p{69.60pt}>{\raggedright\arraybackslash}p{69.60pt}}
\toprule
\textbf{Model} & \textbf{Original20} & \textbf{F15} & \textbf{F13} & \textbf{F8} \\
\midrule\endfirsthead
\multicolumn{5}{l}{\textit{Table S11 (continued)}}\\
\toprule
\textbf{Model} & \textbf{Original20} & \textbf{F15} & \textbf{F13} & \textbf{F8} \\
\midrule\endhead
\midrule\multicolumn{5}{r}{\textit{Continued on next page}}\\\endfoot
\bottomrule\endlastfoot
ST19 RAW 16k & 13.38 & 15.57 & 15.96 & 24.13 \\
ST19 RAW 64k & 13.26 & 14.65 & 15.88 & 24.58 \\
ST40 RAW 64k & 10.60 & 12.94 & 16.99 & 26.43 \\
FM RAW 256k & 10.14 & 12.07 & 16.10 & 25.01 \\
FM LATE 256k & 11.06 & 10.35 & 16.73 & 28.85 \\
Bridge3 fixed-text RAW 64k & 13.35 & 15.30 & 16.63 & 25.56 \\
Bridge3 fixed-text LATE 64k & 14.59 & 14.06 & 19.49 & 29.30 \\
CLIP-X RAW 64k & 8.40 & 13.13 & 14.74 & 20.68 \\
CLIP-X LATE 64k & 9.80 & 15.25 & 16.59 & 22.68 \\
\end{longtable}}
\vspace{-0.6em}\noindent{\footnotesize Eight-second centroid BA (\%). Uniform chances: 5\%, 6.67\%, 7.69\%, 12.5\%. All columns are post-score exploratory evaluations. Fused concepts, their original member classes and participants are successively equal-weighted. Differences across columns are not matched measures of unchanged task difficulty.}\par\medskip

\begin{figure}[!htbp]
\centering
\includegraphics[width=\linewidth,height=0.55\textheight,keepaspectratio]{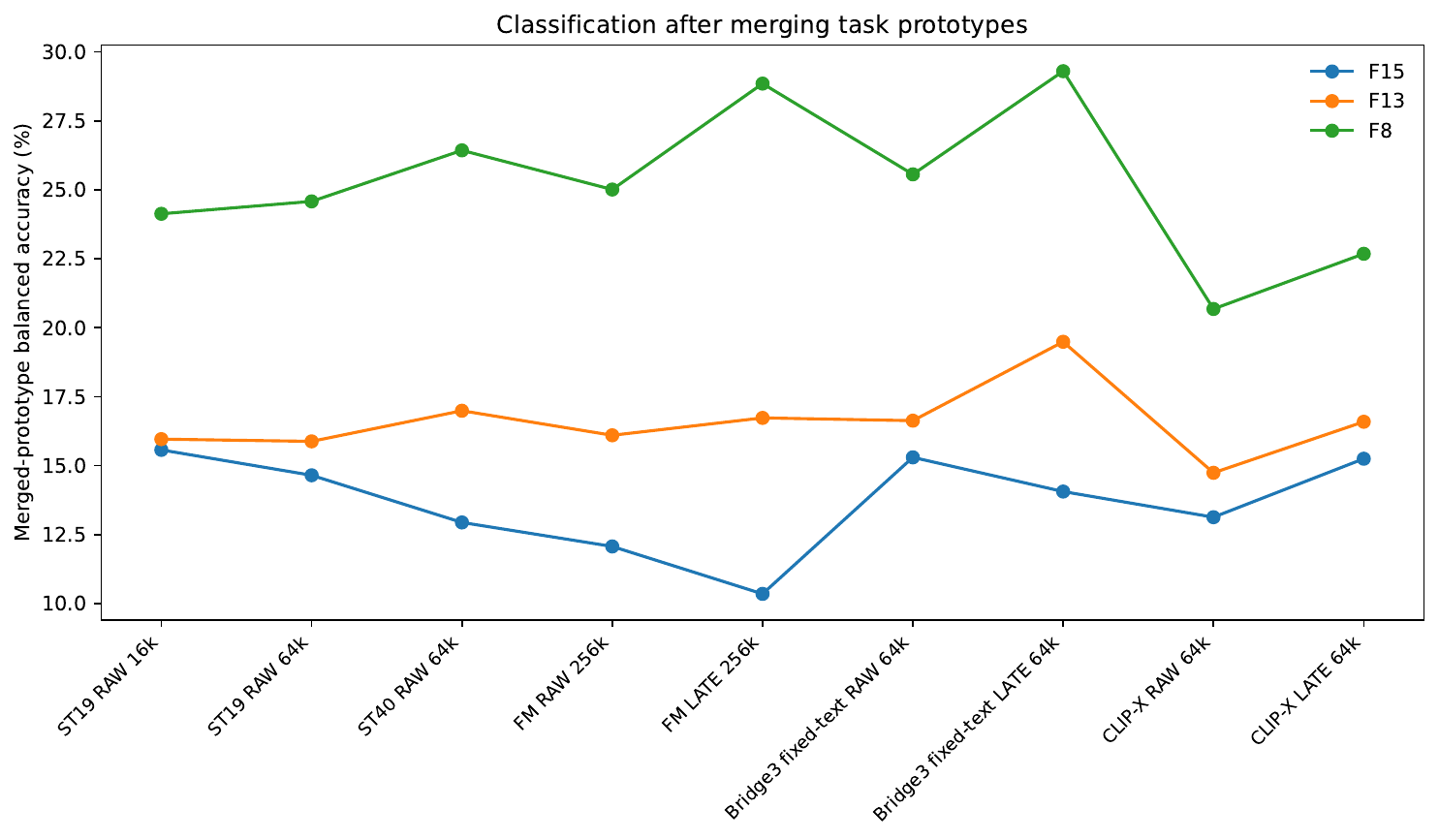}
\caption{\textbf{Merged-prototype classification at three semantic resolutions.} F15, F13 and F8 contain 15, 13 and 8 candidates, with chance levels 6.67\%, 7.69\% and 12.5\%. F15 and F13 retain the thinking/meditation and driving distinctions; F8 merges them. These scores use merged prototypes rather than relabeling the fixed original predictions.}
\label{fig:semantic-fusion}
\end{figure}

Merged-prototype prediction uses a normalized sum of original prototypes. Fixed-argmax relabeling leaves the original prediction intact and changes only its mapped label. Log-mean-exp pooling is a third estimator. Their scores need not agree, and the highest post-score variant does not replace the frozen readout. Original fusion results and temperature sensitivity remain in the archive.

\Needspace{105pt}
\par\medskip\noindent{\small\textbf{Table S12. Source-incidence-preserving semantic partition diagnostic, all conditions.}}
{\fontsize{8.8}{11.2}\selectfont\setlength{\tabcolsep}{3pt}\renewcommand{\arraystretch}{1.18}
\begin{longtable}{>{\raggedright\arraybackslash}p{141.10pt}>{\raggedright\arraybackslash}p{102.62pt}>{\raggedright\arraybackslash}p{102.62pt}>{\raggedright\arraybackslash}p{99.41pt}}
\toprule
\textbf{Condition} & \textbf{F15 observed / null (\%)} & \textbf{F13 observed / null (\%)} & \textbf{F8 observed / null (\%)} \\
\midrule\endfirsthead
\multicolumn{4}{l}{\textit{Table S12 (continued)}}\\
\toprule
\textbf{Condition} & \textbf{F15 observed / null (\%)} & \textbf{F13 observed / null (\%)} & \textbf{F8 observed / null (\%)} \\
\midrule\endhead
\midrule\multicolumn{4}{r}{\textit{Continued on next page}}\\\endfoot
\bottomrule\endlastfoot
ST19 RAW 16k & 17.31 / 13.71 & 17.32 / 14.15 & 24.60 / 21.71 \\
ST19 RAW 64k & 16.40 / 13.51 & 17.29 / 14.11 & 25.27 / 22.24 \\
ST40 RAW 64k & 13.91 / 11.89 & 16.90 / 13.44 & 25.52 / 20.95 \\
FM RAW 256k & 13.47 / 11.62 & 16.72 / 13.22 & 25.16 / 20.48 \\
FM LATE 256k & 13.02 / 11.55 & 16.50 / 12.85 & 28.06 / 21.01 \\
Bridge3 fixed-text RAW 64k & 16.92 / 13.74 & 17.62 / 14.35 & 25.79 / 22.68 \\
Bridge3 fixed-text LATE 64k & 16.26 / 14.07 & 19.89 / 15.71 & 28.14 / 24.23 \\
CLIP-X RAW 64k & 12.78 / 10.62 & 13.62 / 10.97 & 20.26 / 18.30 \\
CLIP-X LATE 64k & 14.09 / 11.67 & 14.65 / 11.36 & 22.04 / 17.83 \\
\end{longtable}}
\vspace{-0.6em}\noindent{\footnotesize Fixed original 20-class combined-bank argmax labels are relabeled. Each null preserves group size and each group's exact original dataset-ID composition. 10,000 draws per partition; distinct partitions encountered: 720 / 3,848 / 2,145. This source-conditioned diagnostic is not the merged-prototype evaluation. Complete tail fractions and size-only controls are in R1\_source\_conditioned\_null.csv; they are post-score descriptive frequencies, not confirmatory p-values.}\par\medskip

The source-conditioned null preserves original dataset identity, rather than treating the shared auditory/meditation facility as a single dataset. This is an exact condition on the benchmark's source-incidence structure, not a complete nuisance model. Its constraints also leave some parts of the F8 grouping unchanged across draws. The residual measures what remains after this particular source-incidence constraint, not generalization to independent laboratories.

\FloatBarrier
\subsection*{S7. Analysis provenance and reproducibility boundaries}

The statistical and interpretive additions to the archived reserved evaluation comprise joint-participant uncertainty from saved D2 scores, nine-condition rank and weighting accounting, class concentration and standardized precision, and source-incidence-preserving partition diagnostics from saved twenty-class confusion matrices. Training indices and loss code were inspected without modifying model files. All these additions are post-score; the original D2 predictions and historical per-class bootstrap summaries were not overwritten.

The text-only geometry analysis uses the frozen task bank and does not read EEG samples or model prediction arrays. Its complete cosine matrix and plotting inputs are included. The historical data-addition and calibration figures are redrawn from the same archived numeric values, without new training or inference.

The record distinguishes three kinds of reproducibility. The included aggregate tables support checking numerical identities, rank disagreement and summary plots. The mathematical utilities and synthetic tests support inspection of metric and partition definitions. Full reproduction of the trained models and participant resampling additionally requires the source-specific data versions, accepted-window masks, participant linkage, exact per-checkpoint prototype banks and original score arrays. The present release supplies the first two levels, not complete end-to-end reproduction.

Outstanding limitations include unresolved entries in the physical-input record, incomplete checkpoint-specific consolidation of bank membership and realized exposures, cross-source participant/family duplication, and incomplete external-evaluation accounting for the 240-checkpoint randomized training sweep. These gaps limit both exact reproduction and claims of independent cross-study validation.

\FloatBarrier
\subsection*{S8. Original-space task similarities and projection sensitivity}

The 44 numbered points in Figure~\ref{fig:taskspace} are the primary-training functional-task entries recorded for the Bridge3 primary training mixture. They are not the full union of tasks used across every historical experiment, and no development or reserved-evaluation points are included in this display. The nearest-neighbor table below retains all entries, not just interpretable examples. Its neighbors describe the geometry of the text bank. The very high similarity between the word- and phoneme-perception entries (0.993), for example, should not be read as evidence that the corresponding operations are neurally indistinguishable.

\begin{figure}[!htbp]
\centering
\includegraphics[width=\linewidth,height=0.68\textheight,keepaspectratio]{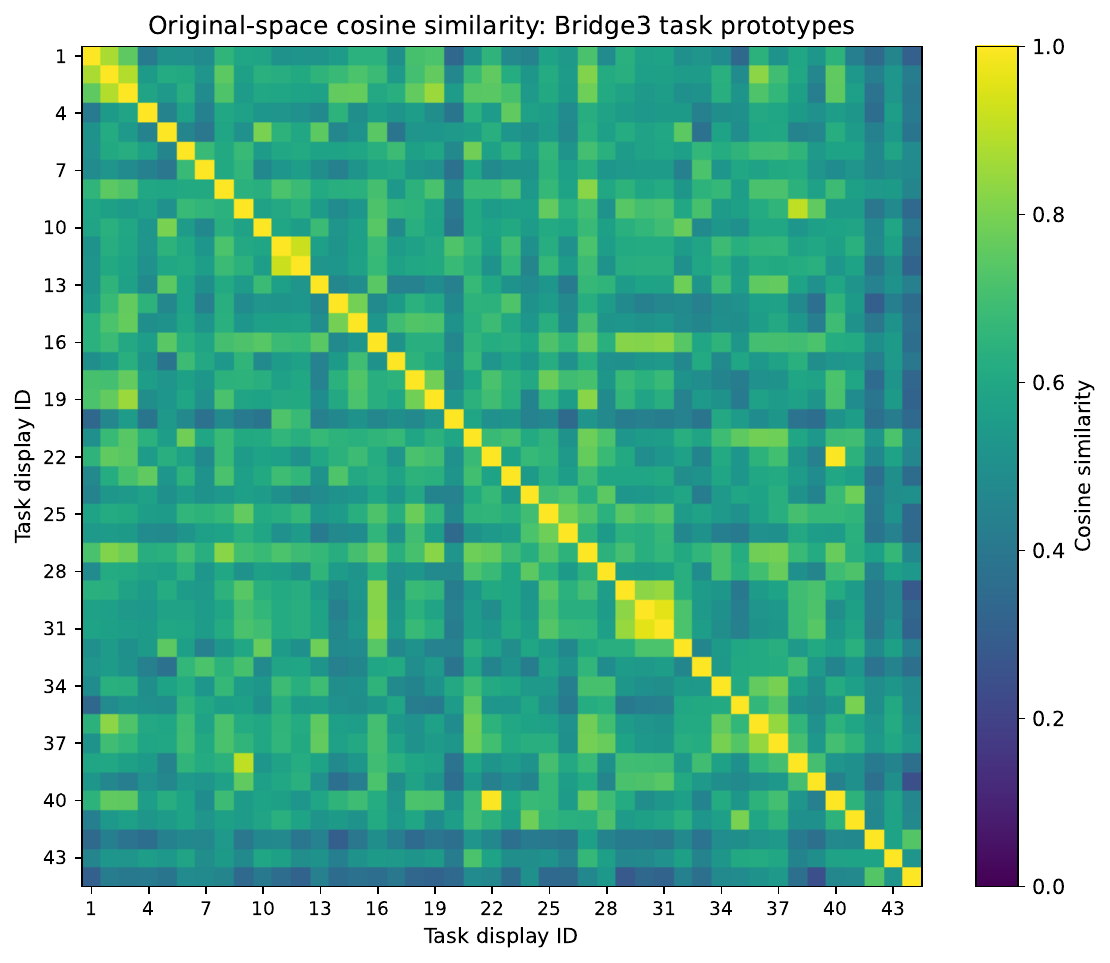}
\caption{\textbf{Cosine similarity among Bridge3 task prototypes.} Pairwise similarities of the 44 centroids in their original 384-dimensional space. The order follows the archived functional-task indices and uses the display identifiers of Figure~\ref{fig:taskspace} and Table S13. The numeric matrix is included in \texttt{anc/data/task\_space/}.}
\label{fig:taskspace-sim}
\end{figure}

\par\medskip\noindent{\small\textbf{Table S13. All displayed tasks and their nearest other task in the original space.}}
{\fontsize{8.8}{11.2}\selectfont\setlength{\tabcolsep}{4pt}\renewcommand{\arraystretch}{1.04}
\begin{longtable}{>{\raggedright\arraybackslash}p{22pt}>{\raggedright\arraybackslash}p{188pt}>{\raggedright\arraybackslash}p{188pt}>{\raggedleft\arraybackslash}p{32pt}}
\toprule
\textbf{ID} & \textbf{Task} & \textbf{Nearest other task} & \textbf{Cosine} \\
\midrule\endfirsthead
\multicolumn{4}{l}{\textit{Table S13 (continued)}}\\
\toprule
\textbf{ID} & \textbf{Task} & \textbf{Nearest other task} & \textbf{Cosine} \\
\midrule\endhead
\midrule\multicolumn{4}{r}{\textit{Continued on next page}}\\\endfoot
\bottomrule\endlastfoot
1 & Auditory local/global regularities & 2: Active auditory oddball & 0.873 \\
2 & Active auditory oddball & 3: Passive auditory oddball & 0.884 \\
3 & Passive auditory oddball & 2: Active auditory oddball & 0.884 \\
4 & Autobiographical recall & 23: Rest & 0.756 \\
5 & Compositional multitasking & 10: Mental arithmetic & 0.800 \\
6 & Face perception & 21: Passive visual stimulation & 0.783 \\
7 & Food valuation & 33: Visual animal go/no-go categorization & 0.718 \\
8 & Go/no-go inhibition & 27: Simple reaction time & 0.822 \\
9 & Intentional visual-memory encoding & 38: Visual search: incidental encoding & 0.903 \\
10 & Mental arithmetic & 5: Compositional multitasking & 0.800 \\
11 & Motor execution & 12: Motor imagery & 0.918 \\
12 & Motor imagery & 11: Motor execution & 0.918 \\
13 & Multi-source interference task & 32: Verbal working memory & 0.778 \\
14 & Music imagery & 15: Music listening & 0.792 \\
15 & Music listening & 14: Music imagery & 0.792 \\
16 & N-back & 31: Verbal memory: retention & 0.827 \\
17 & Naturalistic emotion viewing & 15: Music listening & 0.694 \\
18 & Naturalistic speech listening & 19: Passive auditory cue & 0.786 \\
19 & Passive auditory cue & 3: Passive auditory oddball & 0.849 \\
20 & Passive movement & 11: Motor execution & 0.723 \\
21 & Passive visual stimulation & 36: Active visual oddball & 0.787 \\
22 & Phoneme perception & 40: Word perception & 0.993 \\
23 & Rest & 4: Autobiographical recall & 0.756 \\
24 & Sentence reading & 41: Word semantic choice & 0.780 \\
25 & Sentence reading: control & 26: Sentence reading: prediction & 0.779 \\
26 & Sentence reading: prediction & 25: Sentence reading: control & 0.779 \\
27 & Simple reaction time & 8: Go/no-go inhibition & 0.822 \\
28 & Stroop & 37: Visual search & 0.766 \\
29 & Verbal Sternberg working memory & 31: Verbal memory: retention & 0.844 \\
30 & Verbal memory: manipulation & 31: Verbal memory: retention & 0.960 \\
31 & Verbal memory: retention & 30: Verbal memory: manipulation & 0.960 \\
32 & Verbal working memory & 13: Multi-source interference task & 0.778 \\
33 & Visual animal go/no-go categorization & 7: Food valuation & 0.718 \\
34 & Visual flanker & 37: Visual search & 0.793 \\
35 & Visual object choice & 41: Word semantic choice & 0.798 \\
36 & Active visual oddball & 37: Visual search & 0.843 \\
37 & Visual search & 36: Active visual oddball & 0.843 \\
38 & Visual search: incidental encoding & 9: Intentional visual-memory encoding & 0.903 \\
39 & Visuospatial sequence memory & 9: Intentional visual-memory encoding & 0.756 \\
40 & Word perception & 22: Phoneme perception & 0.993 \\
41 & Word semantic choice & 35: Visual object choice & 0.798 \\
42 & Three-armed bandit reward learning & 44: Risk/ambiguity decision & 0.731 \\
43 & Temporal production/perception & 21: Passive visual stimulation & 0.716 \\
44 & Risk/ambiguity decision & 42: Three-armed bandit reward learning & 0.731 \\
\end{longtable}}
\noindent{\footnotesize Similarities exclude the self-match. Display labels expand working-memory and interference-task shorthand; original source names and functional-task identifiers are preserved in the released CSV.}

\par\medskip\noindent{\small\textbf{Table S14. Sensitivity of local neighborhood preservation in the text-space map.}}
\begin{center}
\begin{tabular}{rrrr}
\toprule
\textbf{Perplexity} & \textbf{Seed 42} & \textbf{Seed 43} & \textbf{Seed 44} \\
\midrule
5 & 0.899 & 0.896 & 0.910 \\
10 & 0.903 & 0.952 & 0.929 \\
20 & 0.725 & 0.538 & 0.548 \\
\bottomrule
\end{tabular}
\end{center}
\noindent{\footnotesize Five-neighbor trustworthiness measures local preservation, not decoding accuracy. The primary setting (perplexity 10, seed 42) was fixed before viewing results. Software: scikit-learn 1.8.0, NumPy 2.3.5, Matplotlib 3.10.8.}

\FloatBarrier
\subsection*{S9. Source inventories and dataset citations}

The late-stage source lists were recovered from the accepted-record indices, whose SHA-256 hashes are supplied in \texttt{anc/configs/dataset\_catalog\_R7.json}. Each column below denotes source membership only. The same accession can contribute a training task and a distinct held-out task, as for ds002680 and ds005189. The lists do not authorize use of all source participants or intervals for training.

\par\medskip\noindent{\small\textbf{Table S15. Primary training-source membership.}}
{\fontsize{8.6}{10.8}\selectfont\setlength{\tabcolsep}{3pt}\renewcommand{\arraystretch}{1.15}
\begin{longtable}{p{100pt}p{105pt}p{95pt}p{120pt}}
\toprule
\textbf{Accession} & \textbf{ST19-MATCHED} & \textbf{ST40 / FM} & \textbf{Bridge3 / CLIP-X} \\
\midrule\endfirsthead
\multicolumn{4}{l}{\textit{Table S15 (continued)}}\\
\toprule
\textbf{Accession} & \textbf{ST19-MATCHED} & \textbf{ST40 / FM} & \textbf{Bridge3 / CLIP-X} \\
\midrule\endhead
\bottomrule\endfoot
ds001810 & -- & yes & -- \\
ds002218 & -- & yes & -- \\
ds002680 & yes & yes & yes \\
ds002718 & -- & yes & -- \\
ds002721 & yes & yes & yes \\
ds003420 & -- & yes & -- \\
ds003458 & -- & yes & yes \\
ds003655 & yes & yes & yes \\
ds003690 & yes & yes & yes \\
ds003751 & yes & yes & yes \\
ds003753 & -- & yes & -- \\
ds003766 & yes & yes & yes \\
ds003774 & yes & yes & yes \\
ds004043 & -- & yes & -- \\
ds004148 & yes & yes & yes \\
ds004166 & -- & yes & -- \\
ds004200 & -- & yes & yes \\
ds004362 & yes & yes & yes \\
ds004408 & yes & yes & yes \\
ds004502 & -- & yes & -- \\
ds004574 & yes & yes & yes \\
ds004602 & yes & yes & yes \\
ds004621 & yes & yes & yes \\
ds004626 & -- & yes & -- \\
ds004883 & yes & yes & yes \\
ds004917 & -- & yes & yes \\
ds004942 & yes & yes & yes \\
ds005089 & -- & yes & -- \\
ds005095 & yes & yes & yes \\
ds005189 & yes & yes & yes \\
ds005383 & -- & yes & -- \\
ds005403 & -- & yes & -- \\
ds005429 & yes & yes & yes \\
ds005540 & -- & yes & -- \\
ds006018 & yes & yes & yes \\
ds006104 & yes & yes & yes \\
ds006466 & yes & yes & yes \\
ds006480 & yes & yes & yes \\
ds006848 & yes & yes & yes \\
ds007012 & yes & yes & yes \\
ds007052 & yes & yes & yes \\
ds007056 & yes & yes & yes \\
ds007058 & -- & yes & -- \\
ds007069 & yes & yes & yes \\
ds007096 & yes & yes & yes \\
ds007137 & yes & yes & yes \\
ds007169 & yes & yes & yes \\
ds007262 & yes & yes & yes \\
ds007554 & yes & yes & yes \\
ds008701 & yes & yes & yes \\
nm000109 & yes & yes & yes \\
nm000136 & -- & yes & -- \\
nm000180 & yes & yes & yes \\
nm000193 & -- & yes & -- \\
nm000203 & -- & yes & -- \\
nm000228 & yes & yes & yes \\
nm000275 & -- & yes & -- \\
nm000278 & yes & yes & yes \\
nm000279 & yes & yes & yes \\
\end{longtable}}
\noindent{\footnotesize Column counts are 38, 59 and 41 source accessions. They are not independent-study counts. Native four-second M3CV data (nm000166) are auxiliary and excluded from these primary counts. Download links and all 73 training/evaluation accessions are in the accompanying CSV.}
\par\medskip\noindent{\small\textbf{Table S16. Evaluation and auxiliary source attribution.}}
{\fontsize{8.6}{10.8}\selectfont\setlength{\tabcolsep}{3pt}\renewcommand{\arraystretch}{1.15}
\begin{longtable}{p{55pt}p{57pt}p{258pt}p{50pt}}
\toprule
\textbf{Source} & \textbf{Role} & \textbf{Dataset / associated publication} & \textbf{Cited distribution} \\
\midrule\endfirsthead
\multicolumn{4}{l}{\textit{Table S16 (continued)}}\\
\toprule
\textbf{Source} & \textbf{Role} & \textbf{Dataset / associated publication} & \textbf{Cited distribution} \\
\midrule\endhead
\bottomrule\endfoot
ds007172 & D1 & EEG-Asymmetries Dataset \citep{data_ds007172} \citep{reinke2025asymmetries} & 1.0.0 \\
erp-core & D1 & ERP CORE \citep{kappenman2021} & Project data distribution \\
ds004117 & D1-ST & Sternberg Working Memory \citep{data_ds004117} \citep{onton2005} & 1.0.1 \\
ds004860 & D1-ST & Investigating the cognitive conflict triggered by moral judgment of accidental harm: an event-related potentials study \citep{data_ds004860} & 1.0.0 \\
ds006142 & D1-ST & Essex EEG Movie Memory dataset \citep{data_ds006142} & 1.0.2 \\
nm000219 & D1-ST & BNCI 2020-002 Attention Shift (Covert Spatial Attention) dataset \citep{data_nm000219} \citep{reichert2020} & NEMAR record \\
ds006446 & D1-ST & Cueing the future to reduce temporal discounting \citep{data_ds006446} & 1.0.0 \\
ds004579 & D1-ST & Interval Timing Task \citep{data_ds004579} & 1.0.0 \\
ds002680 & D2 & Go-nogo categorization and detection task \citep{data_ds002680} & 1.2.0 \\
ds003061 & D2 & EEG data from an auditory oddball task \citep{data_ds003061} & 1.1.0 \\
ds003969 & D2 & Meditation vs thinking task \citep{data_ds003969} \citep{braboszcz2017} & 1.0.0 \\
ds004657 & D2 & Driving with Autonomous Aids \citep{data_ds004657} & 1.0.3 \\
ds005189 & D2 & Search Superiority Recollection Familiarity \citep{data_ds005189} & 1.0.2 \\
ds006979 & D2 & Examining Perceptual Grouping on Stages of Processing in Visual Working Memory: An ERP Study \citep{data_ds006979} & 1.0.1 \\
ds008711 & D2 & RSVP with flankers - sentences with semantic and syntactic violations \citep{data_ds008711} & 1.0.0 \\
nm000166 & auxiliary4s & M3CV: Multi-subject, Multi-session, Multi-task EEG Database \citep{data_nm000166} & NEMAR record \\
\end{longtable}}
\noindent{\footnotesize Distribution versions credit identifiable public records. Their bibliographic resolution is distinct from confirming every original raw-file hash and preprocessing step. ds006979 provider metadata lists Ramzaoui and Beck, while its DOI creator field listed only Ramzaoui at the audit; both provider-listed authors are credited. The ds004579 record did not supply a verified publication year, so the citation uses n.d.}
\par\medskip\noindent{\small\textbf{Table S17. Historical self-supervised pilot sources.}}
{\fontsize{8.6}{10.8}\selectfont\setlength{\tabcolsep}{3pt}\renewcommand{\arraystretch}{1.15}
\begin{longtable}{p{100pt}p{190pt}p{130pt}}
\toprule
\textbf{Source family} & \textbf{Source identifier} & \textbf{Distribution citation} \\
\midrule\endfirsthead
\multicolumn{3}{l}{\textit{Table S17 (continued)}}\\
\toprule
\textbf{Source family} & \textbf{Source identifier} & \textbf{Distribution citation} \\
\midrule\endhead
\bottomrule\endfoot
PEERS / Courier & ds004395, ds004706 & \citep{data_ds004395} \citep{data_ds004706} \\
SparrKULee & KU Leuven RDR & \citep{data_sparrkulee} \\
THINGS-EEG1 & ds003825 & \citep{data_ds003825} \\
EEG Moments & ds008257 & \citep{data_ds008257} \\
Language ERP & Zenodo 6992085 & \citep{data_language_erp} \\
\end{longtable}}
\noindent{\footnotesize The pilot grouped PEERS and Courier into one source family. These six distributions comprise five named families. Their historical accepted intervals are not specified by the late-stage primary-source table.}


\end{document}